\documentclass[longauth]{aa}
\usepackage{graphicx}
\usepackage{gensymb}
\usepackage{txfonts}
\usepackage[flushleft]{threeparttable}
\renewcommand{\arraystretch}{1.1}
\usepackage[hidelinks,colorlinks=true,linkcolor=blue,citecolor=blue]{hyperref}
\usepackage{longtable,tabu}
\usepackage{lipsum}
\usepackage{adjustbox}
\usepackage{stfloats}
\usepackage{scrextend}
\usepackage{subcaption}
\usepackage{amssymb}
\usepackage{caption}
\usepackage{amsmath}
\usepackage{amssymb}
\usepackage{mathrsfs}
\usepackage[letterspace=150]{microtype}
\usepackage{kantlipsum}
\def \teff {$T_{\mathrm{eff}}$}
\newcommand{\juliet}{{\sc \tt juliet}\xspace}
\newcommand{\celerite}{{\sc \tt celerite}\xspace}
\newcommand{\batman}{{\sc \tt batman}\xspace}

\newcommand{\radvel}{{\sc \tt radvel}\xspace}
\newcommand{\dynesty}{{\sc \tt dynesty}\xspace}
\newcommand{\multinest}{{\sc \tt MultiNest}\xspace}

\newcommand{\pymultinest}{{\sc \tt PyMultiNest}\xspace}

\newcommand{\tpfplotter}{{\sc \tt tpfplotter}\xspace}
\newcommand{\ldcu}{{\sc \tt LDCU}\xspace}
\newcommand{\iSpec}{{\sc \tt iSpec}\xspace}
\newcommand{\scikit}{{\sc \tt scikit-learn}\xspace}
\newcommand{\SpecMatch}{{\sc \tt SpecMatch-Emp}\xspace}

\usepackage{array}
\usepackage{amssymb}
\usepackage{mathrsfs}
\usepackage{placeins}
\newcolumntype{T}{>{\tiny}c} 
\providecommand{\bjdtdb}{\ensuremath{\rm {BJD_{TDB}}}}
\providecommand{\feh}{\ensuremath{\left[{\rm Fe}/{\rm H}\right]}}
\providecommand{\teff}{\ensuremath{T_{\rm eff}}}

\providecommand{\fave}{\langle F \rangle}

\newcommand{\logg}{\mbox{$\log g_*$}\xspace}
\newcommand{\vsini}{\mbox{$v \sin \it{i}_{*}$}\xspace}
\newcommand{\kms}{\mbox{km\,s$^{-1}$}\xspace}

\newcommand{\mplanet}{\mbox{$M_{\rm p}$}\xspace}

\newcommand{\mjup}{\mbox{$\mathrm{M_{\rm Jup}}$}\xspace}
\newcommand{\rjup}{\mbox{$\mathrm{R_{\rm Jup}}$}\xspace}

\newcommand{\rearth}{\mbox{$\mathrm{R_{\rm \oplus}}$}\xspace}
\newcommand{\mstar}{\mbox{$M_{*}$}\xspace}
\newcommand{\rstar}{\mbox{$R_{*}$}\xspace}

\newcommand{\msol}{\mbox{$\mathrm{M_\odot}$}\xspace}
\newcommand{\rsol}{\mbox{$\mathrm{R_\odot}$}\xspace}

\renewcommand{\arraystretch}{1.3}

\begin{document} 

\title{Three new brown dwarfs and a massive hot Jupiter revealed by TESS around early-type stars\thanks{The photometric and radial velocity data in this work are only available in electronic form at the CDS via anonymous ftp to cdsarc.u-strasbg.fr (130.79.128.5) or via \url{http://cdsarc.u-strasbg.fr/viz-bin/qcat?J/A+A/}}}

\titlerunning{}
\authorrunning{A. Psaridi, et al.}
   \author{Angelica Psaridi
   \inst{\ref{inst-geneva}}\fnmsep\thanks{\email{angeliki.psaridi@unige.ch}}
     \and Fran\c{c}ois Bouchy \inst{\ref{inst-geneva}}
     \and Monika Lendl \inst{\ref{inst-geneva}}
     \and Nolan Grieves \inst{\ref{inst-geneva}}
     \and Keivan G. Stassun \inst{\ref{inst-vandy}}
    \and Theron Carmichael \inst{\ref{inst-Edinburgh}}
     \and Samuel Gill \inst{\ref{inst-war},\ref{inst-warexo}}
     \and Pablo A. Pe\~{n}a Rojas \inst{\ref{inst-chile}}
    \and Tianjun Gan \inst{\ref{inst-Tsinghua}}
     \and Avi Shporer \inst{\ref{inst-mit}}
     \and Allyson Bieryla \inst{\ref{inst-harvard}}
    \and Rafael Brahm \inst{\ref{inst-adolfo},\ref{inst-mifa}}
    \and Jessie L. Christiansen \inst{\ref{inst-caltech}}
    \and Ian J. M. Crossfield \inst{\ref{inst-Kansas}}
     \and Franck Galland \inst{\ref{inst-geneva},\ref{inst-grenoble}}
     \and Matthew J. Hooton \inst{\ref{inst-Cavendish}}
     \and Jon M. Jenkins \inst{\ref{inst-nasaames}}
     \and James S. Jenkins \inst{\ref{inst-chile},\ref{inst-chiletechologia}}
     \and David W. Latham \inst{\ref{inst-cfa}}
     \and Michael B. Lund \inst{\ref{inst-caltech}}
     \and Joseph E. Rodriguez \inst{\ref{inst-michigan}}
     \and Eric B. Ting \inst{\ref{inst-nasaames}}
     \and Stéphane Udry \inst{\ref{inst-geneva}}
     \and Sol{\`e}ne Ulmer-Moll \inst{\ref{inst-geneva}}
     \and Robert A. Wittenmyer \inst{\ref{inst-queensland}}
     \and Yanzhe Zhang \inst{\ref{inst-Kansas}}
    \and George Zhou \inst{\ref{inst-cfa}}
     \and Brett Addison \inst{\ref{inst-queensland}}
      \and Marion Cointepas \inst{\ref{inst-geneva},\ref{inst-grenoble}}
     \and Karen A.\ Collins \inst{\ref{inst-cambridge}}
     \and Kevin I.\ Collins \inst{\ref{inst-GeorgeMason}}
      \and Adrien Deline \inst{\ref{inst-geneva}}
      \and Courtney D. Dressing \inst{\ref{inst-berkeley}}
     \and Phil Evans \inst{\ref{inst-elsauce}}
     \and Steven Giacalone \inst{\ref{inst-berkeley}}
     \and Alexis Heitzmann \inst{\ref{inst-queensland}}
     \and Ismael Mireles \inst{\ref{inst-newmexico}}
     \and Dany Mounzer \inst{\ref{inst-geneva}}
     \and Jon Otegi \inst{\ref{inst-geneva},\ref{inst-zurich}}
     \and Don J. Radford \inst{\ref{inst-brierfield}}
     \and Alexander Rudat \inst{\ref{inst-mit}}
     \and Joshua E. Schlieder \inst{\ref{inst-nasa}}
     \and Richard P. Schwarz \inst{\ref{inst-Patashnick}}
     \and Gregor Srdoc \inst{\ref{inst-Kotizarovci}}
     \and Chris Stockdale \inst{\ref{inst-hazelwood}}
     \and Olga Suarez \inst{\ref{inst-cotedazur}}
     \and Duncan J. Wright \inst{\ref{inst-queensland}}
     \and Yinan Zhao \inst{\ref{inst-geneva}}}
 \institute{
   Observatoire de Gen{\`e}ve, Universit{\'e} de Gen{\`e}ve, Chemin Pegasi, 51, 1290 Versoix, Switzerland \label{inst-geneva}
    \and
    Vanderbilt University, Department of Physics \& Astronomy, 6301 Stevenson Center Lane, Nashville, TN 37235, USA \label{inst-vandy}
    \and
    Institute for Astronomy, University of Edinburgh, Royal Observatory, Blackford Hill, Edinburgh, EH9 3HJ, UK \label{inst-Edinburgh}
    \and
    Department of Physics, University of Warwick, Gibbet Hill Road, Coventry, CV4 7AL, UK \label{inst-war}
    \and
    Centre for Exoplanets and Habitability, University of Warwick, Gibbet Hill Road, Coventry, CV4 7AL, UK \label{inst-warexo}
    \and
    N\'ucleo de Astronom\'ia, Facultad de Ingenier\'ia y Ciencias, Universidad Diego Portales, Av. Ej\'ercito 441, Santiago, Chile \label{inst-chile}
    \and
    Department of Astronomy, Tsinghua University, Beĳing 100084, People’s Republic of China \label{inst-Tsinghua}
    \and
    Department of Physics and Kavli Institute for Astrophysics and Space Research, Massachusetts Institute of Technology, Cambridge, MA 02139, USA\label{inst-mit}
    \and
    Harvard University, Cambridge, MA 02138 \label{inst-harvard}
    \and
    Facultad de Ingenier{\'i}a y Ciencias, Universidad Adolfo Ib{\'a}\~{n}ez, Av. Diagonal las Torres 2640, Pe\~{n}alol{\'e}n, Santiago, Chile \label{inst-adolfo}
    \and
    Millennium Institute for Astrophysics, Chile \label{inst-mifa}
    \and
    NASA Exoplanet Science Institute, Caltech/IPAC, Mail Code 100-22, 1200 E. California Blvd., Pasadena, CA 91125, USA \label{inst-caltech}
    \and
    Department of Physics and Astronomy, University of Kansas, Lawrence, KS 66045, USA \label{inst-Kansas}
    \and
    Cavendish Laboratory, JJ Thomson Avenue, Cambridge CB3 0HE, UK \label{inst-Cavendish}
    \and
    NASA Ames Research Center, Moffett Field, CA 94035, USA \label{inst-nasaames}
    \and
    University of Grenoble Alpes, CNRS, IPAG, F-38000 Grenoble,
    France \label{inst-grenoble}
    \and
    Centro de Astrof{\'i}sica y Tecnolog{\'i}as Afines (CATA), Casilla 36-D, Santiago, Chile \label{inst-chiletechologia}
    \and
    Center for Astrophysics, Harvard \& Smithsonian, 60 Garden Street, Cambridge, MA 02138, USA \label{inst-cfa}
    \and
    Department of Physics and Astronomy, Michigan State University, East Lansing, MI 48824, USA \label{inst-michigan}
    \and
    Centre for Astrophysics, University of Southern Queensland, Toowoomba, QLD 4350, Australia  \label{inst-queensland}
    \and
    Center for Astrophysics \textbar \ Harvard \& Smithsonian, 60 Garden Street, Cambridge, MA 02138, USA \label{inst-cambridge}
    \and
    George Mason University, 4400 University Drive, Fairfax, VA, 22030 USA \label{inst-GeorgeMason}
    \and
    Department of Astronomy, University of California Berkeley, Berkeley, CA 94720, USA \label{inst-berkeley}
    \and 
    El Sauce Observatory, Coquimbo Province, Chile \label{inst-elsauce} 
    \and 
    Department of Physics and Astronomy, University of New Mexico, 210 Yale Blvd NE, Albuquerque, NM 87106, USA \label{inst-newmexico}
    \and
    University of Zürich, Institute for Computational Science,
    Winterthurerstrasse 190, 8057 Zürich, Switzerland \label{inst-zurich}
    \and 
    Brierfield Observatory, New South Wales, Australia \label{inst-brierfield}
    \and
    NASA Goddard Space Flight Center, 8800 Greenbelt Rd, Greenbelt, MD 20771, USA \label{inst-nasa} 
    \and
    Patashnick Voorheesville Observatory, Voorheesville, NY 12186, USA \label{inst-Patashnick}
    \and
    Kotizarovci Observatory, Sarsoni 90, 51216 Viskovo, Croatia \label{inst-Kotizarovci}
    \and
    Hazelwood Observatory, Australia \label{inst-hazelwood} 
    \and
    Universit\'e C\^ote d'Azur, Observatoire de la C\^ote d'Azur, CNRS, Laboratoire Lagrange, Bd de l'Observatoire, CS 34229, 06304 Nice cedex 4, France \label{inst-cotedazur}
    }
   \date{Received March 03, 2022; Accepted April 25, 2022}

   \abstract{The detection and characterization of exoplanets and brown dwarfs around massive AF-type stars is essential to investigate and constrain the impact of stellar mass on planet properties. However, such targets are still poorly explored in radial velocity (RV) surveys because they only feature a small number of stellar lines and those are usually broadened and blended by stellar rotation as well as stellar jitter. As a result, the available information about the formation and evolution of planets and brown dwarfs around hot stars is limited.}{We aim to increase the sample and precisely measure the masses and eccentricities of giant planets and brown dwarfs transiting early-type stars detected by the Transiting Exoplanet Survey Satellite (TESS).
}{We followed bright (V < 12 mag) stars with \teff~> 6200~K that host giant companions (R > 7 \rearth) using ground-based photometric observations as well as high precision radial velocity measurements from the CORALIE, CHIRON, TRES, FEROS, and MINERVA-Australis spectrographs. }{In the context of the search for exoplanets and brown dwarfs around early-type stars, we present the discovery of three brown dwarf companions, TOI-629b, TOI-1982b, and TOI-2543b, and one massive planet, TOI-1107b. From the joint analysis of TESS and ground-based photometry in combination with high precision radial velocity measurements, we find the brown dwarfs have masses between 66 and 68~\mjup, periods between 7.54 and 17.17 days, and radii between 0.95 and 1.11~\rjup. The hot Jupiter TOI-1107b has an orbital period of 4.08 days, a radius of 1.30~\rjup, and a mass of 3.35~\mjup. As a by-product of this program, we identified four low-mass eclipsing components (TOI-288b, TOI-446b, TOI-478b, and TOI-764b).
}{Both TOI-1107b and TOI-1982b present an anomalously inflated radius with respect to the age of these systems. TOI-629  is among the hottest stars with a known transiting brown dwarf. TOI-629b and TOI-1982b are among the most eccentric brown dwarfs. The massive planet and the three brown dwarfs add to the growing population of well-characterized giant planets and brown dwarfs transiting AF-type stars and they reduce the apparent paucity.} 


   \keywords{brown dwarfs - planetary systems - stars: early-type - techniques: photometric - techniques: radial velocities - binaries: eclipsing}

   \maketitle
%

\section{Introduction}\label{sec:intro}
\begin{figure}
  \centering
  \includegraphics[width=0.45\textwidth]{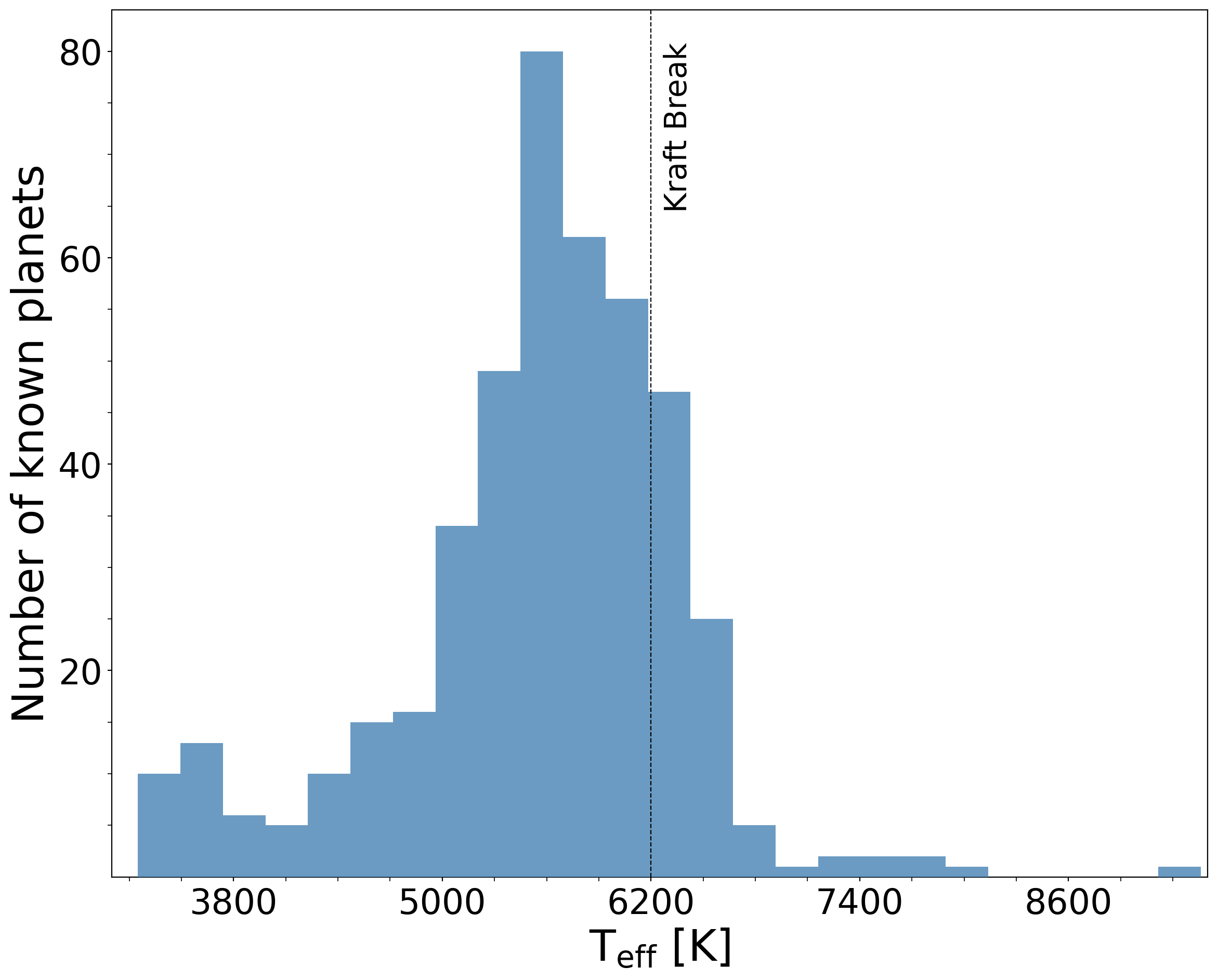}
  \caption{Number of known exoplanets from the PlanetS Catalog \cite[][accessible on the Data \& Analysis Center for Exoplanet DACE\textsuperscript{\ref{dacefootnote}}]{otegi2020} with precise mass and radius measurements ($\sigma_{M}/M$ $\leq$ 25$\%$ and $\sigma_{R}/R$ $\leq$ 8$\%$) as a function of stellar effective temperature. The black vertical line at 6200~K indicates the Kraft break (\citealt{kraft}).
  }
  \label{fig:HistogramTrRv}
\end{figure}

Even though our understanding of giant planet formation and evolution mechanisms has significantly improved over the past twenty years, the impact of the host star’s properties (mass, temperature, age, and metallicity) on the giant planet distribution remains to be thoroughly studied. The timescales and occurrences of the different planet formation scenarios have to be assessed, quantified, and compared. To that extent, detecting hot Jupiters and brown dwarfs orbiting hot stars is essential if we want to constrain giant planet and brown dwarf formation models across various stellar types. Since the discovery of the first exoplanet around a solar-like star in 1995  \citep{mayordidier}, more than 500 planets with well-constrained densities ($\sigma_{M}/M$ $\leq$ 25$\%$ and $\sigma_{R}/R$ $\leq$	8$\%$) have been discovered by transit surveys and validated with radial-velocity (RV) follow-up, according to the PlanetS exoplanet catalog of \cite{otegi2020} with accurate mass and radius determinations (accessible on the Data \& Analysis Center for Exoplanet DACE\protect\footnote{\url{https://dace.unige.ch}\label{dacefootnote}}). These surveys have mainly focused on late-type main-sequence stars because planets transiting hot and usually fast-rotating stars are harder to validate. With effective temperatures higher than 6200~K (Kraft break, \citealt{kraft}), these stars have thin convective envelopes that do not efficiently generate magnetic winds and drive angular momentum loss. As a result, they remain rapidly rotating and due to their high temperature they exhibit fewer spectral lines. Their high rotation velocities lead to a substantial broadening of their observed spectral lines, making RV measurements of hot stars difficult. In addition, stellar activity and pulsations can induce RV variations that can mask or even mimic the RV signature of orbiting exoplanets \citep{vanderburg2016}. In particular, only 20$\%$ (99 exoplanets) of transiting exoplanets with precise densities have been detected to orbit AF-type stars with \teff~> 6200~K ($\lesssim$ F8V).  Figure \ref{fig:HistogramTrRv} summarizes the scarcity of planets detected around host stars with different stellar types. The vertical line at 6200~K indicates the Kraft break where stars retain their high rotational velocities.


Several spectroscopic surveys on volume-limited samples of early-type stars have been performed using the Fourier interspectrum method (\citealt{Chelli2000}), with various spectrographs: the ELODIE (\citealt{Baranne1996}) survey (\citealt{Galland2005a}, \citeyear{Galland2005b}, \citeyear{GallandBD2006}), the SOPHIE (\citealt{SophieTeam}) survey (\citealt{Desort2009}, \citealt{Borgniet2014}, \citeyear{Borgniet2019}) and the HARPS (\citealt{Pepe2002}) survey (\citealt{Desort2008}, \citealt{Borgniet2017}) have led to several discoveries of planets around F-type stars. In this frame, a planet was also discovered around the A5V-type star $\beta$ Pictoris (\citealt{Lagrange2020}). Other interesting detections occurred from transit surveys, such as KELT-9b \citep{Gaudi:2017}, MASCARA-1b \citep{mascara1}, MASCARA-4b \citep{mascara4}, WASP-33b \citep{WASP-33}, and WASP-167b/KELT-13b \citep{WASP-167}. Studying this population is scientifically valuable for several reasons. First, planets orbiting hot stars have a higher probability of being in a misaligned polar or retrograde orbit \citep{winn2010}. The high obliquities of these planets may shed light on their formation since it can be linked to orbital migration processes around stars of high masses \citep{Fabrycky2007}. Atmospheric characterization of these misaligned hot Jupiters provides constraints on the composition of their atmospheres which may in turn reveal clues as to their formation history. Second, these planets receive large amounts of UV radiation, compared to planets around late-type stars, that can drive unique chemical processes in their atmospheres \citep{Casewell2015}. Third, the core-accretion planet formation scenario predicts an increase in gas giant frequency with stellar mass \citep{KennedyKenyon}, therefore enlarging the population of exoplanets transiting massive stars will allow to test such correlation.

The Transiting Exoplanet Survey Satellite (TESS, \citealt{Ricker2015}) is an all-sky survey that has been monitoring bright targets for more than three years. Unlike Kepler, TESS is optimized for bright stars and was designed to search for planets transiting a large variety of stars including early-type, massive stars. TESS has contributed tens of new planets (e.g., TOI-1431b/MASCARA-5b; \citealt{TOI1431}, TOI-2109b; \citealt{TOI2109}, TOI-2685b; \citealt{TOI2685}, TOI-3362b; \citealt{TOI3362}) to this still sparse population and more than 300 candidates released as TESS Objects of Interest (TOIs).

We performed a RV follow-up dedicated to the validation and mass measurement of extrasolar planets and brown dwarfs transiting a sample of AF-type, main-sequence stars from the TESS mission and we present the discovery of TOI-1107b, TOI-629b, TOI-1982b, and TOI-2543b, a massive planet and three massive brown dwarfs, respectively. TOI-1107b, TOI-1982b, and TOI-2543 orbit F-type stars while TOI-629b orbits a \teff~= 9110 $\pm$ 200K, A-type star. In Section \ref{sec:observations} we describe our sample, the TESS photometric data and the ground-based follow-up photometric and spectroscopic observations. Section \ref{sec:analysis} presents the host star characterization and the global photometric and spectroscopic analysis of the systems. We discuss the systems in context in Section \ref{sec:discussionconclusion} and conclude in Section \ref{sec:conclusion}. As a by-product of our RV follow-up, we present four low-mass eclipsing components in the Appendix~\ref{sec:falsepositives}.

%
\section{Observations}\label{sec:observations}
\subsection{Sample selection and observations}
We carried out a RV follow-up focusing on TOIs with spectral types between A0 and $\sim$ F8. The F8 ($\sim$ 6200 K) limit corresponds to the Kraft break and the A0 to the earliest spectral type for which the cross-correlation technique with the appropriate binary mask provides precise RV measurements. The sample consists of 312 TESS giant planet candidates (R > 7 \rearth) brighter than V = 12 mag that were detected during the first three years of the TESS mission (Sectors 1-39). We exclude targets photometrically identified as false positives (transit depth chromaticity measured in different filters) such as blended eclipsing binaries (BEB), spectroscopically identified as false positives such as single- and double-lined spectral binaries (SB1 and SB2) and BEBs and already known planets. Our final sample consists of $\sim$ 120 targets. Thanks to this survey, we confirm the discovery of three new brown dwarfs (TOI-629b, TOI-1982b, and TOI-2543b), a massive planet (TOI-1107b) and four low-mass stars (identified as SB1s). The list and the parameters of the identified SB1 systems can be found in the Appendix~\ref{sec:falsepositives}.

\begin{figure*}
  \centering
  \includegraphics[width=0.35\textwidth]{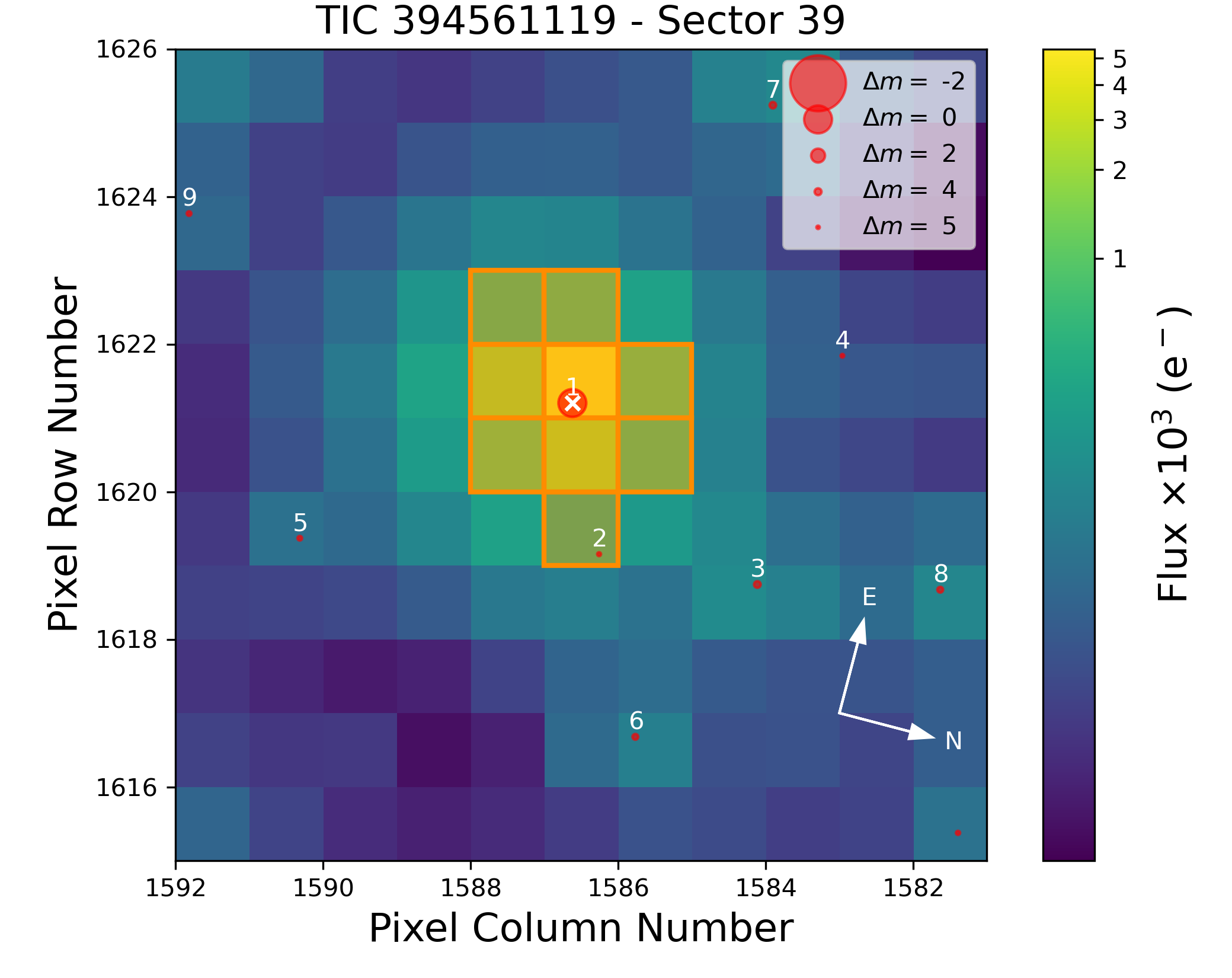}
  \includegraphics[width=0.35\textwidth]{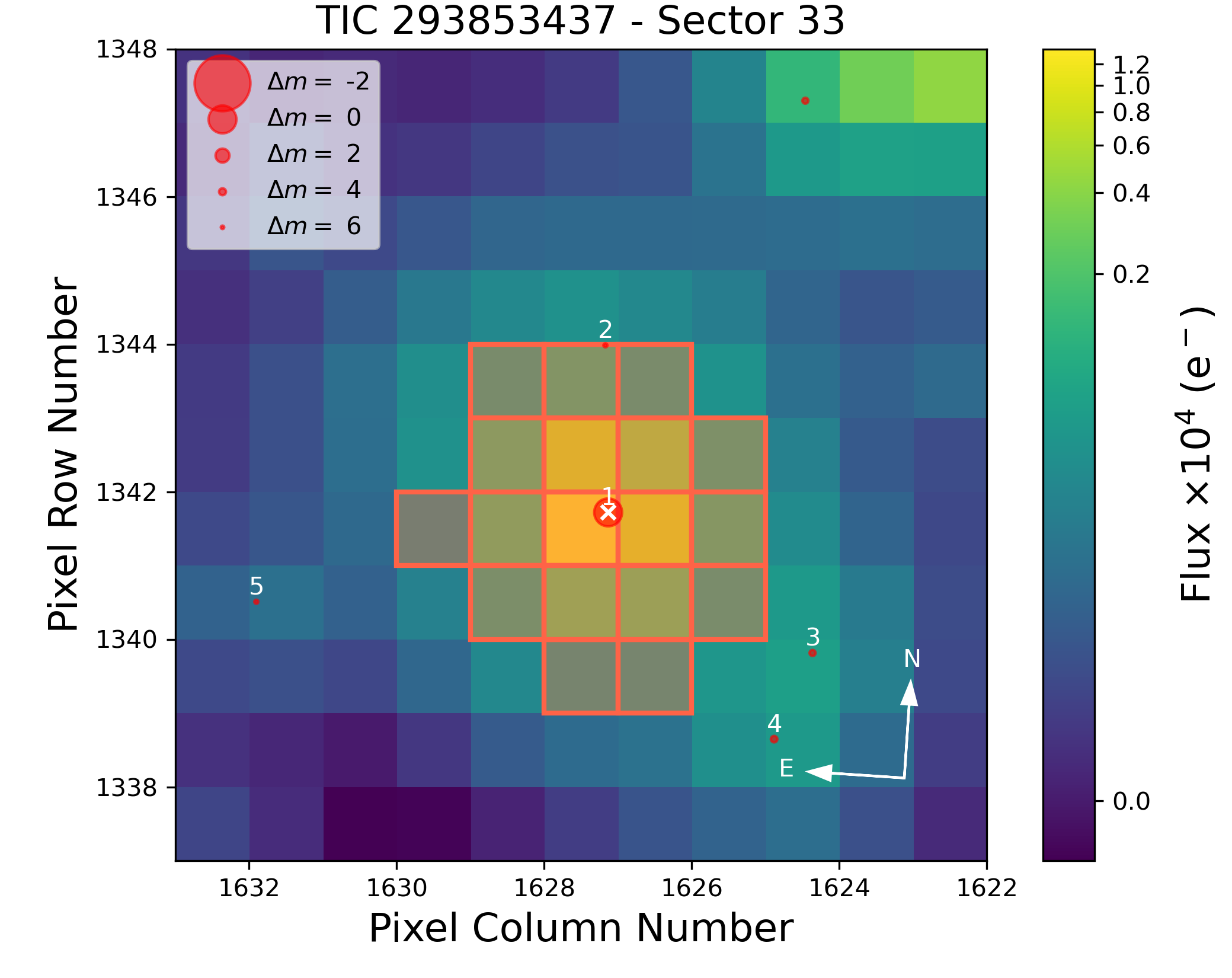}
  \includegraphics[width=0.35\textwidth]{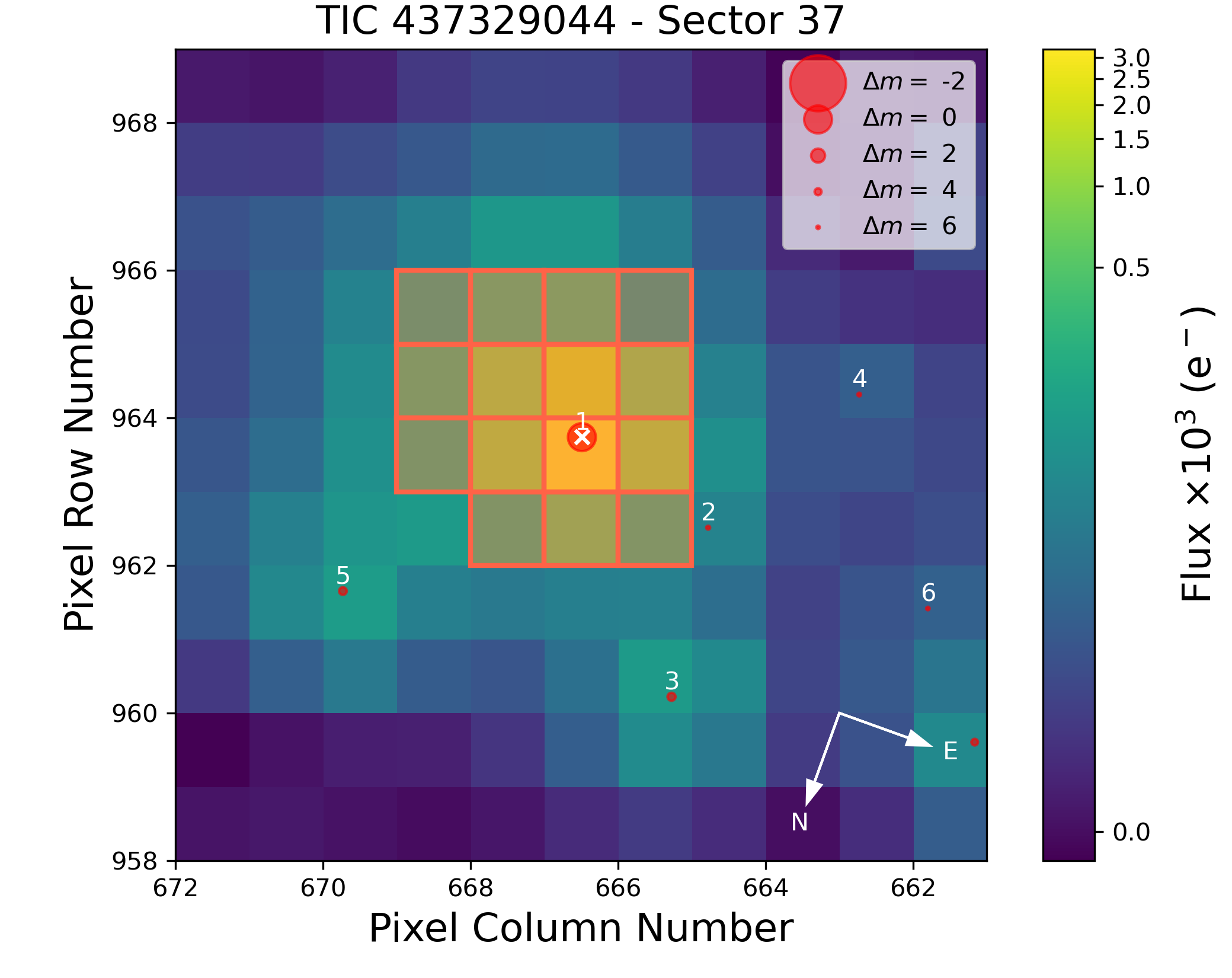}
  \includegraphics[width=0.35\textwidth]{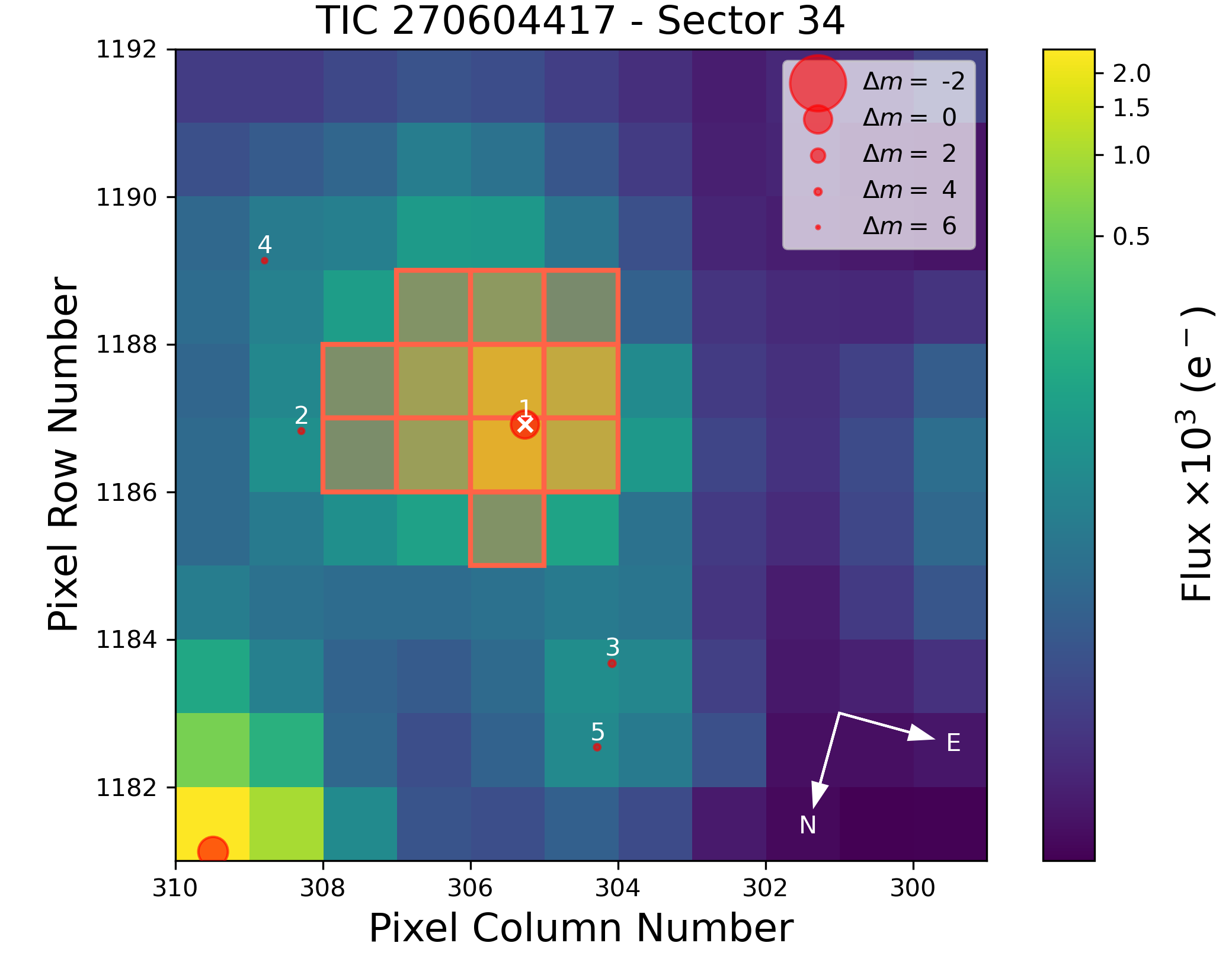}
  \caption{TESS Target Pixel File (TPF) of TOI-1107 (TIC 394561119; \textit{top left}), TOI-629 (TIC 293853437; \textit{top right}), TOI-1982 (TIC 437329044; \textit{bottom left}), and TOI-2543 (TIC 270604417; \textit{bottom right}) created with \tpfplotter (\citealt{TPFplotter}). Orange pixels define the aperture mask that was used to extract the photometry. Red circles indicate the neighbor \textit{Gaia} DR2 catalog objects. The size is proportional to the brightness difference of the sources with the targets (see legend). Our targets are marked with a white cross. Pixel scale is 21 $\arcsec$/pixel.}
  \label{fig:TPF}
\end{figure*}

\subsection{TESS and ground-based photometry}
All four targets were initially observed and detected by the NASA TESS space mission. Afterward, ground-based time-series follow-up photometry was performed for TOI-1107 and TOI-1982 as part of the TESS Follow-up Observing Program (TFOP, \citealt{karen2018}) to confirm the TESS photometric ephemerides and exclude blending scenarios. The summary of the observations can be found in Table~\ref{tab:Photometry_info} and their description below. All light curves can be found in the Appendix~\ref{sec:lightcurves}.

\begin{table}
\tiny
\centering
\setlength{\tabcolsep}{6pt}
\begin{tabular}{lccc}
        \hline
        \hline
        Date(s) & Facility &  Detrending \\
        \hline
         \textbf{TOI-1107} & &\\
         2021 Apr 28 - Jun 24 & TESS 2-min  & GPs \\
         2019 Apr 22 - Jul 18 & TESS FFI & GPs \\
         2020 May 07 & Brierfield  (B) & $\textit{p(t)}$, $\textit{p(sky)}$ \& $\textit{p(AM)}$\\
         2020 Aug 13 & ASTEP ($R_c$) & $\textit{p(t)}$ \& $\textit{p(AM}^\textit{2}\textit{)}$ \\
         2021 Mar 25 & LCO-SAAO (B) & $\textit{p(sky)}$, $\textit{p(FWHM)}$ \& $\textit{p(AM)}$\\
         2021 Mar 25 & LCO-SAAO ($z^\prime$) & $\textit{p(FWHM)}$ \\
        \textbf{TOI-629} &&\\  
         2020 Dec 18 - 2021 Jan 13 & TESS 2-min & GPs \\
         2018 Dec 11 - 2019 Jan 07 & TESS FFI & GPs \\
        \textbf{TOI-1982} &&\\  
        2021 Apr 02 - Apr 28 & TESS 2-min & GPs\\
        2019 Apr 22 - May 21 & TESS FFI & GPs\\
       	2020 Jul 01 & Hazelwood ($R_c$) & $\textit{p(sky)}$ \& $\textit{p(FWHM)}$ \\
       	2021 Apr 02 & El Sauce ($R_c$) & $\textit{p(sky)}$ \& $\textit{p(AM)}$ \\
       	2021 Apr 19 & LCO-CTIO (B) & $\textit{p(sky)}$ \& $\textit{p(AM)}$ \\
       	2021 Apr 19 & LCO-CTIO ($z\prime$) & $\textit{p(sky)}$ \& $\textit{p(AM)}$ \\
        \textbf{TOI-2543} &&\\  
        2021 Jan 13 - Feb 09 & TESS 2-min & - \\
        2019 Feb 02 - Feb 28 & TESS FFI & GPs \\
        \hline
    \end{tabular}
    \caption{Summary of the TESS and ground photometric observations of TOI-1107, TOI-629, TOI-1982 and TOI-2543.}
\begin{tablenotes}
\item
\textbf{Notes:} The notation of the baseline models for, $\textit{p(j}^\textit{i}\textit{)}$, refers to a polynomial of degree \textit{i} in parameter \textit{j} (t:time, AM:airmass, FWHM:stellar FWHM and sky:sky background). The notation of GPs refers to Gaussian Processes. The low level of correlated noise in TOI-2543 2-minute cadence light curves did not necessitate further detrending.
\end{tablenotes}
\label{tab:Photometry_info}
\end{table}

\subsubsection{TESS light curves}\label{sec:TESSLCs}

The transit signatures were initially detected by TESS during the first year of the survey (July 2018 - July 2019) in the Full Frame Image (FFI) observations with a cadence of 30-minutes and flagged as promising TOIs for follow-up observations. Afterward, they were reobserved at 2-minute cadence during the third year of the mission (July 2020 - July 2021). The photometry of the short cadence data is extracted by the Science Processing Operations Center (SPOC; \citealt{Jenkins2016}) and consists of the Simple Aperture Photometry (SAP) flux and the Presearch Data Conditioning SAP (PDCSAP; \citealt{Smith2012},\citealt{Stumpe2012},\citeyear{Stumpe2014}) flux. For our analysis, we made use of the PDCSAP light curves since they are corrected for long term systematic trends, such as instrumental artifacts and dilution. The observations at 30-minute cadence were processed by the MIT Quick Look Pipeline (QLP; \citealt{QLP}). When visually comparing the various light curves, we noticed that the detrended QLP light curves showed a slighly increased flux during transit compared to the un-detrended data. Therefore we decided to use the raw normalized simple aperture photometry from the best aperture (SAP flux) and account for systematic noise using Gaussian Process regression. All data are publicly available and were downloaded from the Mikulski Archive
for Space Telescopes (MAST$\footnote{\url{http://archive.stsci.edu/tess/}}$).

TOI-1107 (TIC 394561119) was observed in Sectors 11-13 with 30-minute cadence and Sectors 38 and 39 with 2-minute cadence. We detected 17 transits in the FFIs and 13 transits in 2-minute cadence light curves with a transit depth of $\sim$ 0.54$\%$ and an average period of 4.08 days. TOI-1107 passed all the Data Validation (DV) tests (\citealt{Twicken2018}, \citealt{Li2019}) and the difference image centroiding test that placed the transit signature
source within 0.2 $\pm$ 2.5$\arcsec$ of the target star.

TOI-629 (TIC 293853437) was monitored in Sector 6 at 30-minute and Sector 33 at 2-minute cadence. Three transits with a depth of $\sim$ 0.23$\%$ and a period of 8.72 days were identified in both QLP and 2-minute data. The candidate passed the DV test and the difference image centroid test located the source of the transits to within 2.0 $\pm$ 2.5$\arcsec$ of the target star.

TOI-1982 (TIC 437329044) was observed in Sector 11 at 30-minute and Sector 37 at 2-minute cadence. The 30-minute cadence data were processed by the QLP that identified two $\sim$ 0.53$\%$ deep transits 17.1 days apart located at the beginning and at the end of the light curve. During perigee passage while downloading data, the data collection was paused for 1.18 days near the middle of Sector 11. As a result, the period of the target was initially reported to be half (8.58 days). Later, the 2-minute cadence data confirmed the 17.1 days period with only one transit visible in Sector 37.

Finally, TOI-2543 (TIC 270604417) was observed in Sector 8 at 30-minute cadence and in Sector 34 at 2-minute cadence. Two transits were detected in the QLP data and three transits in the 2-minute data with a depth of $\sim$ 0.28$\%$ and an orbital period of 7.54 days. The transit signature passed the DV test and the difference image report located the source of the transits to within 0.8 $\pm$ 2.5$\arcsec$ of the target star.

The Target Pixel Files (TPFs) of our targets and the aperture masks used by the pipeline to extract the photometry (orange pixels), generated with \tpfplotter, can be found in Figure ~\ref{fig:TPF}. The position of nearby stars from the Gaia DR2 catalog is marked with red circles. The size of the red circles indicates the different magnitudes in contrast with the target star that is marked with a white cross. For TOI-629, TOI-1982 and TOI-2543 no nearby sources are identified within 6 mag of the target in the aperture mask. For TOI-1107, we identified one star within the aperture at an angular distance of 43.69$\arcsec$ and with magnitude contrast $\Delta$m = 5.6 mag in the G-band. The complementary ground-based seeing-limited photometry with several instruments (see Table \ref{tab:Photometry_info}) confirms that the signal originates from TOI-1107. Moreover, PDCSAP flux light curves account for the dilution caused by nearby stars in order to prevent underestimation of the transiting object radius. For TOI-1107, the estimated flux contamination ratio (ratio of the total contaminant flux over the target star flux in TESS-band) is 0.01. 

\subsubsection{LCOGT photometric follow-up}
We observed full transits of TOI-1107b in B-band and Pan-STARRS $z$-short band on March 25, 2021 from the Las Cumbres Observatory Global Telescope \citep[LCOGT;][]{Brown:2013} 1.0-m network node at South Africa Astronomical Observatory (SAAO). We observed ingresses of TOI-1982b in B-band and Pan-STARRS $z$-short band on April 19, 2021 from the LCOGT 1.0-m network node at Cerro Tololo Inter-American Observatory (CTIO). The data that are affected by poor sky transparency have been removed from the light curves. We use the {\tt TESS Transit Finder}, which is a customized version of the {\tt Tapir} software package \citep{Jensen:2013}, to schedule our transit observations. The 1.0-m telescopes are equipped with $4096\times4096$ SINISTRO cameras having an image scale of $0.389\arcsec$ per pixel, resulting in a $26\arcmin\times26\arcmin$ field of view. The images were calibrated by the standard LCOGT {\tt BANZAI} pipeline \citep{McCully:2018}, and photometric data were extracted using {\tt AstroImageJ} \citep{Collins:2017}. The images were focused and have mean stellar point-spread-functions with a FWHM of $\sim 2\arcsec$, and circular photometric apertures with radius 3.9$\arcsec$ (SAAO) and 5.8$\arcsec$ (CTIO) were used to extract the differential photometry. A larger aperture was used for the CTIO observations due to the poor skies and highly variable seeing. The photometric apertures exclude all flux from the nearest Gaia EDR3 neighbors. Achromatic transits were detected on-target in both systems.

\subsubsection{Hazelwood photometric follow-up}
The Hazelwood Observatory is a private backyard observatory with a 0.32-m Planewave CDK telescope working at f/8, a SBIG STT3200 2148$\times$1472 CCD, giving a 20$\arcmin\times$13$\arcmin$ field of view and 0.55\arcsec. The Hazelwood Observatory, operated by Chris Stockdale in Victoria, Australia, observed a 13 ppt egress of TOI-1982 in $R_c$ filter on July 01, 2020 and checked all nearby stars confirming that none of them show an eclipse that could explain the transit seen in TESS data.

\subsubsection{El Sauce photometric follow-up}
We observed a full transit of TOI-1982 on April 02, 2021 in Johnson-Cousins $R_c$-band using the Evans 0.36-m telescope at El Sauce Observatory in Coquimbo Province, Chile. The telescope is equipped with a STT1603-3 CCD camera with 1536$\times$1024 pixels binned 2$\times$2 in-camera resulting in an image scale of 1.47$\arcsec$/pixel. The photometric data for TOI-1982 were obtained from 455$\times$60~sec exposures using a circular 8.8$\arcsec$ aperture, processed in AstroImageJ (\citealt{Collins:2017}).

\subsubsection{ASTEP photometric follow-up}
TOI-1107 was observed on August 13, 2020, with the Antarctica Search for Transiting ExoPlanets (ASTEP) program on the East Antarctic plateau at a latitude of $-75.1 \deg$ and elevation of $3\,233~\rm m$ \citep{guillot2015, mekarnia2016}. ASTEP is a custom 0.4\,m Newtonian telescope equipped with a 5-lens Wynne coma corrector and a 4k$\times$4k front-illuminated FLI Proline KAF-16801E CCD. The camera has an image scale of 0.93$\arcsec$/pixel resulting in a 1$^{o}\times 1^{o}$ corrected field of view. The focal instrument dichroic plate splits the beam into a blue wavelength channel for guiding, and a nonfiltered red science channel roughly matching an $R_c$ transmission curve \citep{abe}. The telescope is automated or remotely operated when needed. Due to the extremely low data transmission rate at the Concordia Station, the data are processed on-site using an automated IDL-based aperture photometry pipeline. The calibrated light curve is reported via email and the raw light curves of about 1\,000 stars of the field are  transferred to Europe on a server in Rome, Italy and are then available for deeper analysis. These data files contain each star’s flux computed through $10$ fixed circular apertures radii, so that optimal light curves can be extracted. For TOI-1107, an 11.2  pixel radius aperture gave the best result. TOI-1107 was observed under good weather conditions  with a  non windy ($\sim$ 3 ms$^{-1}$) clear sky, and air temperatures ranged between
$-65^o$C and $-75^o$C. 
\subsubsection{Brierfield photometric follow-up}
We observed a transit of TOI-1107 on May 06, 2020 in the B-band using a 0.36-m telescope (PlaneWave CDK14) at the Brierfield Observatory, a home observatory in Brierfield, New South Wales, Australia. The detector was a Moravian 16803 camera, which provided a pixel scale of 1.47$\arcsec$/pixel. Seeing conditions were average, with a full moon. We observed a continuous transit using 143 images of 180~sec each over 558 minutes. The images were reduced and measured with AstroImageJ using a photometric aperture of 10.29$\arcsec$.

\begin{figure*}[t]
  \centering
  \includegraphics[width=0.36\textwidth]{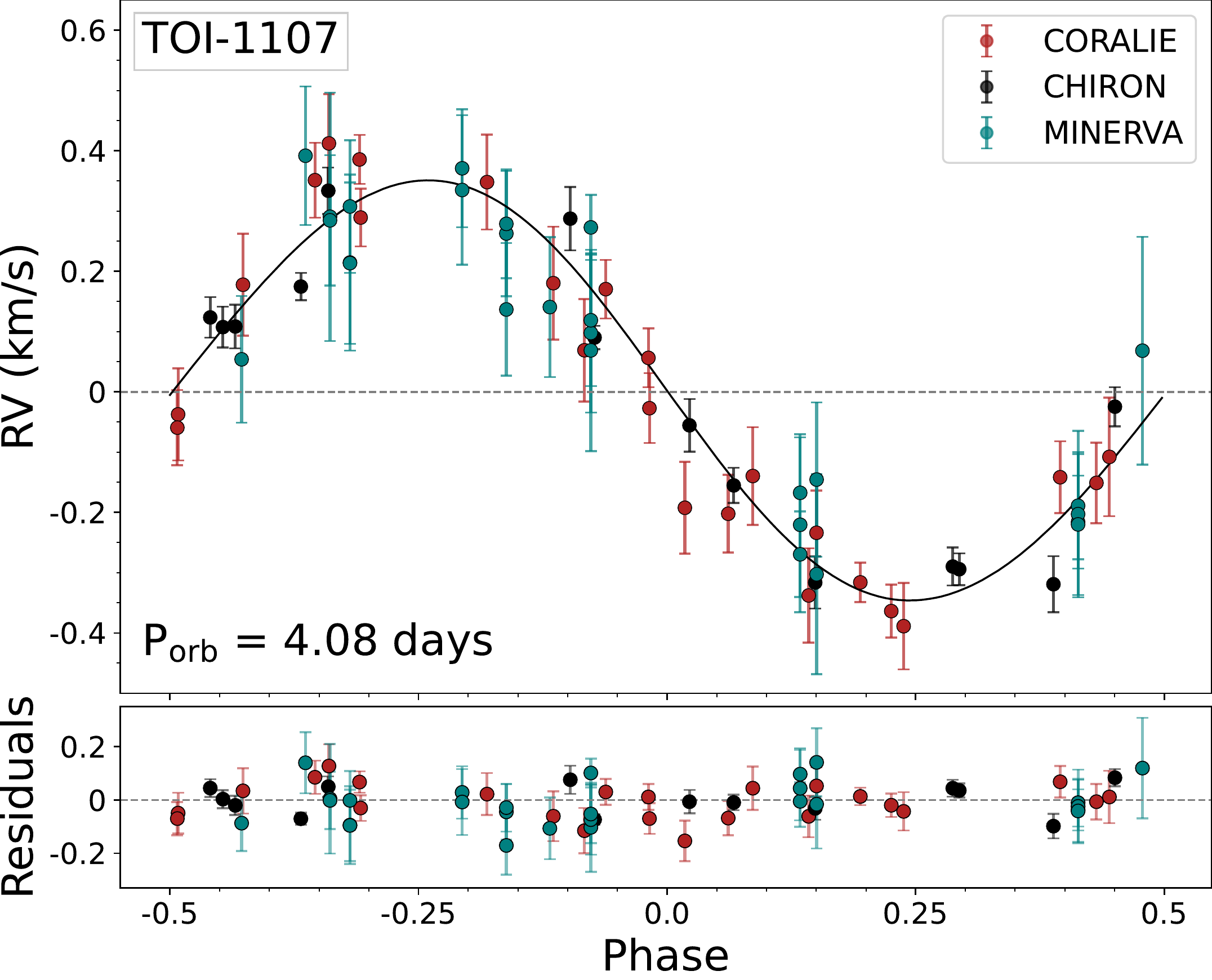}
  \hspace{0.7cm}
  \includegraphics[width=0.36\textwidth]{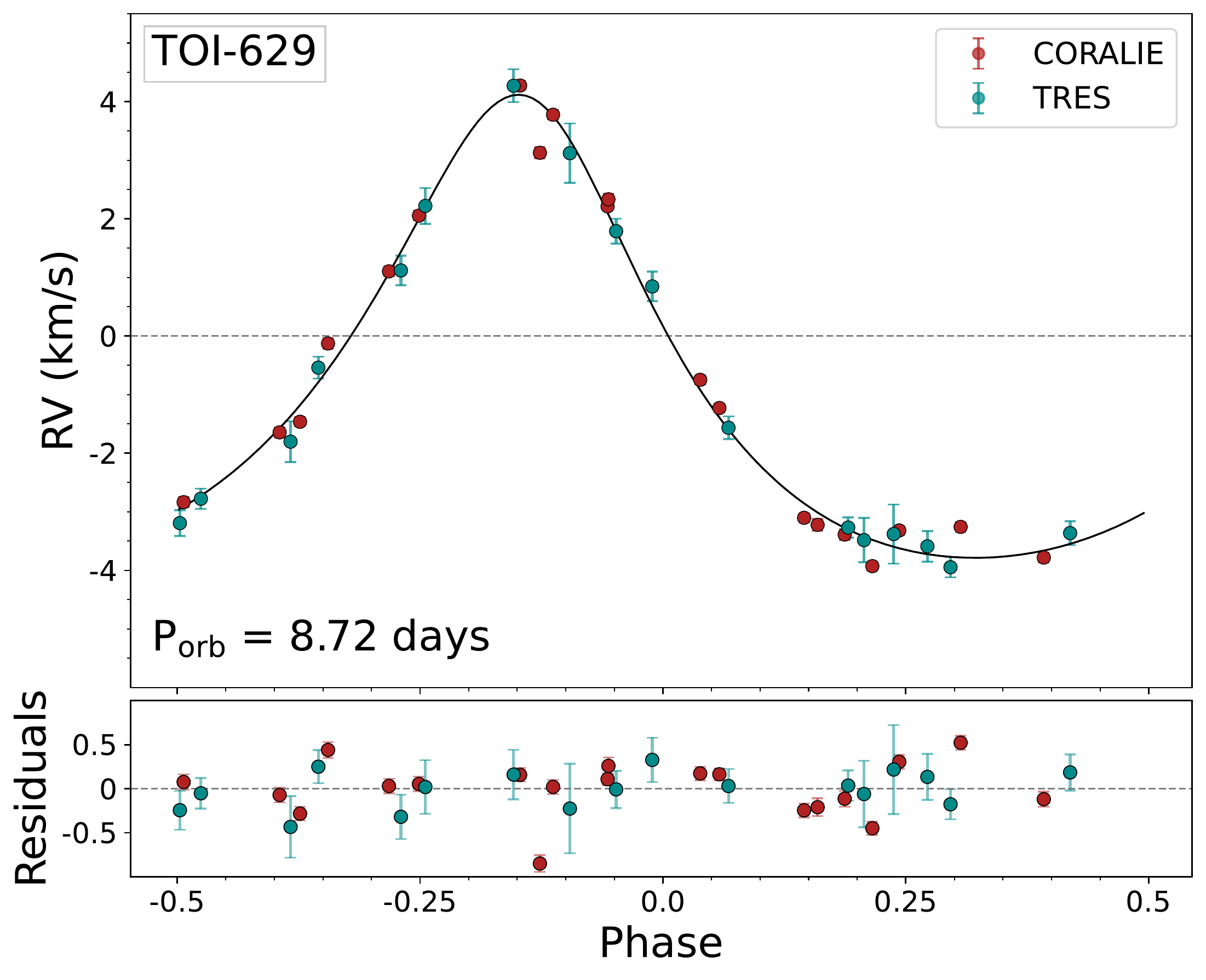}
  \includegraphics[width=0.36\textwidth]{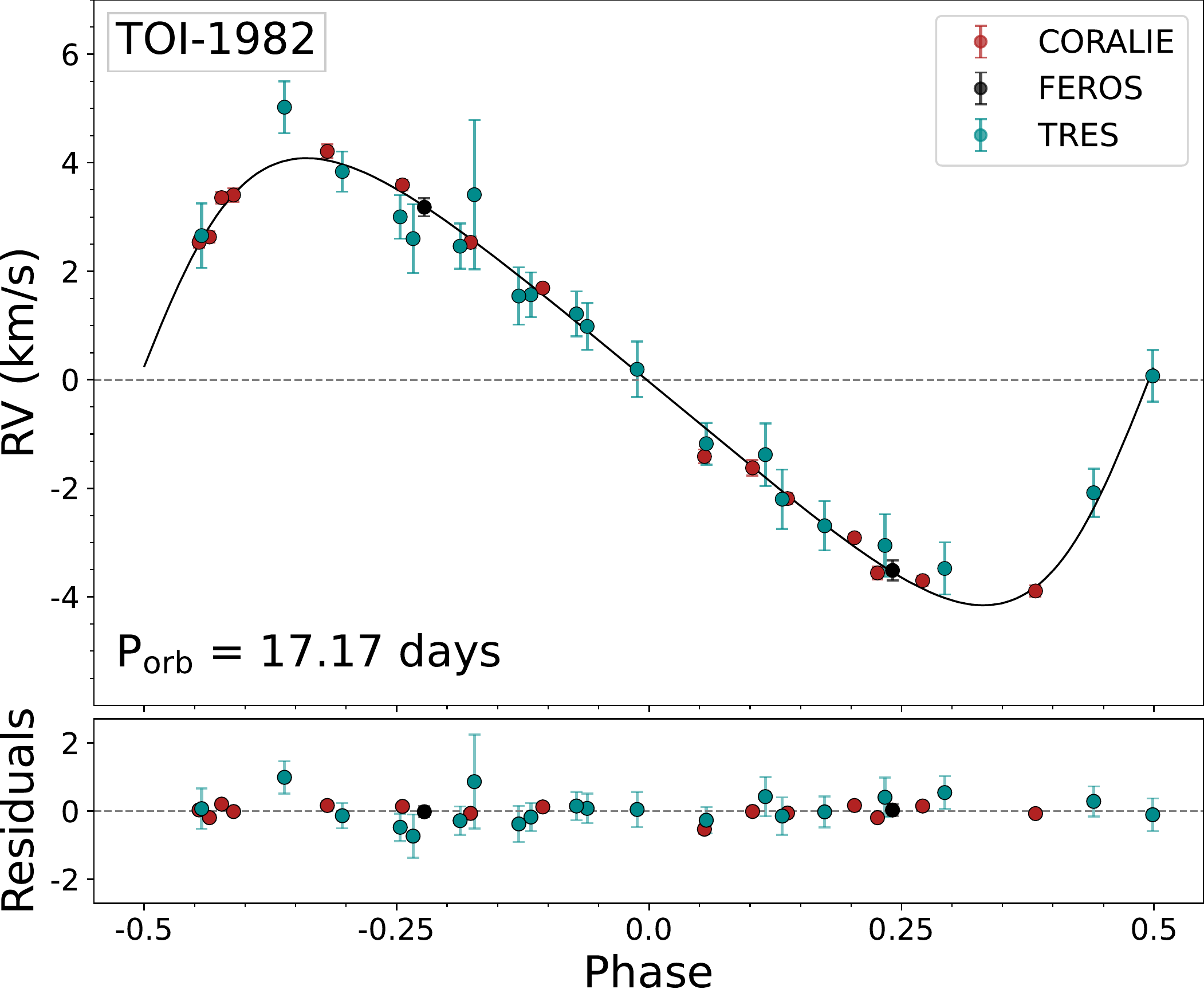}
  \hspace{0.7cm}
  \includegraphics[width=0.36\textwidth]{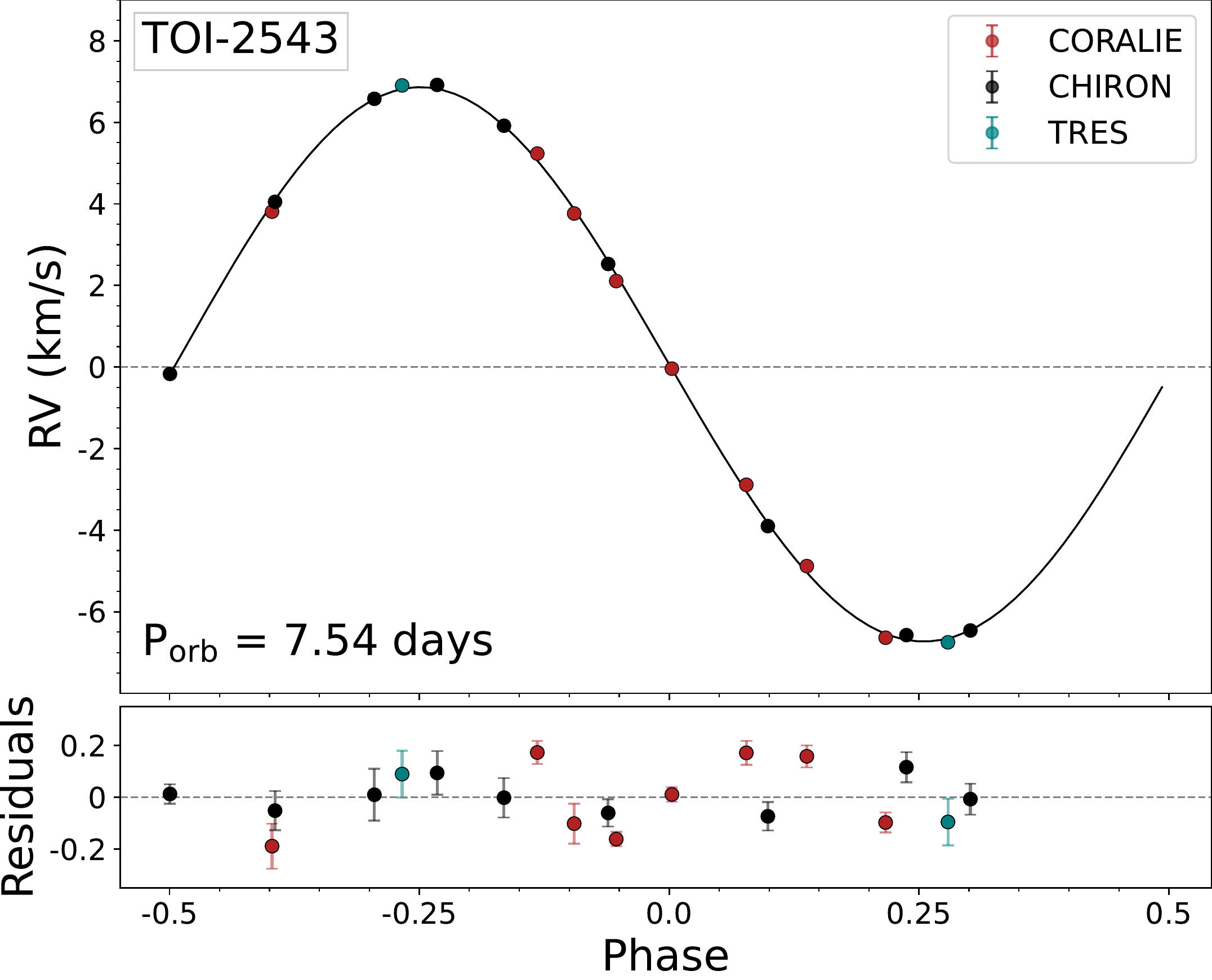}
  \caption{Relative RVs of TOI-1107 (\textit{top left}), TOI-629 (\textit{top right}), TOI-1982 (\textit{bottom left}) and TOI-2543 (\textit{bottom right}). The different colors indicate different instruments. The black line shows the model fit derived with \juliet. The residuals of the model fit are shown in the bottom panels.}
  \label{fig:Radial_velocities}
\end{figure*}

\subsection{High resolution spectroscopy}
The four targets were monitored with different spectrographs, CORALIE, CHIRON, TRES, FEROS, and MINERVA-Australis, in order to identify the nature of the candidate, determine its mass and measure its orbital parameters. These observations are described
in the following paragraphs and the radial velocities are displayed in Figure \ref{fig:Radial_velocities}.

\subsubsection{CORALIE}
For all the targets we obtained spectroscopic data  with the high resolution CORALIE spectrograph that is installed at the Swiss 1.2-m Leonhard Euler Telescope at ESO’s La Silla Observatory \citep{Queloz2001}. CORALIE has a resolution of R $\sim$ 60, 000 and is fed by a 2$\arcsec$ fiber.
A total of 24 RVs were obtained for TOI-1107 from 2021 April 03 to 2021 June 10 using exposure times of 1800~s which translated in spectra with a signal-to-noise ratio per
resolution element (S/N) between 15 and 34. TOI-629 was observed 20 times with CORALIE from 2020 November 26 to 2021 March 28 with exposure times of 1800 - 2400~s resulting in a S/N between 39 and 65. We collected 15 RVs for TOI-1982 from 2021 April 12 to 2021 June 12 using 1800 - 2400~s exposures giving a S/N between 15 and 33. Finally, we obtained eight spectra for TOI-2543 from 2021 November 23 to 2021 December 25 using exposure times of 1800~s resulting in a S/N between 13 and 25.

The RV of each epoch was derived by cross-correlating the spectrum with a binary mask that matches the stellar spectral type \citep{Baranne1996}. Afterward, a Gaussian is fit to the cross-correlation function (CCF). The RV measurement of each spectrum corresponds to the central position of the Gaussian fit. The bisector-span (BIS), the contrast (depth) and the full width at half-maximum (FWHM) were recorded with the standard CORALIE data reduction pipeline. For TOI-629, our hottest target, we used an A0 binary mask. The RVs of TOI-1107 and TOI-1982 were computed by cross-correlating with several binary masks and selected a G2 mask \citep{Pepe2002} that reached higher precision in the mass measurement.

For each star, we also compute the RVs with the Fourier interspectrum method (\citealt{Chelli2000}), consisting of correlating, in the Fourier space, the spectra of the star with a reference spectrum built by averaging all the spectra acquired for this star (\citealt{Galland2005a}). The resulting RVs are consistent with those obtained with the CCF method described above.

\subsubsection{TRES}
We used the TRES instrument on Mt. Hopkins, Arizona, USA to obtain spectra for TOI-629 and TOI-1982. TRES has a resolving power of $R\sim 44, 000$ and covers a wavelength range of $\rm$ 3900 $\AA$ to $\rm$ 9100 $\AA$. We used multiple echelle orders within this wavelength range for each spectrum to measure a relative RV at each phase in the orbit of the transiting BD. We visually review the orders to remove cosmic rays and omit those with low S/N. Each order is cross-correlated with the highest observed signal-to-noise spectrum of the target star and then the average RV of all the orders per spectrum is taken as the RV of the star for that observation.
For TOI-629, we took a total of 18 spectra with TRES. These spectra were taken between 2019 September 17 and 2019 October 28 with exposure times ranging between 600~s and 2200~s and S/N between 64 and 165.
For TOI-1982, we took a total of 20 spectra with TRES. The first of these spectra was taken on 2021 January 8, but the remaining spectra were taken months later from 2021 April 15 to 2021 June 2. The exposure times for these spectra range from 720\,s to 3600\,s and a S/N range of 22 to 55. Finally, for TOI-2543, we collected two spectra on 2021 March 22 and on April 2 with exposure times of 1900~s and 2400~s, respectively. The S/N of the spectra is 30 and 38.

\begin{figure*}[t]
  \centering
  \includegraphics[width=0.7\textwidth]{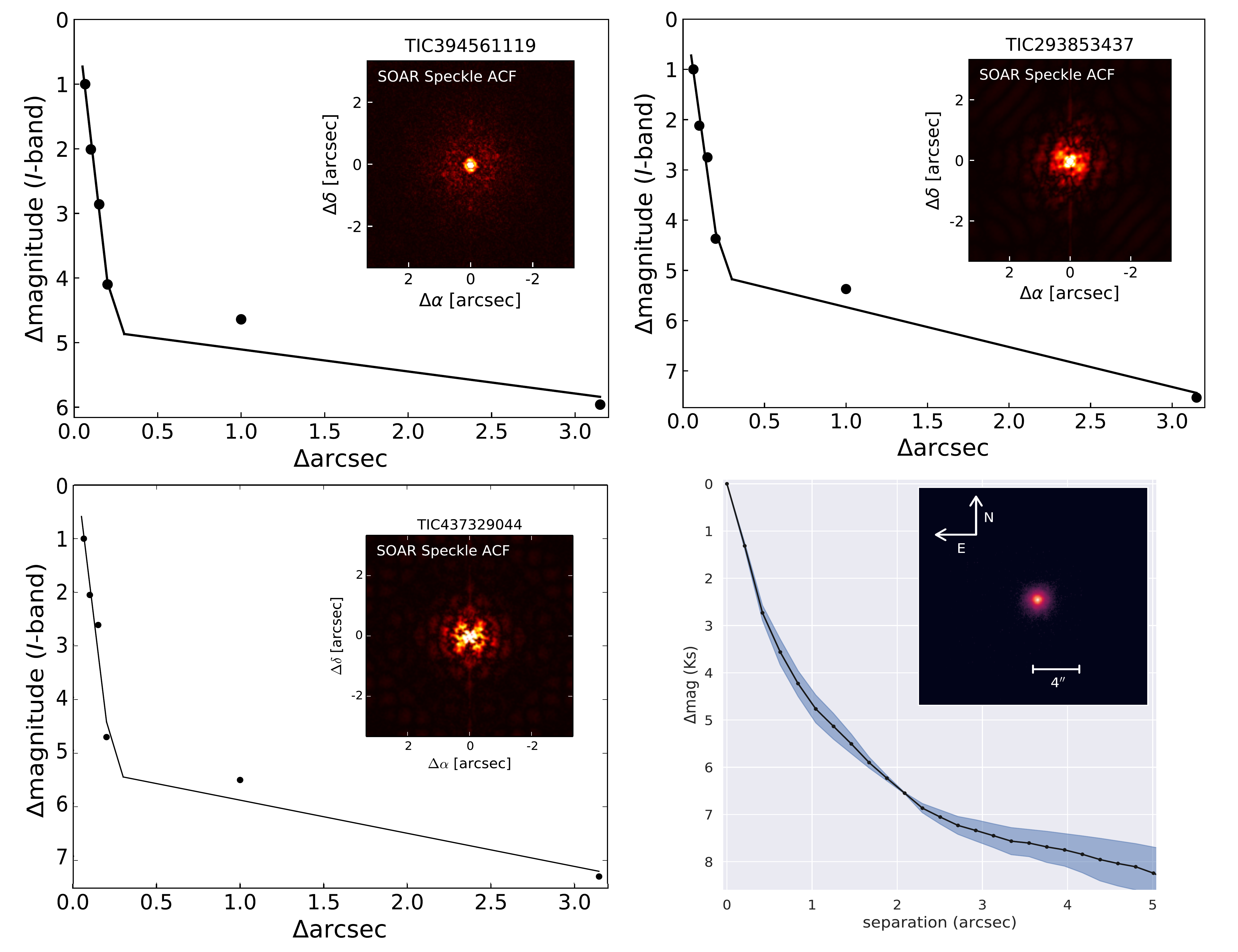}
  \caption{SOAR speckle imaging (5$\sigma$ upper limits) of TOI-1107 (TIC 394561119; \textit{top left}), TOI-629 (TIC 293853437; \textit{top right}), and TOI-1982 (TIC 437329044; \textit{bottom left}) that rules out stellar companions within 3 $\arcsec$. The image inset represents the speckle auto-correlation function. Adaptive optics image and contrast curve for TOI-2543 (TIC 270604417; \textit{bottom right}) taken with the ShARCS camera on the Shane 3-m telescope at Lick Observatory. The contrast curve was generated by calculating the median values (solid lines) and root-mean-square errors (blue, shaded regions) in annuli centered on the target, where the bin width of each annulus is equal to the full width at half maximum of the point spread function.}
  \label{fig:soar_shane}
\end{figure*}

\subsubsection{CHIRON}
TOI-1107 and TOI-2543 were observed with the CHIRON fiber fed echelle spectrograph at the SMARTS 1.5-m telescope located at Cerro Tololo Inter-American Observatory, Chile \citep{2013PASP..125.1336T}. CHIRON is a high resolution echelle, with a resolving power of R = 80, 000 and a spectral coverage of 4100 to 8700\,\AA{}. Observations of TOI-2543 were obtained in the \emph{fiber} mode, yielding a spectral resolving power of $R\sim$ 28, 000. For TOI-1107, we took a total of 14 spectra from 2021 January 15 to 2021 March 31 with an exposure time of 1800~s and S/N between 20 and 39. A total of 9 RVs were obtained for TOI-2543 from 2021 April 06 to 2021 May 09 using a median exposure time of 1800~s, resulting in a median S/N of $\sim$120. We make use of spectra extracted via the official CHIRON pipeline as per \citet{2021AJ....162..176P}. Radial velocities were derived from a least-squares deconvolution between the observation and a nonrotating synthetic template, generated using the ATLAS9 atmosphere models \citep{Castelli:2004} at the spectral parameters of the targets. The derived broadening profile is fit with a kernel accounting for the effects of rotational, macroturbulent, and instrumental broadening and radial velocity shift. In addition, we also estimate stellar atmosphere parameters for the targets. We match the observed spectra against a library of $\sim 10,000$ observed spectra classified by the Spectroscopic Classification Pipeline \citep{buchhave_2012}. The library is interpolated and the observed spectrum classified via a Gradient Boosting Classifier from the \scikit package. The uncertainties in our classification is the scatter in the results from each observed epoch, and do not account for systematic uncertainties that may be present in the template matching. 
\subsubsection{FEROS}
We obtained two spectra of TOI-1982 with the Fiber-fed Extended Range Optical Spectrograph (FEROS, R = 50, 000) mounted on the MPG 2.2-m telescope installed at La Silla Observatory, in Chile. These two observations were executed on January 28, 2020 and March 07, 2020 close to orbital phases 0.25 and 0.75 according to the photometric ephemeris provided by TESS. The adopted exposure time was of 600~s which translated into a S/N of $\sim$90. These observations were performed with the simultaneous calibration mode using a ThAr lamp in the secondary fiber, which allows to trace some instrumental velocity drifts. FEROS data was processed with the CERES pipeline \citep{ceres}, which performs the optimal extraction, wavelength calibration and barycentric correction of the raw data. Precision RVs and line bisector spans were also computed with CERES by using the cross-correlation technique. 
\subsubsection{MINERVA-Australis}
We obtained 26 observations of TOI-1107 from 2020 October 29 to 2021 January 30 using the Minerva-Australis telescope array \citep{minervaaddison}, located at Mt Kent observatory, Australia. Minerva-Australis is composed of four identical 0.7-m CDK700 telescopes feeding into a single environment controlled KiwiSpec spectrograph. The spectrograph has a resolving power of $R\sim$ 80, 000 over the wavelength region of 5000 – 6300~\AA. The adopted exposure time was of 3600~s which translated into a S/N of $\sim$30. Simultaneous wavelength calibration is provided by two additional fibers that bracket the telescope fibers, each illuminated by a quartz lamp through an iodine cell. We optimally extract the spectra from each telescope fiber individually. RVs are measured from each telescope independently via a cross-correlation between individual exposures and a spectral template generated from the median combined observed target star spectra. 
\begin{figure*}[t]
  \centering
  \includegraphics[width=0.7\textwidth]{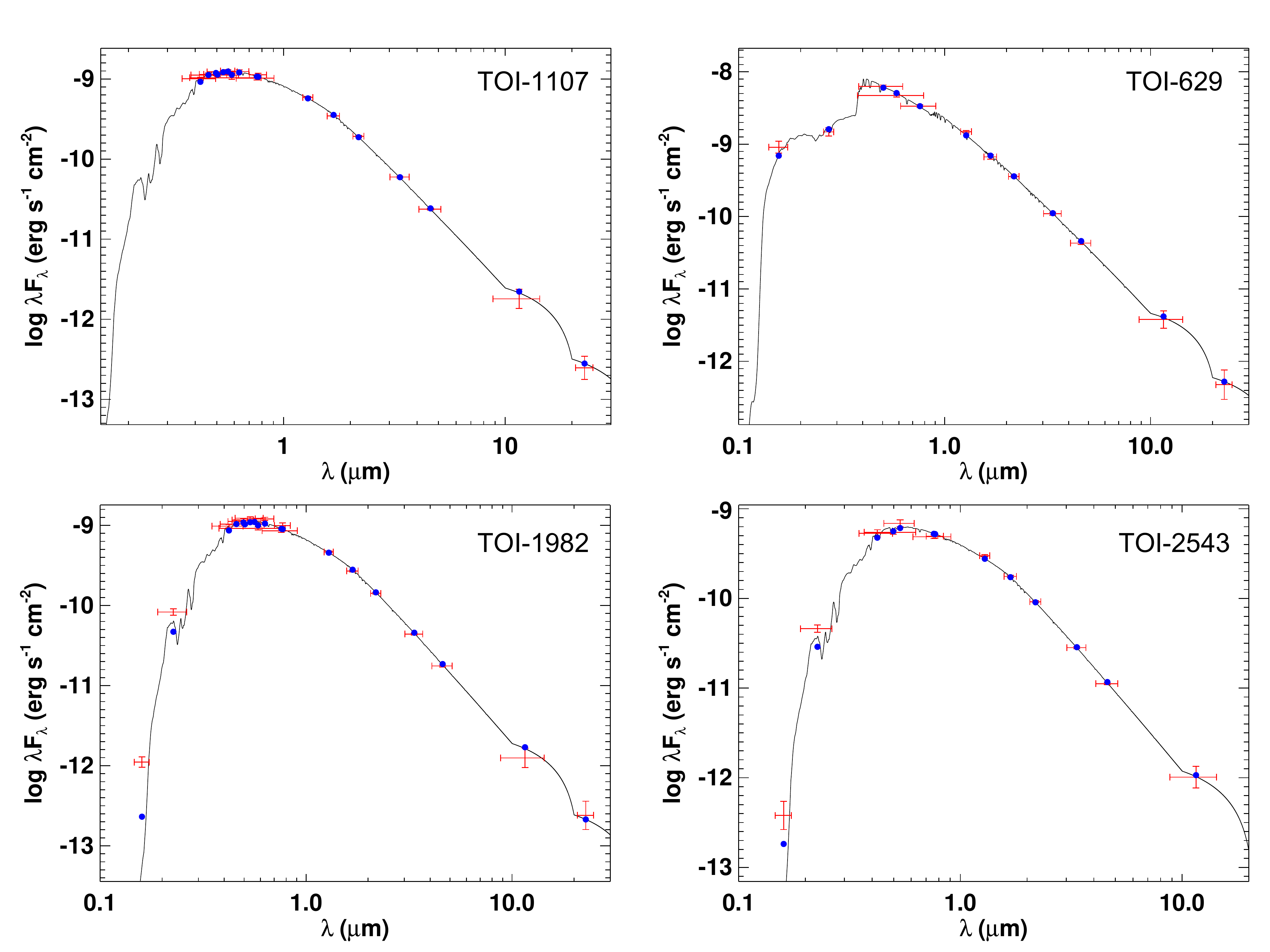}
  \caption{Spectral energy distributions (SEDs) for the four targets studied in this paper. The red symbols represent the observed photometric measurements, where the horizontal bars represent the effective width of the bandpass, and the blue symbols represent the model fluxes from the best-fit Kurucz stellar atmosphere model (black).}
  \label{fig:SEDs}
\end{figure*}
\subsection{High resolution imaging}
TESS's large pixel size (21$\arcsec$) makes it necessary to consider possible flux contamination by nearby stars that can result in flux dilution and therefore underestimation of the observed transit depth or even false-positive transit signals. Stellar neighbors can be ruled out with high angular resolution imaging.

 In this work we used speckle imaging with the 4.1-m Southern Astrophysical Research (SOAR) telescope \citep{SOAR} to confirm that the periodic signals originate from our targets. SOAR speckle imaging was obtained for three of our four targets in I-band, a similar near-IR bandpass as TESS. TOI-1107 was observed on January 08, 2020, TOI-629 on November 11, 2019 and TOI-1982 on July 17, 2021. No nearby sources were detected within 3$\arcsec$. Figure~\ref{fig:soar_shane} shows the contrast curve with the 5$\sigma$ detection limit marked with a black line. The inset images zoomed and centered to the targets that represent the speckle auto-correlation function do not show any stellar companions.

We observed TOI-2543 on March 28, 2021 using the ShARCS camera on the Shane 3-m telescope at Lick Observatory \citep{2012SPIE.8447E..3GK, 2014SPIE.9148E..05G, ShARCS}. The observation was taken with the Shane adaptive optics system in natural guide star mode. The final image was constructed using a sequence of images taken in a 4-point dither pattern with a separation of 4$\arcsec$ between each dither position. The image at each dither position was taken using a $Ks$ filter ($\lambda_0 = 2.15$ $\mu$m, $\Delta \lambda = 0.32$ $\mu$m) with an exposure time of 15~s. A more detailed description of the observing strategy and reduction procedure can be found in \cite{2020AJ....160..287S}. Our reduced image and corresponding contrast curve are shown in Figure ~\ref{fig:soar_shane}. We find no nearby stellar companions within our detection limits.

\section{Analysis}\label{sec:analysis}
\subsection{Host star analysis}\label{sec:stellaranalysis}
\subsubsection{Spectral analysis}\label{sec:spectralanalysis}

In order to determine suitable priors for our later spectral energy distribution (SED) fitting that provides us with the final stellar parameters, we use the CORALIE spectra. Each spectrum was shifted to a common wavelength axis and combined to a 1D coadded spectrum to increase the signal-to-noise.

For the stellar characterization of TOI-1107, TOI-1982, and TOI-2543 we use \SpecMatch \citep{specmatch}. \SpecMatch is able to extract the fundamental properties of a star, including the effective temperature (\teff), surface gravity (log \textit{g}), and metallicity ([Fe/H]), using its optical spectra. These spectra are compared to a library of 404 high resolution (R$\sim$ 60, 000) and high signal-to-noise (S/N > 100) stellar spectra obtained with Keck/HIRES. The code is able to achieve accuracies of 100~K in effective temperature, 15$\%$ in stellar radius and 0.09 dex in metallicity. In order to match our spectra to the library via $\chi^{2}$ minimization we use the spectral region from 5100-5400 $\AA$ that includes the Mgb triplet. For TOI-1107, TOI-1982, and TOI-2543 we ran \SpecMatch on 24, 15, and eight stacked CORALIE spectra, respectively. For TOI-2543, TRES spectra are also used to derive stellar parameters using the Stellar Parameter Classification tool (SPC; \citealt{buchhave_2012}). SPC cross-correlates each observed spectrum against a grid of synthetic spectra based on Kurucz atmospheric model (\citealt{Kurucz1992}) to determine the stellar effective temperature, surface gravity, metallicity, and rotational velocity. The parameters are in agreement with the CORALIE spectral analysis.

Since \SpecMatch is not applicable to stars with effective temperatures higher than $\sim$7000 K, we use a different method to obtain the stellar atmospheric parameters of the A-type star TOI-629. In particular we use the analysis package \iSpec \citep{Blanco-Cuaresma2014} to determine the \teff, \logg, and [Fe/H] using the coadded CORALIE spectra for this target. To generate synthetic spectra we use SPECTRUM \citep{Gray1994} which is a radiative transfer code and ATLAS9 model atmospheres \citep{Castelli:2004}. We estimate macroturbulence with Equation 5.10 from \citet{Doyle2014} and microturbulence is accounted for at the synthesis stage using Equation 3.1 from the same source. We estimate the effective temperature \teff~and gravity \logg using the H$\alpha$, Ca H, and K lines while other available metals are used to determine [Fe/H] and the projected rotational velocity \vsini (Table $\ref{tab:Stellar parameters}$). We fit trial synthetic model spectra until we reach an acceptable match to the data. For the uncertainties estimation we vary individual parameters until the model spectrum is no longer compatible with the data.

\subsubsection{Spectral energy distribution}\label{sec:SED}
As an independent check on the derived stellar parameters, we perform an analysis of the broadband spectral energy distribution (SED) together with the {\it Gaia\/} eDR3 parallax we determined an empirical measurement of the stellar radius, following the procedures described in \citet{Stassun:2016,Stassun:2017,Stassun:2018}. As available, we pulled the $B_T V_T$ magnitudes from {\it Tycho-2}, the $BVgri$ magnitudes from APASS, the $JHK_S$ magnitudes from {\it 2MASS}, the W1--W4 magnitudes from {\it WISE}, and the $G$ magnitude from {\it Gaia}. We also use the {\it GALEX} NUV and/or FUV fluxes when available. Together, the available photometry spans the full stellar SED over the wavelength range 0.35--22~$\mu$m, and extends down to 0.15~$\mu$m when {\it GALEX} data are available (see Figure~\ref{fig:SEDs}). We perform a fit using Kurucz stellar atmosphere models, with the priors on effective temperature (\teff), surface gravity ($\log g$), and metallicity ([Fe/H]) from the spectroscopically determined values. The remaining free parameter is the extinction ($A_V$), which we restrict to the maximum line-of-sight value from the dust maps of \citet{Schlegel:1998}. In the following paragraphs, we summarize the resulting fit parameters for each system in turn.

TOI-1107: The resulting fit (Figure \ref{fig:SEDs}, \textit{top left}) has small reduced $\chi^2$ of 1.2. The best fit extinction is $A_V = 0.20 \pm 0.05$ and the bolometric flux at Earth is $F_{\rm bol} = 1.88 \pm 0.02 \times 10^{-9}$ erg~s$^{-1}$~cm$^{-2}$. This gives the stellar radius as \rstar $= 1.81 \pm 0.06$~R$_\odot$, and the stellar mass from the empirical relations of \citet{Torres:2010} is \mstar $= 1.35 \pm 0.08~M_\odot$, consistent with that obtained from $\log g$ and \rstar of \mstar $= 1.31 \pm 0.18~M_\odot$. In Section \ref{sec:stellarrotation} we provide an estimate of the stellar age using the above stellar radius \rstar, the rotation period found in the light curves and the empirical activity-age relations of \citet{Mamajek2008}.

TOI-629:
The resulting fit (Figure \ref{fig:SEDs}, \textit{top right}) has small reduced $\chi^2$ of 1.2. The best fit extinction is $A_V = 0.37 \pm 0.05$. Integrating the (unreddened) model SED gives the bolometric flux at Earth of $F_{\rm bol} = 10.61 \pm 0.04 \times 10^{-9}$ erg~s$^{-1}$~cm$^{-2}$. Taking the $F_{\rm bol}$ and \teff~together with the {\it Gaia\/} eDR3 parallax \citep[with no adjustment for systematic offset; see][]{StassunTorres:2021}, gives the stellar radius as  \rstar $= 2.37 \pm 0.11$~R$_\odot$. Finally, estimating the stellar mass from the empirical relations of \citet{Torres:2010} gives \mstar $ = 2.16 \pm 0.13~M_\odot$, which is roughly consistent with that obtained empirically via the spectroscopic $\log g$ together with \rstar, giving \mstar $= 3.25 \pm 0.72~M_\odot$. In addition, \rstar together with the spectroscopic \vsini gives an estimate of the projected rotation period, $P_{\rm rot}/\sin \it{i_{*}}$ = 1.77 $\pm$ 0.10~days. We can integrate the SED at $\mathrm{\lambda}$ < 912~$\AA$ to obtain an estimate of the XUV irradiation from the star at a distance of 1~AU, $F_{\rm XUV,1AU} = 1.2 \pm 0.4$ erg~s$^{-1}$~cm$^{-2}$, though we adopt a factor of 10 systematic error on this flux due to uncertainties in the model atmospheres in the UV \citep[see, e.g.,][for the example case of KELT-9]{Gaudi:2017}. Finally, the system appears to be a kinematic member of the Theia~838 ``stellar string''. The Gaia distance and proper motions (in both right ascension and declination) for this star are consistent with the distributions in distance and proper motion of other stellar members of that association, as determined by \citet{Kounkel:2020}, giving an estimated age of $\tau_* = 0.32 \pm 0.13$~Gyr.

TOI-1982:
The resulting fit (Figure \ref{fig:SEDs}, \textit{bottom left}) has small reduced $\chi^2$ of 1.5, excluding the {\it GALEX} data which suggests chromospheric activity. The best fit extinction is $A_V = 0.01 \pm 0.01$, and the bolometric flux at Earth is $F_{\rm bol} = 1.40 \pm 0.02 \times 10^{-9}$ erg~s$^{-1}$~cm$^{-2}$. This gives the stellar radius as \rstar $= 1.51~\pm~0.05~$R$_\odot$, and the stellar mass from the empirical relations of \citet{Torres:2010} is \mstar $= 1.41 \pm 0.08~M_\odot$, which is somewhat inconsistent with that obtained from $\log g$ and \rstar of \mstar $= 0.71 \pm 0.10~M_\odot$, suggesting the spectroscopic $\log g$ in this case is underestimated. We can use the {\it GALEX} excess to estimate the chromospheric activity via the empirical relations of \citet{Findeisen2011}, giving $\log R'_{HK} = -4.83 \pm 0.10$, which from the empirical activity-age relations of \citet{Mamajek2008} gives the system age to be $\tau_* = 3.6 \pm 1.5$~Gyr.

TOI-2543:
The resulting fit (Figure \ref{fig:SEDs}, \textit{bottom right}) has small reduced $\chi^2$ of 0.8, excluding the {\it GALEX} photometry which suggests chromospheric activity. The best fit extinction is $A_V = 0.02 \pm 0.02$, and the bolometric flux at Earth is $F_{\rm bol} = 0.77 \pm 0.02 \times 10^{-10}$ erg~s$^{-1}$~cm$^{-2}$. This gives the stellar radius as \rstar $= 1.86~\pm~0.15$~R$_\odot$, and the stellar mass from the empirical relations of \citet{Torres:2010} is \mstar $= 1.29 \pm 0.08~M_\odot$. In addition, \rstar together with the spectroscopic \vsini gives an estimate of the projected rotation period, $P_{\rm rot}/\sin \it{i_{*}}$ = 8.73 $\pm$ 0.82 days. Finally, as with TOI-1982 we can estimate the chromospheric activity from the {\it GALEX} excess, which gives $\log R'_{\rm HK}$ = -4.95~$\pm$~0.05, yielding an estimated age of $\tau_*$ = 5.6 $\pm$ 0.9~Gyr.

All the stellar parameters are listed in Table $\ref{tab:Stellar parameters}$. The stellar mass that we present in the table and use for our analysis is the one estimated from the empirical relations of \citet{Torres:2010}. The stellar luminosity is calculated using the stellar radius and mass from the table.

\begin{figure}[h]
  \centering
 \includegraphics[width=0.46\textwidth]{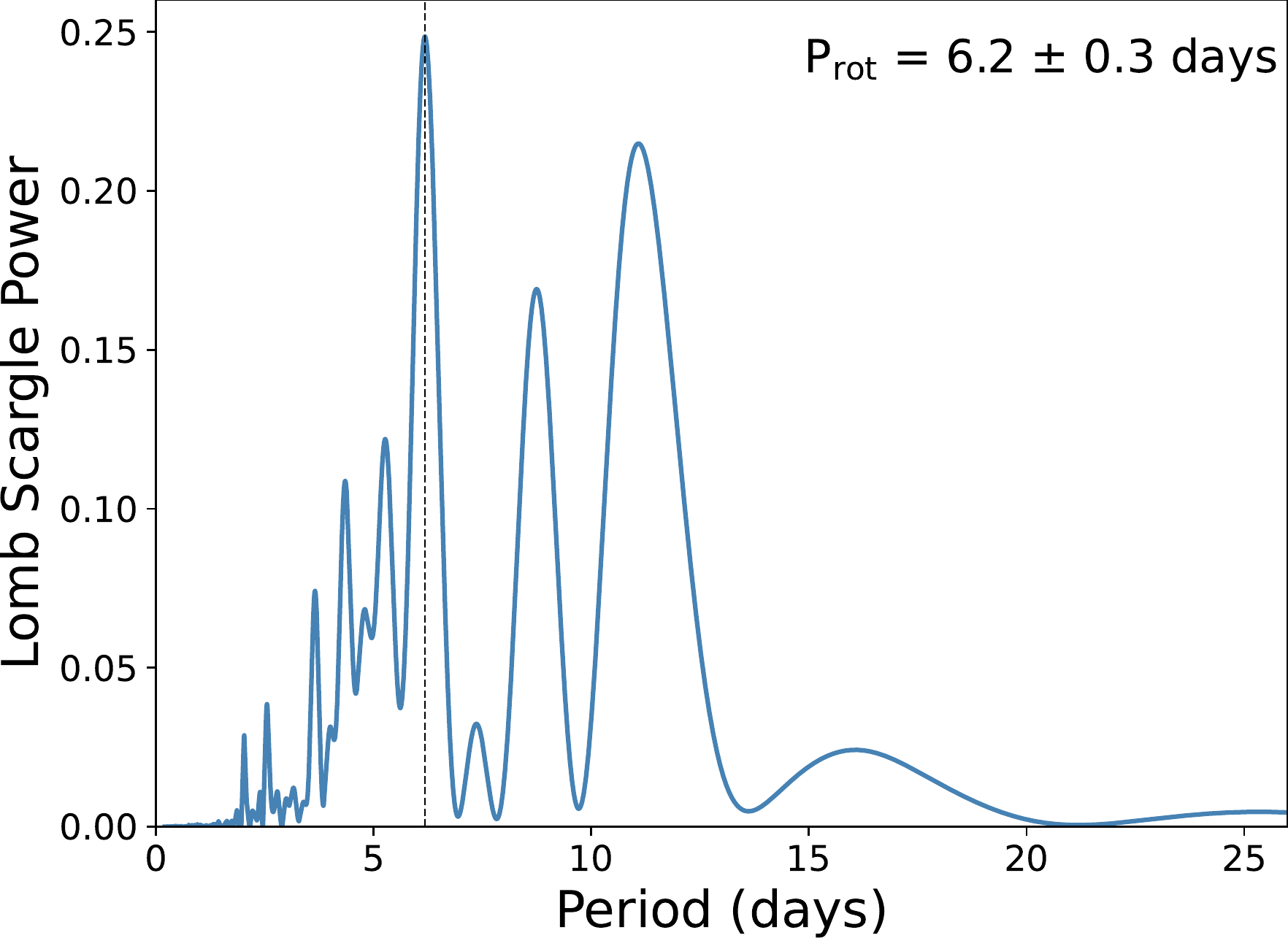}
  \caption{Lomb-Scargle periodogram from the TESS light SAP-flux Sector 38 and 39 light curves of TOI-1107. The periodogram indicates a peak frequency at 6.2 $\pm$ 0.3 days which is in agreement with the rotational period obtained from the
CORALIE calibration.}
  \label{fig:TOI1107_LombScarge}
\end{figure}

\subsection{Joint transit and RV analysis}\label{sec:GlobalAnalysis}
We retrieved the companion parameters of the four systems using a joint photometric and radial-velocity modeling with \juliet (\citealt{juliet}). \juliet is a publicly available tool that allows to fit both transit models (via \batman package, \citealt{batman}) and RVs (via \radvel package, \citealt{radvel}) simultaneously and model the noise using Gaussian processes (GPs) (via \celerite package, \citealt{Foreman:2017}). The code uses nested sampling algorithms in order to explore the parameter space with \dynesty \citep{dynesty} and performs a proper model comparison via Bayesian evidences (lnZ) with \multinest \citep{multinest} using the \pymultinest \citep{pymultinest} Python software package. For the transit model, \juliet performs an efficient parametrization by fitting for the parameters $r_{1}$ and $r_{2}$ to ensure uniform exploration of the $p$ (planet-to-star ratio) and $b$ (impact parameter) space. Also, \juliet fits for the stellar density $\rho_{*}$ that is combined with the orbital period to provide the scaled semi-major axis ($\it{a/\rstar}$).

For our joint analyses we use the TESS light curves, the ground based photometry and the follow-up RV measurements from CORALIE, CHIRON, TRES, FEROS, and MINERVA-Australis. As first step we isolate each TESS transit by applying a cut of 1.5 days before and after each transit midpoint. For our analysis we use the stellar density and its uncertainty derived from our SED fit as a Gaussian prior. The quadratic stellar limb-darkening coefficients and their uncertainties are derived using the \ldcu$\footnote{\url{https://github.com/delinea/LDCU}}$ code, a modified version of the python routine implemented by \citet{limb_espinoza}, for each
photometric filter used. \ldcu computes the limb-darkening coefficients and their corresponding uncertainties using a set of stellar intensity profiles accounting for the uncertainties on the stellar parameters. For all TESS data the determined limb-darkening coefficients are used as Gaussian priors in our global analysis. The precision of the ground based photometry does not allow us to fit the limb darkening parameters and therefore we keep them fixed. A white-noise jitter term is added in quadrature to the error bars of both photometry and RV data to account for underestimated uncertainties and additional noise that was not captured by our modeling. The jitter terms are first fit using large log-uniform priors. Considering that they are neither significantly different than zero, we decided to fix them to zero. Since there are no significant contamination sources nearby TOI-629, TOI-1982, and TOI-2543 we fix the dilution factor of both TESS and ground photometry to one. For TOI-1107, as we mentioned in Section \ref{sec:TESSLCs}, the contamination from its nearby star (1$\%$) is much smaller than the transit depth uncertainty (5$\%$) therefore the affect is negligible and the dilution factor is fixed to one as well. 

We model our TESS data using a Gaussian process (GP) fit independently to the 2-minute and 30-minute cadence data. To do this, we use a celerite (approximate) Mat\'ern-3/2 kernel (\citealt{Foreman:2017}) with hyperparameters amplitude ($\sigma_{GP}$) and timescale ($\rho_{GP}$). The low level of correlated noise in TOI-2543 2-minute cadence light curves does not necessitate further detrending. The ground-based photometric data are affected by correlated noise that originates from atmospheric, instrumental or stellar effects. To account for it we use a combination of polynomials in different variables (time, airmass, FWHM, and sky background). To understand the origin of the systematic and find a good combination of polynomials to model each light curve we use the minimization of the Bayesian Information and selected models with the smallest Bayesian information criterion (BIC) and the smallest number of free parameters (less complex model). 

For our joint analysis we first fit our radial velocity data with simple Keplerian models on DACE\textsuperscript{\ref{dacefootnote}} in order to determine mean values for the radial velocity semi-amplitude (K) and the systemic radial velocity ($\mu$). Afterward, we fit wide uniform priors centered on the predicted K and of $\mu$ of each companion. The results of the global MCMC analysis of TOI-1107, TOI-629, TOI-1982, and TOI-2543 can be found in Tables ~\ref{tab:Juliet6292543}, ~\ref{tab:Juliet11071982}.

\citet{Barnes2009} demonstrated how the gravity darkening induced by rapid stellar rotation can distort the light curves of any transiting companions, and how this can be used to study the orbital alignment. As our host star analysis reveals that TOI-629 has a \vsini of 56.93 $\pm$ 3.61, we test the effect of fitting for a gravity darkening signature in the light curve following the strategy of \citet{Hooton2022}. Although a comparison between this fit and \juliet fit showed no clear evidence for these distortions, a measurement of the orbital alignment from the transit light curve may be possible in the future when more data have been acquired.

\begin{table*}
\caption{Stellar and companion parameters for TOI-1107, TOI-629, TOI-1982, and TOI-2543 systems.}
\centering
\tiny
\renewcommand{\arraystretch}{1.2}
\begin{tabular}{lccccc}
\hline\hline
\textbf{Property} & \textbf{TOI-1107}& \textbf{TOI-629} & \textbf{TOI-1982}& \textbf{TOI-2543}& \textbf{Source} \\
 \hline
    \textbf{Identifiers}                & & \\
    TIC ID      & 394561119             & 293853437& 437329044& 270604417 & TICv8 \\
    2MASS ID    & J10222579-8213080    &J06231868+0053000&J13502038-2323001& J09062327+0334036 & 2MASS \\
    Gaia ID     & 5192364501633274880 & 3123364264204562944 & 6287361053627037696 & 579084083968560256 & GAIA DR2 \\
    & & \\
    \textbf{Astrometric properties}     & & \\
    R.A. (J2000)  & 10:22:25.82         &06:23:18.69&13:50:20.36& 09:06:23.26
 & GAIA DR2 \\
    Dec (J2000)   & -82:13:07.97        & 00:53:00.11 &-23:23:00.16&03:34:03.61& GAIA DR2 \\
    Parallax (mas)  & 3.53 $\pm$ 0.03     & 2.97 $\pm$ 0.07 & 3.72 $\pm$ 0.04 & 2.33 $\pm$ 0.18 & GAIA DR2 \\
    Distance (pc)   &  283.61 $\pm$ 2.03   & 336.29 $\pm$ 8.09 & 268.54 $\pm$ 2.87 & 429.61 $\pm$ 33.42 & GAIA DR2 \\
    $\mu_{\rm{R.A.}}$ (mas yr$^{-1}$)     & 9.033 $\pm$ 0.015 & 6.521 $\pm$ 0.020 & -29.579 $\pm$ 0.021 & -7.922 $\pm$ 0.194  & GAIA DR2 \\
    $\mu_{\rm{Dec}}$ (mas yr$^{-1}$)     & 3.101 $\pm$ 0.014 & -3.758 $\pm$ 0.018 & -6.484 $\pm$ 0.018  & -5.917 $\pm$ 0.147 & GAIA DR2 \\
    & & \\
    \textbf{Photometric properties} & & \\
    TESS (mag)  & 10.007 $\pm$ 0.006  & 8.694 $\pm$ 0.006 & 10.192 $\pm$ 0.006 & 10.816 $\pm$ 0.006 & TICv8 \\
    B (mag)     & 11.038 $\pm$ 0.085   & 8.838 $\pm$ 0.036 & 11.064 $\pm$ 0.088&  11.718 $\pm$ 0.181 & Tycho-2 \\
    V (mag)     & 10.553 $\pm$ 0.006    & 8.737 $\pm$ 0.003 & 10.551 $\pm$ 0.006& 11.178 $\pm$ 0.012 & Tycho-2 \\
    G (mag)     & 10.377 $\pm$ 0.003  & 8.7320 $\pm$ 0.0004 & 10.5063 $\pm$ 0.0005 & 11.1730 $\pm$ 0.0004 & GAIA DR2 \\
    J (mag)     & 9.545 $\pm$ 0.024     & 8.540 $\pm$ 0.037 & 9.798 $\pm$ 0.028 & 10.280 $\pm$ 0.024 & 2MASS \\
    H (mag)     & 9.322 $\pm$ 0.023     & 8.607 $\pm$ 0.075 & 9.592 $\pm$ 0.022 & 10.052 $\pm$ 0.023 & 2MASS \\
    K (mag)     & 9.205 $\pm$ 0.019    & 8.533 $\pm$ 0.021 & 9.54 $\pm$ 0.019 & 10.010 $\pm$ 0.023 & 2MASS \\
        & & \\
    \textbf{Stellar parameters} & & \\
    \teff~(K)  &  6311 $\pm$ 98 & 9100 $\pm$ 200 & 6325 $\pm$ 110 &  6060 $\pm$ 82 & Sect. \ref{sec:spectralanalysis}  \\ 
    $[Fe/H]$ dex & -0.10 $\pm$ 0.09 & 0.10 $\pm$ 0.15 & -0.10 $\pm$ 0.09 & -0.28 $\pm$ 0.10 & Sect. \ref{sec:spectralanalysis}  \\   
    log $g$ (cgs)  & 4.05 $\pm$ 0.04 &  4.02 $\pm$ 0.05 & 4.23 $\pm$ 0.04 & 4.01 $\pm$ 0.08 & Sect. \ref{sec:spectralanalysis}  \\  
    \vsini ($\kms$) &  12.31 $\pm$ 0.79 & 56.93 $\pm$ 3.61 &  37.58 $\pm$ 2.38 & 8.2 $\pm$ 0.5 & Sect. \ref{sec:stellarrotation} \\    
    \rstar (R$_\odot$)  & 1.81 $\pm$ 0.06 & 2.37 $\pm$ 0.11  & 1.51 $\pm$ 0.05 & 1.86 $\pm$ 0.15 & Sect. \ref{sec:SED}  \\    
    \mstar (M$_\odot$)  &  1.35 $\pm$ 0.08  & 2.16 $\pm$ 0.13  & 1.41 $\pm$ 0.08 & 1.29 $\pm$ 0.08 & Sect. \ref{sec:SED}  \\ 
     $Age$ (Gyr)  & 2.6 $\pm$ 0.2$^*$ & 0.32 $\pm$ 0.13 &\textbf{ 3.6 $\pm$ 1.5 }& 5.6 $\pm$ 0.9  & Sect. \ref{sec:stellarrotation}$^*$, \ref{sec:SED}   \\ 
    $A_{v}$ (mag)  &  0.20 $\pm$ 0.05 & 0.37 $\pm$ 0.05 & 0.01 $\pm$ 0.01 & 0.02 $\pm$ 0.02 & Sect. \ref{sec:SED}  \\ 

    $L_{*}$ (L$_\odot$)  & 4.69 $\pm$ 0.43 & 34.78 $\pm$ 4.45 & 3.29 $\pm$ 0.32 & 4.21 $\pm$ 0.72 & \ref{sec:SED}  \\ 
    
    $\chi^{2}_{v}$   & 1.2 & 1.2 & 1.5 & 0.8 & Sect. \ref{sec:SED} \\ 
    $F_{bol}$ (10$^{-9}$ erg s$^{-1}$ cm$^{-2}$) & 1.88 $\pm$ 0.02 & 10.61 $\pm$ 0.04 & 1.40 $\pm$ 0.02 & 0.77 $\pm$ 0.02 & Sect. \ref{sec:SED}  \\ 
    $P_{rot}$/$\sin{\it i_*}$ (days)  & - & 1.77 $\pm$ 0.10 & - & 8.73 $\pm$ 0.82 & Sect. \ref{sec:SED}  \\ \\ 
    \textbf{Companion parameters} & & \\
        P (days) & 4.0782387 $^{+0.0000024}_{-0.0000025}$& 8.717993 $^{+0.000012}_{-0.000013}$& 17.172446 $^{+0.000043}_{-0.000044}$& 7.542776 $^{+0.000031}_{-0.000031}$ & Sect. \ref{sec:GlobalAnalysis} \\[1.1ex] 
        p = $R_{b}/R_{*}$ &  0.07379 $^{+0.00033}_{-0.00030}$ & 0.04790 $^{+0.00043}_{-0.00042}$ & 0.06914 $^{+0.00095}_{-0.00096}$& 0.05299 $^{+0.00077}_{-0.00076}$ & Sect. \ref{sec:GlobalAnalysis}\\[1.1ex]   
        b = ($\it{a/\rstar}$) cos $i$ & 0.154 $^{+0.107}_{-0.099}$ &0.203 $^{+0.128}_{-0.134}$  & 0.819 $^{+0.012}_{-0.013}$& 0.185 $^{+0.133}_{-0.122}$& Sect. \ref{sec:GlobalAnalysis}\\[1.1ex]   
        $i$ &  88.63 $^{+0.89}_{-0.96}$ & 88.65 $^{+0.90}_{-0.91}$ & 88.21 $^{+0.07}_{-0.08}$ & 88.85 $^{+0.76}_{-0.88}$& Sect. \ref{sec:GlobalAnalysis}\\[1.1ex]   
        $\textit{e}$ &0.025 $^{+0.023}_{-0.016}$ & 0.298$^{+0.008}_{-0.008}$ & 0.272 $^{+0.014}_{-0.014}$ & 0.009 $^{+0.003}_{-0.002}$ & Sect. \ref{sec:GlobalAnalysis}\\[1.1ex]   
        M$_{b}$ (\mjup) & 3.35 $^{+0.18}_{-0.18}$  & 66.98 $^{+2.96}_{-2.95}$ & 65.85 $^{+2.75}_{-2.72}$ & 67.62 $^{+3.45}_{-3.45}$& Sect. \ref{sec:GlobalAnalysis}\\[1.1ex]   
        R$_{b}$ (\rjup) & 1.30 $^{+0.05}_{-0.05}$  & 1.11 $^{+0.05}_{-0.05}$& 1.08 $^{+0.04}_{-0.04}$ & 0.95 $^{+0.09}_{-0.09}$& Sect. \ref{sec:GlobalAnalysis}\\[1.1ex]   
        $\rho_{b}$ (g cm$^{-3}$)& 1.89 $^{+0.23}_{-0.22}$  & 61.16 $^{+9.34}_{-9.33}$& 66.06 $^{+8.10}_{-8.09}$ &97.53 $^{+29.35}_{-29.36}$& Sect. \ref{sec:GlobalAnalysis}\\[1.1ex]   
        $\textit{a}$ (AU) &  0.0561 $^{+0.0024}_{-0.0025}$  & 0.1090 $^{+0.0058}_{-0.0064}$&0.1457 $^{+0.0066}_{-0.0066}$  &0.0788 $^{+0.0079}_{-0.0081}$& Sect. \ref{sec:GlobalAnalysis} \\[1.1ex]   
        T$_{eq}$ (K) & 1728 $^{+27}_{-27}$  & 2047 $^{+45}_{-45}$ & 982 $^{+17}_{-17}$ & 1419 $^{+26}_{-26}$& This work$^{\it{(\alpha)}}$\\[1.1ex] 
        $\fave$ (10$^{8}$ erg s$^{-1}$ cm$^{-2}$) & 7.24 $^{+1.25}_{-1.26}$  & 1.92 $^{+0.35}_{-0.36}$& 1.07 $^{+0.19}_{-0.19}$& 3.67 $^{+0.92}_{-0.93}$& This work$^{\it{(b)}}$\\ [1.1ex]
    \hline

\end{tabular}
\\
\begin{tablenotes}
\item Sources: TICv8 \citep{Stassun2019}, 2MASS \citep{Skrutskie2006}, Gaia Data Release 2 \citep{GaiaCollaboration2018}, Tycho-2 \citep{Tycho}
\\\textbf{}
\textbf{Notes:} $^{\it{(\alpha)}}$ The equilibrium temperature is calculated using T$_{eq}$ = \teff(1-A)$^{1/4}$$\sqrt{\frac{R_{*}}{2\textit{a}}}$, assuming an albedo A = 0. $^{\it{(b)}}$ The incident flux is calculated using $\fave$ = $\frac{L_{*}}{4\pi\textit{a}^{2}}$.
\end{tablenotes}
\label{tab:Stellar parameters}
\end{table*}

\section{Discussion}\label{sec:discussionconclusion}

\subsection{TOI-1107b: A massive planet transiting a F-type star}
TOI-1107b is a massive hot Jupiter with an estimated mass of 3.35 $\pm$ 0.18~\mjup and a radius of 1.30 $\pm$ 0.05~\rjup that transits a F6V star with a period of 4.0782387 $^{+0.0000024}_{-0.0000025}$ days. To date, only 17 confirmed planets more massive than 3~\mjup with well-constrained densities ($\sigma_{M}/M$ $\leq$ 25$\%$ and $\sigma_{R}/R$ $\leq$ 8$\%$) orbiting stars with \teff~> 6200~K have been identified (PlanetS catalog of \citet{otegi2020}, accessible on the Data \& Analysis Center for Exoplanet DACE\textsuperscript{\ref{dacefootnote}}). \par
Massive planets are scientifically interesting since they can provide insight into the planet formation processes at the transition between giant massive planets and brown dwarfs. \citet{Santos2017} studied the statistical properties of giant planets, together with those of their host stars in a search for clues about the process of planet formation and evolution. The results suggested that there are two distinct giant planet populations with masses above and below $\sim$ 4~\mjup. More specifically, they found that lower-mass planets may form by the core-accretion process and orbit metal-rich stars, while more massive giant planets may form via disk instability and their hosts have lower average metallicity values. This theory was corroborated by \citet{Schlaufman2018} who found that planets with masses $\lesssim$ 4~\mjup orbit preferentially more metal-rich stars, unlike planets with masses $\gtrsim$ 10~\mjup that do not share this property. With a metallicity of [Fe/H] = -0.10 $\pm$ 0.09 the TOI-1107 system is not consistent with the theory presented in the above studies.

\subsection{Massive planet and BD eccentricity distribution}\label{sec:EccentricityMass}

\begin{figure}[ht]
  \centering
 \includegraphics[width=0.5\textwidth]{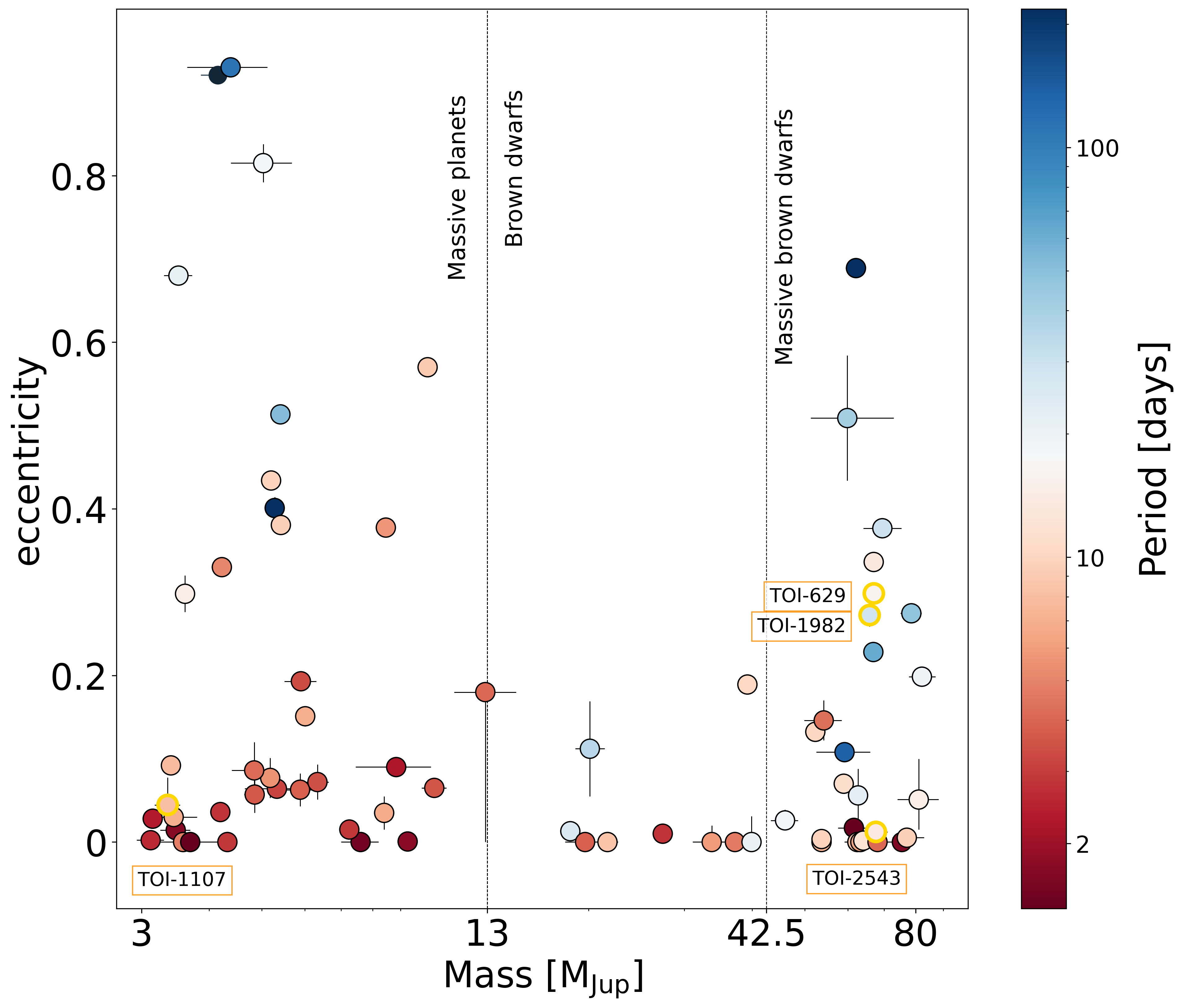}
  \caption{Eccentricity-companion mass distribution of known massive planets (from the updated exoplanet catalog of \citealt{otegi2020}) and transiting brown dwarfs (Table \ref{tab:browndwarf_list}). The different colors indicate the orbital period. The massive planet (TOI-1107b) and three brown dwarfs (TOI-629b, TOI-1982b, and TOI-2543b) presented in this work are displayed with yellow edge color.}
  \label{fig:eccmass}
\end{figure}

Figure $\ref{fig:eccmass}$ (\textit{bottom left}) shows the eccentricity distribution for massive planets and transiting brown dwarfs. The color of each point denotes the orbital period. The first vertical line (13~\mjup) represents the transition between massive planets and low-mass brown dwarfs and the second (42.5~\mjup) the transition between low-mass and high-mass brown dwarfs. 

TOI-629b is a 67.0 $\pm$ 3.0~\mjup brown dwarf with a radius of 1.11 $\pm$ 0.05~\rjup that transits a A2V star with a period of 8.717993 $^{+0.000012}_{-0.000013}$ days. TOI-1982b is a brown dwarf with a mass of 65.8 $\pm$ 2.7~\mjup and a radius of 1.08 $\pm$ 0.04~\rjup transiting a F7V star with a period of 17.172446  $^{+0.000043}_{-0.000044}$ days. TOI-2543b is a 67.6 $\pm$ 3.4~\mjup massive brown dwarf with a radius of 0.95 $\pm$ 0.09~\rjup that transits a F9V star with a period of 7.542776 $\pm$ 0.000031 days. The diagram places TOI-629 and TOI-1982 among the most eccentric systems with eccentricities of 0.298 $\pm$ 0.008 and 0.272 $\pm$ 0.014, respectively. The detections of TOI-629b and TOI-1982b are consistent with the \cite{maandge} observations regarding the different eccentricity distribution of brown dwarfs with masses greater and lower than 42.5~\mjup. More specifically, there is strong evidence that brown dwarfs less massive than 42.5~\mjup tend to have lower eccentricities while brown dwarfs with masses above this threshold have higher and more diverse eccentricities. This could be explained by the different formation mechanisms with the less massive being formed via disk gravitational instability \citep{Boss1997} and the more massive through molecular cloud fragmentation \citep{HennebelleChabrier2008}, similar to
the formation of stellar binaries. Since the eccentricity distribution of these targets is consistent with that of binary stars, we see that there are no eccentric brown dwarfs at short periods ($\lesssim$ 10 days), which can be explained by the tidal circularization effect (\citealt{Laughlin06}). 
Massive planets show more diversity in their eccentricities compared to the brown dwarf population. TOI-1107b has a low, typical eccentricity which is compatible with planets with similar masses.
\begin{table*}
\centering
\tiny
\renewcommand{\arraystretch}{1.35}
\caption{List of published transiting brown dwarfs as of December 2021.}
\begin{adjustbox}{width=\textwidth}
\begin{tabular}{lccrccccccc}
\hline\hline
Name & M$_{\rm2}$ (\mjup) & R$_{\rm2}$ (\rjup) & $P$ (days) & $D$ (mmag) &$ecc$ & M$_{\rm1}$ (\msol) & R$_{\rm1}$ (\rsol) & \teff~(K) & \feh & Ref. \\

\hline

HATS-70b & 12.9$_{-1.6}^{+1.8}$ & 1.38$_{-0.07}^{+0.08}$ & 1.89 & 5.928 & <0.18 & 1.78\,$\pm$\,0.12 & 1.88$_{-0.07}^{+0.06}$ & 7930$_{-820}^{+630}$ & 0.04$_{-0.11}^{+0.10}$ & (1) \\ 
TOI-1278b & 18.5\,$\pm$\,0.5 & 1.09$_{-0.20}^{+0.24}$ & 14.48 & 40.041 & 0.013\,$\pm$\,0.004 & 0.54\,$\pm$\,0.02 & 0.57\,$\pm$\,0.01 & 3799\,$\pm$\,42 & -0.01\,$\pm$\,0.28 & (2) \\ 
GPX-1b & 19.7\,$\pm$\,1.6 & 1.47\,$\pm$\,0.10 & 1.74 & 9.723 & 0.000\,$\pm$\,0.000 & 1.68\,$\pm$\,0.10 & 1.56\,$\pm$\,0.10 & 7000\,$\pm$\,200 & 0.35\,$\pm$\,0.10 & (3) \\ 
Kepler-39b & 20.1$_{-1.2}^{+1.3}$ & 1.24$_{-0.10}^{+0.09}$ & 21.09 & 8.590 & 0.112\,$\pm$\,0.057 & 1.29$_{-0.07}^{+0.06}$ & 1.40\,$\pm$\,0.10 & 6350\,$\pm$\,100 & 0.10\,$\pm$\,0.14 & (4) \\ 
CoRoT-3b & 21.7\,$\pm$\,1.0 & 1.01\,$\pm$\,0.07 & 4.26 & 4.590 & 0 (fixed) & 1.37\,$\pm$\,0.09 & 1.56\,$\pm$\,0.09 & 6740\,$\pm$\,140 & -0.02\,$\pm$\,0.06 & (5) \\ 
KELT-1b & 27.4\,$\pm$\,0.9 & 1.12$_{-0.03}^{+0.04}$ & 1.22 & 6.302 & 0.010$_{-0.007}^{+0.010}$ & 1.33\,$\pm$\,0.06 & 1.47\,$\pm$\,0.04 & 6516\,$\pm$\,49 & 0.05\,$\pm$\,0.08 & (6) \\ 
NLTT 41135b & 33.7$_{-2.6}^{+2.8}$ & 1.13$_{-0.17}^{+0.27}$ & 2.89 & 317.042 & <0.02 & 0.19$_{-0.02}^{+0.03}$ & 0.21$_{-0.01}^{+0.02}$ & 3230\,$\pm$\,130 & -0.25\,$\pm$\,0.25 & (7) \\ 
WASP-128b & 37.2\,$\pm$\,0.8 & 0.94\,$\pm$\,0.02 & 2.21 & 7.244 & <0.007 & 1.16\,$\pm$\,0.04 & 1.15\,$\pm$\,0.02 & 5950\,$\pm$\,50 & 0.01\,$\pm$\,0.12 & (8) \\ 
CWW 89Ab & 39.2$_{-1.1}^{+0.9}$ & 0.94\,$\pm$\,0.02 & 5.29 & 9.157 & 0.189\,$\pm$\,0.002 & 1.10\,$\pm$\,0.04 & 1.03\,$\pm$\,0.02 & 5755\,$\pm$\,49 & 0.20\,$\pm$\,0.09 & (9) (10) \\ 
KOI-205b & 39.9\,$\pm$\,1.0 & 0.81\,$\pm$\,0.02 & 11.72 & 10.082 & <0.031 & 0.93\,$\pm$\,0.03 & 0.84\,$\pm$\,0.02 & 5237\,$\pm$\,60 & 0.14\,$\pm$\,0.12 & (11) \\ 
TOI-1406b & 46.0$_{-2.7}^{+2.6}$ & 0.86\,$\pm$\,0.03 & 10.57 & 4.423 & 0.026$_{-0.010}^{+0.013}$ & 1.18$_{-0.09}^{+0.08}$ & 1.35\,$\pm$\,0.03 & 6290\,$\pm$\,100 & -0.08\,$\pm$\,0.09 & (12) \\ 
EPIC 212036875b & 52.3\,$\pm$\,1.9 & 0.87\,$\pm$\,0.02 & 5.17 & 3.727 & 0.132\,$\pm$\,0.004 & 1.29$_{-0.06}^{+0.07}$ & 1.50\,$\pm$\,0.03 & 6238$_{-60}^{+59}$ & 0.01\,$\pm$\,0.10 & (10) (13) \\ 
TOI-503b & 53.7\,$\pm$\,1.2 & 1.34$_{-0.15}^{+0.26}$ & 3.68 & 6.803 & 0 (fixed) & 1.80\,$\pm$\,0.06 & 1.70$_{-0.04}^{+0.05}$ & 7650$_{-160}^{+140}$ & 0.30$_{-0.09}^{+0.08}$ & (14) \\ 
TOI-852b & 53.7$_{-1.3}^{+1.4}$ & 0.83\,$\pm$\,0.04 & 4.95 & 2.573 & 0.004$_{-0.003}^{+0.004}$ & 1.32$_{-0.04}^{+0.05}$ & 1.71\,$\pm$\,0.04 & 5768$_{-81}^{+84}$ & 0.33\,$\pm$\,0.09 & (15) \\ 
AD 3116b & 54.2\,$\pm$\,4.3 & 1.02\,$\pm$\,0.28 & 1.98 & 124.696 & 0.146\,$\pm$\,0.024 & 0.28\,$\pm$\,0.02 & 0.29\,$\pm$\,0.08 & 3184\,$\pm$\,29 & 0.16\,$\pm$\,0.10 & (16) \\ 
CoRoT-33b & 59.0$_{-1.7}^{+1.8}$ & 1.10\,$\pm$\,0.53 & 14.994 & 5.82 & 0.070\,$\pm$\,0.002 & 0.86\,$\pm$\,0.04 & 0.94$_{-0.08}^{+0.14}$ & 5225\,$\pm$\,80 & 0.44\,$\pm$\,0.10 & (17) \\ 
RIK 72b & 59.2$_{-6.7}^{+6.8}$ & 3.10\,$\pm$\,0.31 & 97.76 & 113.94 & 0.108$_{-0.006}^{+0.012}$ & 0.44\,$\pm$\,0.04 & 0.96\,$\pm$\,0.10 & 3349\,$\pm$\,142 & 0.00\,$\pm$\,0.10 & (18) \\ 
TOI-811b & 59.9$_{-8.6}^{+13.0}$ & 1.26\,$\pm$\,0.06 & 25.17 & 10.812 & 0.509\,$\pm$\,0.075 & 1.32$_{-0.07}^{+0.05}$ & 1.27$_{-0.09}^{+0.06}$ & 6107\,$\pm$\,77 & 0.40$_{-0.09}^{+0.07}$ & (15) \\ 
TOI-263b & 61.6\,$\pm$\,4.0 & 0.91\,$\pm$\,0.07 & 0.56 & 47.264 & 0.017$_{-0.010}^{+0.009}$ & 0.44\,$\pm$\,0.04 & 0.44\,$\pm$\,0.03 & 3471\,$\pm$\,33 & 0.00\,$\pm$\,0.10 & (19) \\ 
KOI-415b & 62.1\,$\pm$\,2.7 & 0.79$_{-0.07}^{+0.12}$ & 166.79 & 4.374 & 0.689\,$\pm$\,0.000 & 0.94\,$\pm$\,0.06 & 1.25$_{-0.10}^{+0.15}$ & 5810\,$\pm$\,80 & -0.24\,$\pm$\,0.11 & (20) \\ 
WASP-30b & 62.5\,$\pm$\,1.2 & 0.95$_{-0.02}^{+0.03}$ & 4.16 & 5.133 & 0 (fixed) & 1.25$_{-0.04}^{+0.03}$ & 1.39\,$\pm$\,0.03 & 6202$_{-51}^{+42}$ & 0.08$_{-0.05}^{+0.07}$ & (21) \\ 
LHS 6343c & 62.7\,$\pm$\,2.4 & 0.83\,$\pm$\,0.02 & 12.71 & 53.175 & 0.056\,$\pm$\,0.032 & 0.37\,$\pm$\,0.01 & 0.38\,$\pm$\,0.01 & 3130\,$\pm$\,20 & 0.04\,$\pm$\,0.08 & (22) \\ 
CoRoT-15b & 63.3\,$\pm$\,4.1 & 1.12$_{-0.15}^{+0.30}$ & 3.06 & 6.444 & 0 (fixed) & 1.32\,$\pm$\,0.12 & 1.46$_{-0.14}^{+0.31}$ &
6350\,$\pm$\,200 & 0.10\,$\pm$\,0.20 & (23) \\ 
TOI-569b & 64.1$_{-1.4}^{+1.9}$ & 0.75\,$\pm$\,0.02 & 6.56 & 2.812 & 0.002$_{-0.001}^{+0.002}$ & 1.21\,$\pm$\,0.05 & 1.48\,$\pm$\,0.03 & 5768$_{-92}^{+110}$ & 0.29$_{-0.08}^{+0.09}$ & (12) \\ 
\textbf{TOI-1982b }& 65.8 $\pm$ 2.8 & 1.08 $\pm$ 0.04 & 17.17 & 5.801 & 0.272 $\pm$ 0.014 & 1.41 $\pm$ 0.08 & 1.51 $\pm$ 0.05  & 6325 $\pm$ 110 & -0.10 $\pm$ 0.09 & this work \\ 
EPIC 201702477b & 66.9\,$\pm$\,1.7 & 0.76\,$\pm$\,0.07 & 40.74 & 7.729 & 0.228\,$\pm$\,0.003 & 0.87\,$\pm$\,0.03 & 0.90\,$\pm$\,0.06 & 5517\,$\pm$\,70 & -0.16\,$\pm$\,0.05 & (24) \\ 
\textbf{TOI-629b }& 67.0 $\pm$ 3.0 & 1.11 $\pm$ 0.05 & 8.72 & 2.503 & 0.299 $\pm$ 0.008 & 2.16 $\pm$ 0.13 & 2.37 $\pm$ 0.11 & 9100 $\pm$ 200 & 0.10 $\pm$ 0.15 & this work \\ 
TOI-2119b & 67\,$\pm$\,2 & 1.11\,$\pm$\,0.03 & 7.2 & 51.869 & 0.3362 $\pm$ 0.0005 & 0.53 $\pm$ 0.02 & 0.51 $\pm$ 0.01 & 3553 $\pm$ 67 & 0.1 $\pm$ 0.1 & (25) \\ 
\textbf{TOI-2543b}& 67.6 $\pm$ 3.5 & 0.95 $\pm$ 0.09 & 7.543 & 3.064 & 0.009 $\pm$ 0.002 & 1.29 $\pm$ 0.08 & 1.86 $\pm$ 0.15 & 6060 $\pm$ 82 &  -0.28 $\pm$ 0.10 & this work \\
LP261-75b & 68.1\,$\pm$\,2.1 & 0.90\,$\pm$\,0.01 & 1.88 & 90.071 & <0.007 & 0.30\,$\pm$\,0.01 & 0.31\,$\pm$\,0.00 & 3100\,$\pm$\,50 & ... & (26) \\ 
NGTS-19b & 69.5$_{-5.4}^{+5.7}$ & 1.03\,$\pm$\,0.05& 17.84 & 23.044 & 0.377\,$\pm$\,0.006 & 0.73\,$\pm$\,0.08 & 0.71\,$\pm$\,0.10 & 4500\,$\pm$\,110 & 0.09\,$\pm$\,0.09 & (27) \\ 
NGTS-7Ab & 75.5$_{-13.7}^{+3.0}$ & 1.38$_{-0.14}^{+0.13}$ & 0.68 & 56.04 & 0 (fixed) & 0.24\,$\pm$\,0.03 & 0.61\,$\pm$\,0.06 & 3359$_{-89}^{+106}$ & 0.00\,$\pm$\,0.10 & (28) \\ 
KOI-189b & 78.6\,$\pm$\,3.5 & 1.00\,$\pm$\,0.02 & 30.36 & 20.257 & 0.275\,$\pm$\,0.004 & 0.76\,$\pm$\,0.05 & 0.73\,$\pm$\,0.02 & 4952\,$\pm$\,40 & -0.12\,$\pm$\,0.10 & (29) \\ 
TOI-148b & 77.1$_{-4.6}^{+5.8}$ & 0.81$_{-0.06}^{+0.05}$ & 4.87 & 4.989 & 0.005$_{-0.004}^{+0.006}$ & 0.97$_{-0.09}^{+0.12}$ & 1.20\,$\pm$\,0.07 & 5990\,$\pm$\,140 & -0.24\,$\pm$\,0.25 & (30) \\ 
TOI-587b & 81.1$_{-7.0}^{+7.1}$ & 1.32$_{-0.06}^{+0.07}$ & 8.04 & 4.722 & 0.051$_{-0.036}^{+0.049}$ & 2.33\,$\pm$\,0.12 & 2.01\,$\pm$\,0.09 & 9800\,$\pm$\,200 & 0.08$_{-0.12}^{+0.11}$ & (30) \\ 
TOI-746b & 82.2$_{-4.4}^{+4.9}$ & 0.95$_{-0.06}^{+0.09}$ & 10.98 & 10.503 & 0.199\,$\pm$\,0.003 & 0.94$_{-0.08}^{+0.09}$ & 0.97$_{-0.03}^{+0.04}$ & 5690\,$\pm$\,140 & -0.02\,$\pm$\,0.23 & (30) \\
\hline
\end{tabular}
\end{adjustbox}
\begin{tablenotes}
\item \textbf{References. }\tiny
(1) \citealt{Zhou2019};
(2) \citealt{Artigau2021}; 
(3) \citealt{Benni2020}; 
(4) \citealt{Bonomo2015}; 
(5) \citealt{Deleuil2008}; 
(6) \citealt{Siverd2012}; 
(7) \citealt{Irwin2010}; 
(8) \citealt{Hodzic2018}; 
(9) \citealt{Nowak2017}; 
(10) \citealt{Carmichael2019}; 
(11) \citealt{Diaz2013}; 
(12) \citealt{Carmichael2020}; 
(13) \citealt{Persson2019}; 
(14) \citealt{Subjak2020}; 
(15) \citealt{Carmichael2021}; 
(16) \citealt{Gillen2017}; 
(17) \citealt{Csizmadia2015}; 
(18) \citealt{David2019}; 
(19) \citealt{Palle2021}; 
(20) \citealt{Moutou2013}; 
(21) \citealt{Triaud2013}; 
(22) \citealt{Johnson2011}; 
(23) \citealt{Bouchy2011}; 
(24) \citealt{Bayliss2017}; 
(25) \citealt{toi2119}; 
(26) \citealt{Irwin2018}; 
(27) \citealt{ngts19}; 
(28) \citealt{Jackman2019}; 
(29) \citealt{Diaz2014}; 
(30) \citealt{Grieves2021_FiveHb}; 
\\
\textbf{Notes:} We do not include the brown dwarf binary system by \citet{Stassun2006}, triple system by \citet{Triaud2020}, and white dwarf companions.
\end{tablenotes}

\label{tab:browndwarf_list}
\end{table*}
\subsection{Stellar rotation}\label{sec:stellarrotation}
From the full width at half maximum (FWHM) of the CORALIE cross-correlation function (CCF) we manage to estimate the projected rotational velocities of the four targets. With their color index B-V and stellar radius $\rstar$ we estimate an upper limit on the stellar rotation period (P$_{rot}$ $\leq$ 2$\pi\rstar$/v $\sin{\it{i_{*}}}$) and compare it with the orbital period in order to examine possible spin-orbit synchronization. 

For TOI-1107 we estimate a projected rotational velocity of \vsini = 12.31 $\pm$ 0.79~\kms using the CORALIE spectra. Since the PDCSAP light curve for TOI-1107 shows stellar variability we use a Lomb-Scargle periodogram and estimate a rotation period of P$_{rot}$ = 6.2 $\pm$ 0.3 days (Figure \ref{fig:TOI1107_LombScarge}). With the derived stellar radius \rstar and projected rotational velocity we estimate the upper limit of TOI-1107 rotation period P$_{rot}$/$\sin{\it{i_{*}}}$ = 7.44 $\pm$ 0.65 days which is in agreement with the 6.2 days modulation found in the light curves. Compared to its orbital period (P$_{orb}$ = 4.0782387 $^{+0.0000024}_{-0.0000025}$ days), TOI-1107b does not provide any conclusive evidence of spin-orbit synchronization. Using the rotation period from the periodogram, the stellar radius \rstar and gyrochronology relations of \citet{Mamajek2008} we find an age estimate of 2.6 $\pm$ 0.2~Gyr which indicates that TOI-1107 is a relatively old F-type star. Our global analysis gives an eccentricity $\textit{e}$ = 0.025 $^{+0.023}_{-0.016}$, which we find not significant using the \citealt{Lucy1971} test and compatible with a circular orbit.
Using Equation 3 from \citet{Laughlin06} we find that TOI-1107b has a circularization timescale of $\sim$ 0.2~Gyr for a tidal quality factor $Q_p$ = 10$^5$ and $\sim$ 2~Gyr for a tidal quality factor $Q_p$ = 10$^6$. That suggests that the massive planet orbit might have undergone tidal circularization but not spin-orbit synchronization. TOI-629 is an A-type, rapidly rotating star (P$_{rot}$/$\sin{\it{i_{*}}}$ = 1.77 $\pm$ 0.10 days). Given its young age (0.32 $\pm$ 0.13~Gyr) and its orbital period P$_{orb}$ = 8.717993 $^{+0.000012}_{-0.000013}$ days we find no evidence of spin-orbit synchronization. TOI-1982 is a relatively old (3.6 $\pm$ 1.5~Gyr) and fast rotating star (P$_{rot}$/$\sin{\it{i_{*}}}$ = 2.03 $\pm$ 0.15). Given its long orbital period of 17.172446 $^{+0.000043}_{-0.000044}$ days, TOI-1982b is unlikely to have reached spin-orbit synchronization. TOI-2543 is the oldest of our targets, with an estimated age of 5.6 $\pm$ 0.9~Gyr and a companion on a nearly circular $\textit{e}$ = 0.009 $^{+0.003}_{-0.002}$ orbit. From the SED analysis we estimate a projected rotation period of $P_{\rm rot}$/$\sin$ $\it{i_{*}}$ = 8.73 $\pm$ 0.82 days. Taking into account that it is approximately one day larger than the orbital period (7.542776 $\pm$ 0.000031 days), we suspect that TOI-2543b can be in a state of spin-orbit synchronization. Assuming a lower bound on the tidal quality factor $Q_p$ = 10$^{4.5}$ (\citealt{Beatty2018}), the circularization timescale for TOI-2543b corresponds to $\sim$ 60~Gyr indicating that we would not expect the system
to have been tidally circularized yet.

\subsection{Mass-radius relation}
\begin{figure*}[t]
  \centering
  \includegraphics[width=0.7\textwidth]{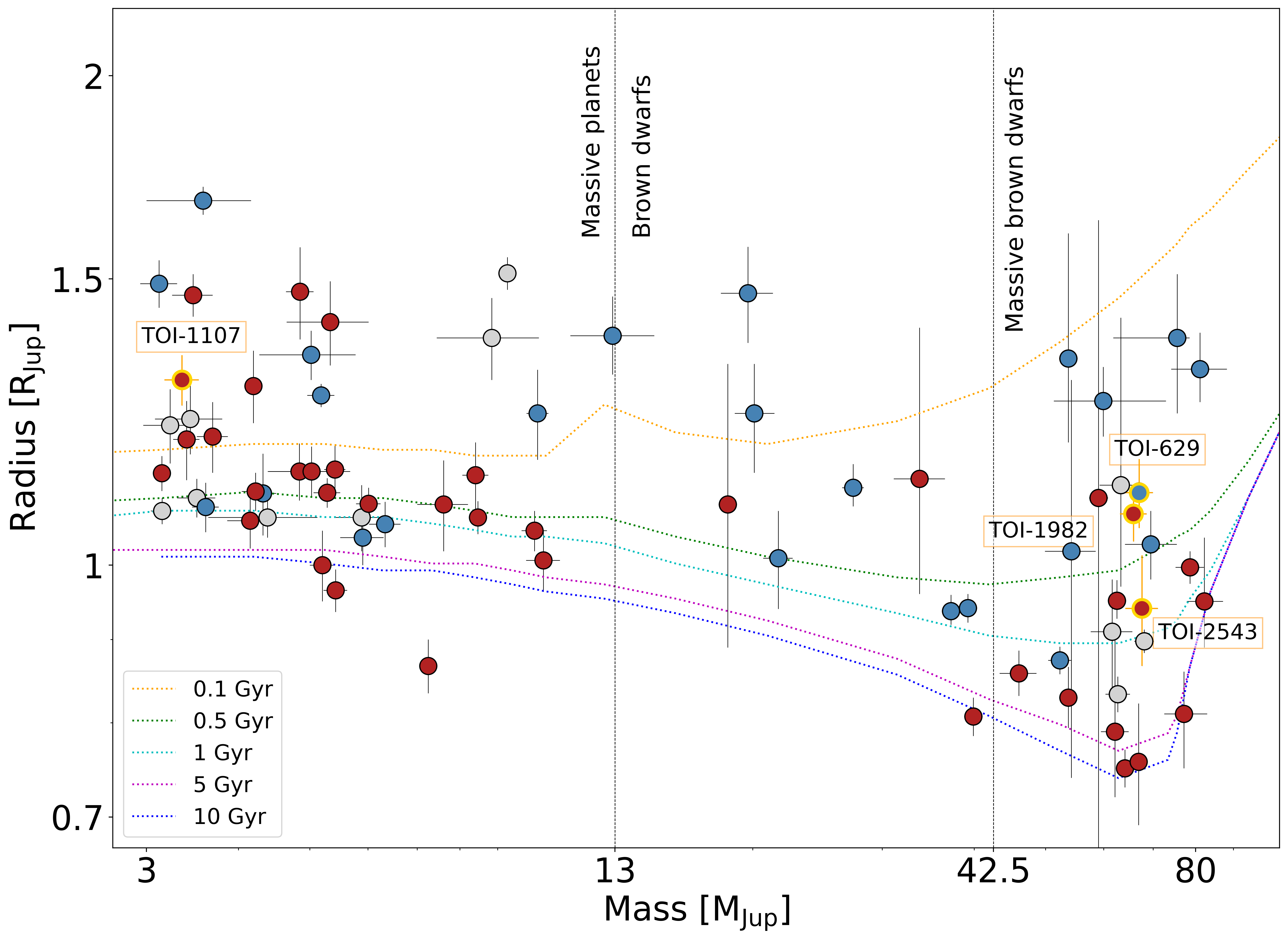}
  \caption{Radius-mass diagram of all massive planets (\mplanet $>$ 3~\mjup, from the updated exoplanet catalog of \citealt{otegi2020}) and transiting brown dwarfs (Table \ref{tab:browndwarf_list}). The colored lines indicate the isochrone models (\citealt{Baraffe2003}, \citeyear{Baraffe2015}). The red color of the points denotes old stars ($\geq$ 2~Gyr), the blue young stars ($<$ 2~Gyr) and the gray not reported age. The brown dwarf RIK 72b (\citealt{David2019}) is not plotted because of its large radius (3.1\,\rjup). The massive planet and three brown dwarfs presented in this work are displayed with yellow edge color.}
  \label{fig:massradius}
\end{figure*}

We place TOI-1107b in the mass-radius diagram of transiting massive planets (> 3~\mjup) and brown dwarfs (Figure $\ref{fig:massradius}$). We color the symbols based on the stellar age where the blue color indicates young stars ($<$ 2~Gyr), the red color indicates old stars ($\geq$ 2~Gyr) and the gray color indicates stars with no reported age. Also, we plot the theoretical evolutionary models from \citet{Baraffe2003,Baraffe2015} for ages of 0.1, 0.5, 1, 5, and 10~Gyr with different colors.
As described in Section $\ref{sec:stellarrotation}$ we derived an age of 2.6 $\pm$ 0.2~Gyr for TOI-1107. Compared to the standard evolutionary tracks (\citealt{Baraffe2003}, \citeyear{Baraffe2015}), TOI-1107b with a radius of 1.30 $\pm$ 0.05~\rjup, appears to be inflated by $\sim$ 20$\%$. Accordingly, we determine the luminosity with the derived stellar radius and effective temperature of our analysis (Table $\ref{tab:Stellar parameters}$). Using the semi-major axis of the orbit (Table $\ref{tab:Juliet11071982}$) we calculate an incident flux of $\fave$ = (2.02 $^{+0.16}_{-0.26}$) $\times$ 10$^{6}$ W~m$^{-2}$ (1480 $^{+115}_{-191}$ S$_\oplus$), which is approximately 10 times larger than the empirical inflation boundary of $\fave$ = 2 $\times$ 10$^{5}$ W~m$^{-2}$ (\citealt{Miller_2011}; \citealt{demoryandseager}). Thus, we expect that the radius of this planet is larger than theoretically predicted. \citet{Sest2018} performed a detailed analysis of the relationship between the planetary radius, mass, and stellar irradiation by studying the population of transiting gas giants with four different mass regimes. With a mass of 3.35~\mjup, TOI-1107b lies above the highest mass boundary (M$_{p}$ >~2.5~\mjup) and the predicted excess radius is $\Delta$R = 0.221 $^{+0.007}_{-0.011}$~\rjup (Equation 11, \citealp{Sest2018}). Given a mass of $\sim$ 3.35~\mjup, an age of $\sim$ 2.6~Gyr and following the theoretical models of planet evolution (\citealt{Baraffe2003}, \citeyear{Baraffe2015}) the predicted radius of TOI-1107b ranges from 1-1.1~\rjup. With an observed radius of 1.30 $\pm$ 0.05~\rjup, TOI-1107b has an inflated radius that is in agreement with the empirical relation proposed by \citet{Sest2018}. The radius excess of hot Jupiters can be explained by different mechanisms that arise from the strong stellar irradiation such as vertical advection of the
potential temperature (\citealt{Tremblin2017}), thermal tidal effects (\citealt{ArrasSocrates2010}), and Ohmic dissipation through magnetohydrodynamic effects (\citealt{OhmicHeating}).

We place the three brown dwarfs in the mass-radius diagram and compare their estimated age from the spectral analysis with the theoretical isochrones. Both TOI-629b and TOI-2543b have masses and radii that are consistent with the theoretical models (\citealt{Baraffe2003}, \citeyear{Baraffe2015}) taking into account their radius uncertainties. With an age of 3.6 $\pm$ 1.5~Gyr and a radius of 1.08 $\pm$ 0.04~\rjup, TOI-1982b appears to be significantly inflated with respect to the theoretical radius of 0.82 $\pm$ 0.04~\rjup (\citealt{Baraffe2003}, \citeyear{Baraffe2015}). Such effect has been detected in massive, transiting brown dwarfs before, for example in CoRoT-15b (\citealt{Bouchy2011}) and NGTS-19b (\citealt{ngts19}).

\subsection{Transiting massive planet and BD population}
\begin{figure*}[t]
  \centering
   \includegraphics[width=0.7\textwidth]{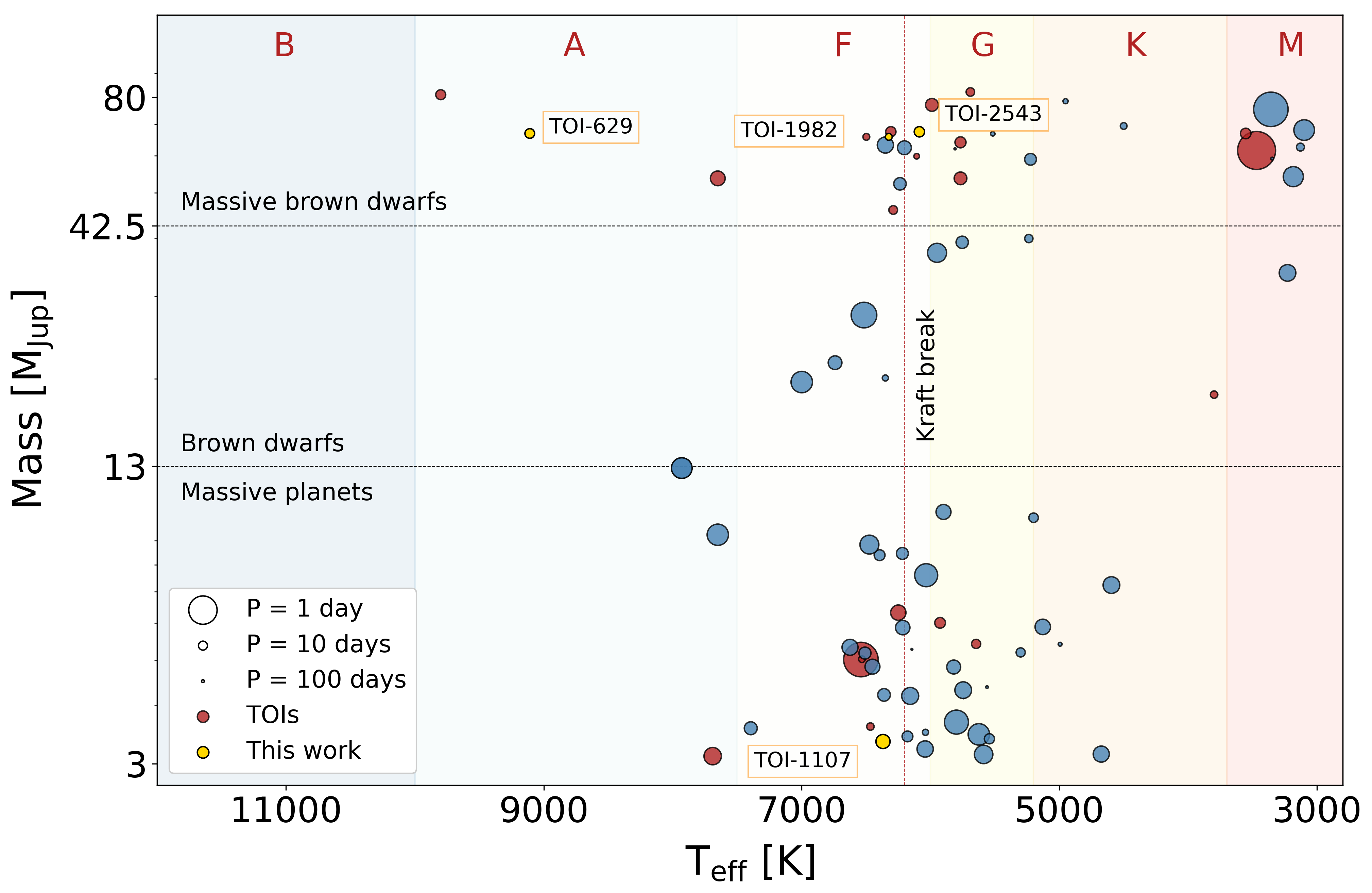}
  \caption{Companion mass over stellar effective temperature T$_{\rm eff}$ diagram for massive planets ($M_{\mathrm{P}}$ > 3~\mjup, from the updated exoplanet catalog of \citealt{otegi2020}) and transiting brown dwarfs (Table \ref{tab:browndwarf_list}) shown in blue. The four new companions are shown in yellow. The red circles display confirmed companions detected by TESS. The size scales inversely with the period of the companion. The red, vertical line corresponds to the Kraft break (6200 K) while the horizontal lines correspond to the approximate boundaries of the brown dwarf regime (13~\mjup) and the transition to high-mass brown dwarfs (42.5~\mjup).}
  \label{fig:massteff}
\end{figure*}

Figure \ref{fig:massteff} shows the mass distribution over the stellar effective temperature for all known transiting massive planets and brown dwarfs, including the four new companions presented in this work. The symbol size is inversely proportional to the brown dwarf's period. Based on the lists compiled by \citet{Carmichael2021} and \citet{Grieves2021_FiveHb} and including the discoveries of TOI-629b, TOI-1982b, and TOI-2543b, the total number of transiting brown dwarfs is 36 (Table \ref{tab:browndwarf_list}). TESS has strongly contributed to the number of these relatively rare objects with 13 detections (red symbols) over the first three years of its operation.    

The current brown dwarf population is sparse and distributes over a large stellar effective temperature range with a significant low occurrence of lighter brown dwarfs (13 - 42.5~\mjup) compared to more massive brown dwarfs (42.5 - 80~\mjup). As discussed in Section \ref{sec:EccentricityMass}, this suggests two different formation mechanisms between the two populations. The sample is not large enough for meaningful comparison, however we see that 20$\%$ of the population is hosted by M-dwarf stars, $\sim$ 8$\%$ by K-type stars, $\sim$ 29$\%$ by G-type stars and $\sim$ 31$\%$ by F-type.
To date there are only four detected brown dwarfs transiting A stars: HATS-70b \citep{Zhou2019}, TOI-503b \citep{Subjak2020}, TOI-587b \citep{Grieves2021_FiveHb}, and TOI-629b (this work). TOI-629 with \teff~= 9100 $\pm$ 200~K is the second hottest star to host a brown dwarf after TOI-587b, a brown dwarf or very low-mass star transiting an A-type star with \teff~= 9800 $\pm$ 200~K. This paucity could possibly be an observational bias in the spectroscopic follow-up strategy toward such hot candidates. No transiting brown dwarfs have been detected orbiting B or hotter type stars. The planet and brown dwarf occurrence rate around OB stars is unknown due to observational difficulties. Gaia \citep{GaiaCollaboration2018} will allow the coverage of this stellar population by revealing astrometric signals of massive companions at intermediate orbital periods. As shown in Figure \ref{fig:massteff} and mentioned by \citet{Udry2003}, \citet{Bouchy18Mjup2011} and \citet{Grieves2021_FiveHb}, there is a significant lack of short period (P $<$ 10 days), massive companions orbiting K and G-type stars. This could be possibly explained by the magnetic braking of G and K-type stars that is is expected to spin down the star and drive ongoing decay of the companion (\citealt{BarkerOgilvie2009}). This effect will result in inward migration and eventually engulfment of the companion by its host star. In contrast, F-type stars undergo less magnetic braking, avoid losing their angular momentum and the decay timescale is expected to be longer, resulting in larger number of massive companions around these stars.

\subsection{Four EBLM systems}\label{sec:EBLMs}
As a by-product of this RV follow-up, we identified four low mass, single-lined spectral eclipsing binaries observed with CORALIE spectrograph. All targets were initially detected photometrically by TESS over the first three years of its operation. The stellar parameters of the primary star were obtained from \SpecMatch on
the stacked CORALIE spectra as described in Section \ref{sec:spectralanalysis}. Our RV measurements reveal that these four F-type primary stars host companions with masses ranging from $\sim$ 156 - 332~\mjup. The stellar properties of the primary and secondary stars are listed in the Appendix in Table $\ref{tab:sb1}$ and their phase-folded radial velocities with the best-fit Keplerian model in Figure $\ref{fig:SB1s}$.

TOI-288 (TIC 47316976) was observed in TESS Sector 2 and 28 and two shallow transits with a depth of $\sim$ 0.17$\%$ were identified with an interval of 711.74 days. The RV measurements with CORALIE revealed a 332.4 $\pm$ 41.8~\mjup low-mass companion with orbital period P$_{orb}$ = 79.08 days and time of conjunction consistent with secondary eclipse. TESS did not observe the main eclipse of the long period eccentric binary system but only the secondary eclipse. As described by \citet{SanterneSecondary}, long-period candidates with eccentric orbit that present a secondary-only eclipse, are more likely to be false positives. Also, 26 out of 37 of our candidates with a transit depth larger than $\sim$ 1.5$\%$ reveal to be eclipse binaries which is in agreement with the follow-up of Kepler giant planet candidates described by \citet{Santerne2016EBLM}. Low-mass stars are ideal for Earth-sized planet detection and characterization. However, this sample is limited when compared to more massive stars because they are intrinsically dimmer. These four targets add to this population and can be used for future stellar characterization and enable an empirical mass-radius exploration.

\section{Conclusion}\label{sec:conclusion}
In the context of the search for exoplanets and brown dwarfs around early-type main-sequence stars, we present the discovery and characterization of three high-mass brown dwarfs, TOI-629b, TOI-1982b, TOI-2543b, and one massive planet, TOI-1107b, identified by the TESS mission. Our analysis is based on both 2-min and QLP TESS data from several sectors in the first and third years of operations. The detections were confirmed via numerous ground-based photometric observations and RV observations with the CORALIE, CHIRON, TRES, FEROS, and MINERVA-Australis spectrographs. The stellar and companion parameters are summarized in Table $\ref{tab:Stellar parameters}$. 

TOI-629 is the second hottest main-sequence star (\teff~= 9100 $\pm$ 200~K) known to host a brown dwarf (M$_{b}$ = 66.98 $^{+2.96}_{-2.95}$~\mjup) and among the most eccentric systems ($\textit{e}$ = 0.298 $\pm$ 0.008). Its high projected stellar rotation velocity required investigation of the stellar rotation-induced gravity darkening effect in the TESS light curves with no evidence of distortion.

TOI-1982b is a massive brown dwarf (M$_{b}$ = 65.85 $^{+2.75}_{-2.72}$~\mjup) on an eccentric orbit ($\textit{e}$ = 0.272 $\pm$ 0.014) around an F-type (\teff~= 6325 $\pm$ 110~K) star. The systems age determination from the SED analysis relative to theoretical predictions indicates that TOI-1982b is significantly inflated. 

TOI-2543 is an F-type star, (\teff~= 6060 $\pm$ 82~K) that hosts a 67.62 $\pm$ 3.45~\mjup massive brown dwarf on a nearly circular orbit ($\textit{e}$ = 0.009 $^{+0.003}_{-0.002}$). The SED analysis estimates that its projected rotation period is one day larger that the brown dwarfs orbital period, indicating that TOI-2543b can be in a state of spin-orbit synchronization but the system has not been tidally circularized yet. 

TOI-1107b is a 3.35 $\pm$ 0.18~\mjup hot Jupiter with an orbital period of 4.08 days, transiting an F-type (\teff~= 6311 $\pm$ 98~K) star. Based on the host star's rotation period that was identified in its light curve we estimated the age of the system ($\tau_*$ = 2.6 $\pm$ 0.2 Gyr). When compared to theoretical isochrone models, TOI-1107b presents an inflated radius due to irradiation from their host star. 

The discoveries of TOI-1107b, TOI-629b, TOI-1982b, and TOI-2543b add to the population of exoplanets and brown dwarfs with well-determined radii and masses (i.e., $\sigma_{R}/R$ $<$ 10 $\%$ and $\sigma_{M}/M$ $<$ 5 $\%$), transiting AF-type stars. Finally, we present four binary systems with low-mass eclipsing components (TOI-288b, TOI-446b, TOI-478b, and TOI-764b) transiting F-type stars that were detected with CORALIE.
Continuation of photometric and radial velocity surveys around early-type stars will allow further investigation on how stellar mass impacts planet and brown dwarf formation and evolution.

\begin{acknowledgements}
We thank the Swiss National Science Foundation (SNSF) and the Geneva University for their continuous support. This work was carried out in the frame of the Swiss National Centre for Competence in Research (NCCR) $PlanetS$ supported by the Swiss National Science Foundation (SNSF). This publication makes use of The Data \& Analysis Center for Exoplanets (DACE), which is a facility based at the University of Geneva (CH) dedicated to extrasolar planet data visualization, exchange, and analysis. DACE is a platform of NCCR $PlanetS$ and is available at https://dace.unige.ch. This paper includes data collected by the TESS mission. Funding for the TESS mission is provided by the NASA Explorer Program.
JSJ gratefully acknowledges support by FONDECYT grant 1201371 and from the ANID BASAL projects ACE210002 and FB210003. Resources supporting this work were provided by the NASA High-End Computing (HEC) Program through the NASA Advanced Supercomputing (NAS) Division at Ames Research Center for the production of the SPOC data products. We acknowledge the use of public TESS data from pipelines at the TESS Science Office and at the TESS Science Processing Operations Center. 
We acknowledge the efforts of Michael Calkin, Perry Berlind, and Gil Esquerdo for their diligent observations using the TRES spectrograph at Mount Hopkins, Arizona. Data presented herein were obtained at the MINERVA-Australis facility from telescope time allocated under the NN-EXPLORE program with support from the National Aeronautics and Space Administration.
MINERVA-Australis is supported by Australian Research Council LIEF Grant LE160100001, Discovery Grants DP180100972 and DP220100365, Mount Cuba Astronomical Foundation, and institutional partners University of Southern Queensland, UNSW Sydney, MIT, Nanjing University, George Mason University, University of Louisville, University of California Riverside, University of Florida, and The University of Texas at Austin.We respectfully acknowledge the traditional custodians of all lands throughout Australia, and recognise their continued cultural and spiritual connection to the land, waterways, cosmos, and community. We pay our deepest respects to all Elders, ancestors and descendants of the Giabal, Jarowair, and Kambuwal nations, upon whose lands the Minerva-Australis facility at Mt Kent is situated. 

\end{acknowledgements}

\bibliographystyle{aa}
\bibliography{bib}

\begin{thebibliography}{140}
\expandafter\ifx\csname natexlab\endcsname\relax\def\natexlab#1{#1}\fi

\bibitem[{{Abe} {et~al.}(2013){Abe}, {Gon{\c{c}}alves}, {Agabi}, {Alapini},
  {Guillot}, {M{\'e}karnia}, {Rivet}, {Schmider}, {Crouzet}, {Fortney}, {Pont},
  {Barbieri}, {Daban}, {Fante{\"\i}-Caujolle}, {Gouvret}, {Bresson}, {Roussel},
  {Bonhomme}, {Robini}, {Dugu{\'e}}, {Bondoux}, {P{\'e}ron}, {Petit},
  {Szul{\'a}gyi}, {Fruth}, {Erikson}, {Rauer}, {Fressin}, {Valbousquet},
  {Blanc}, {Le van Suu}, \& {Aigrain}}]{abe}
{Abe}, L., {Gon{\c{c}}alves}, I., {Agabi}, A., {et~al.} 2013, \aap, 553, A49

\bibitem[{{Acton} {et~al.}(2021){Acton}, {Goad}, {Burleigh}, {Casewell},
  {Breytenbach}, {Nielsen}, {Smith}, {Anderson}, {Battley}, {Bayliss},
  {Bouchy}, {Bryant}, {Csizmadia}, {Eigm{\"u}ller}, {Gill}, {Gillen},
  {Grieves}, {G{\"u}nther}, {Henderson}, {Hodgkin}, {Jackman}, {Jenkins},
  {Lendl}, {McCormac}, {Moyano}, {Nelson}, {Sefako}, {Smith}, {Stalport},
  {Thomas}, {Tilbrook}, {Udry}, {West}, {Wheatley}, {Worters}, {Vines}, \&
  {Alves}}]{ngts19}
{Acton}, J.~S., {Goad}, M.~R., {Burleigh}, M.~R., {et~al.} 2021, \mnras, 505,
  2741

\bibitem[{{Adams} \& {Laughlin}(2006)}]{Laughlin06}
{Adams}, F.~C. \& {Laughlin}, G. 2006, \apj, 649, 1004

\bibitem[{{Addison} {et~al.}(2019){Addison}, {Wright}, {Wittenmyer}, {Horner},
  {Mengel}, {Johns}, {Marti}, {Nicholson}, {Soutter}, {Bowler}, {Crossfield},
  {Kane}, {Kielkopf}, {Plavchan}, {Tinney}, {Zhang}, {Clark}, {Clerte},
  {Eastman}, {Swift}, {Bottom}, {Muirhead}, {McCrady}, {Herzig}, {Hogstrom},
  {Wilson}, {Sliski}, {Johnson}, {Wright}, {Johnson}, {Blake}, {Riddle}, {Lin},
  {Cornachione}, {Bedding}, {Stello}, {Huber}, {Marsden}, \&
  {Carter}}]{minervaaddison}
{Addison}, B., {Wright}, D.~J., {Wittenmyer}, R.~A., {et~al.} 2019, \pasp, 131,
  115003

\bibitem[{{Addison} {et~al.}(2021){Addison}, {Knudstrup}, {Wong},
  {H{\'e}brard}, {Dorval}, {Snellen}, {Albrecht}, {Bello-Arufe}, {Almenara},
  {Boisse}, {Bonfils}, {Dalal}, {Demangeon}, {Hoyer}, {Kiefer}, {Santos},
  {Nowak}, {Luque}, {Stangret}, {Palle}, {Tronsgaard}, {Antoci}, {Buchhave},
  {G{\"u}nther}, {Daylan}, {Murgas}, {Parviainen}, {Esparza-Borges}, {Crouzet},
  {Narita}, {Fukui}, {Kawauchi}, {Watanabe}, {Rabus}, {Johnson}, {Otten}, {Jan
  Talens}, {Cabot}, {Fischer}, {Grundahl}, {Fredslund Andersen},
  {Jessen-Hansen}, {Pall{\'e}}, {Shporer}, {Ciardi}, {Clark}, {Wittenmyer},
  {Wright}, {Horner}, {Collins}, {Jensen}, {Kielkopf}, {Schwarz}, {Srdoc},
  {Yilmaz}, {Senavci}, {Diamond}, {Harbeck}, {Komacek}, {Smith}, {Wang},
  {Eastman}, {Stassun}, {Latham}, {Vanderspek}, {Seager}, {Winn}, {Jenkins},
  {Louie}, {Bouma}, {Twicken}, {Levine}, \& {McLean}}]{TOI1431}
{Addison}, B.~C., {Knudstrup}, E., {Wong}, I., {et~al.} 2021, \aj, 162, 292

\bibitem[{{Aller} {et~al.}(2020){Aller}, {Lillo-Box}, {Jones}, {Miranda}, \&
  {Barcel{\'o} Forteza}}]{TPFplotter}
{Aller}, A., {Lillo-Box}, J., {Jones}, D., {Miranda}, L.~F., \& {Barcel{\'o}
  Forteza}, S. 2020, \aap, 635, A128

\bibitem[{{Arras} \& {Socrates}(2010)}]{ArrasSocrates2010}
{Arras}, P. \& {Socrates}, A. 2010, \apj, 714, 1

\bibitem[{{Artigau} {et~al.}(2021){Artigau}, {H{\'e}brard}, {Cadieux},
  {Vandal}, {Cook}, {Doyon}, {Gagn{\'e}}, {Moutou}, {Martioli}, {Frasca},
  {Jahandar}, {Lafreni{\`e}re}, {Malo}, {Donati}, {Cort{\'e}s-Zuleta},
  {Boisse}, {Delfosse}, {Carmona}, {Fouqu{\'e}}, {Morin}, {Rowe}, {Marino},
  {Papini}, {Ciardi}, {Lund}, {Martins}, {Pelletier}, {Arnold}, {Bouchy},
  {Forveille}, {Santos}, {Bonfils}, {Figueira}, {Fausnaugh}, {Ricker},
  {Latham}, {Seager}, {Winn}, {Jenkins}, {Ting}, {Torres}, \& {Gomes da
  Silva}}]{Artigau2021}
{Artigau}, {\'E}., {H{\'e}brard}, G., {Cadieux}, C., {et~al.} 2021, \aj, 162,
  144

\bibitem[{{Baraffe} {et~al.}(2003){Baraffe}, {Chabrier}, {Barman}, {Allard}, \&
  {Hauschildt}}]{Baraffe2003}
{Baraffe}, I., {Chabrier}, G., {Barman}, T.~S., {Allard}, F., \& {Hauschildt},
  P.~H. 2003, \aap, 402, 701

\bibitem[{{Baraffe} {et~al.}(2015){Baraffe}, {Homeier}, {Allard}, \&
  {Chabrier}}]{Baraffe2015}
{Baraffe}, I., {Homeier}, D., {Allard}, F., \& {Chabrier}, G. 2015, \aap, 577,
  A42

\bibitem[{{Baranne} {et~al.}(1996){Baranne}, {Queloz}, {Mayor}, {Adrianzyk},
  {Knispel}, {Kohler}, {Lacroix}, {Meunier}, {Rimbaud}, \& {Vin}}]{Baranne1996}
{Baranne}, A., {Queloz}, D., {Mayor}, M., {et~al.} 1996, \aaps, 119, 373

\bibitem[{{Barker} \& {Ogilvie}(2009)}]{BarkerOgilvie2009}
{Barker}, A.~J. \& {Ogilvie}, G.~I. 2009, \mnras, 395, 2268

\bibitem[{{Barnes}(2009)}]{Barnes2009}
{Barnes}, J.~W. 2009, \apj, 705, 683

\bibitem[{{Batygin} \& {Stevenson}(2010)}]{OhmicHeating}
{Batygin}, K. \& {Stevenson}, D.~J. 2010, \apjl, 714, L238

\bibitem[{{Bayliss} {et~al.}(2017){Bayliss}, {Hojjatpanah}, {Santerne},
  {Dragomir}, {Zhou}, {Shporer}, {Col{\'o}n}, {Almenara}, {Armstrong},
  {Barrado}, {Barros}, {Bento}, {Boisse}, {Bouchy}, {Brown}, {Brown},
  {Cameron}, {Cochran}, {Demangeon}, {Deleuil}, {D{\'\i}az}, {Fulton}, {Horne},
  {H{\'e}brard}, {Lillo-Box}, {Lovis}, {Mawet}, {Ngo}, {Osborn}, {Palle},
  {Petigura}, {Pollacco}, {Santos}, {Sefako}, {Siverd}, {Sousa}, \&
  {Tsantaki}}]{Bayliss2017}
{Bayliss}, D., {Hojjatpanah}, S., {Santerne}, A., {et~al.} 2017, \aj, 153, 15

\bibitem[{{Beatty} {et~al.}(2018){Beatty}, {Morley}, {Curtis}, {Burrows},
  {Davenport}, \& {Montet}}]{Beatty2018}
{Beatty}, T.~G., {Morley}, C.~V., {Curtis}, J.~L., {et~al.} 2018, \aj, 156, 168

\bibitem[{{Benni} {et~al.}(2021){Benni}, {Burdanov}, {Krushinsky}, {Bonfanti},
  {H{\'e}brard}, {Almenara}, {Dalal}, {Demangeon}, {Tsantaki}, {Pepper},
  {Stassun}, {Vanderburg}, {Belinski}, {Kashaev}, {Barkaoui}, {Kim}, {Kang},
  {Antonyuk}, {Dyachenko}, {Rastegaev}, {Beskakotov}, {Mitrofanova},
  {Pozuelos}, {Kuznetsov}, {Popov}, {Kiefer}, {Wilson}, {Ricker}, {Vanderspek},
  {Latham}, {Seager}, {Jenkins}, {Sokov}, {Sokova}, {Marchini}, {Papini},
  {Salvaggio}, {Banfi}, {Ba{\c{s}}t{\"u}rk}, {Torun}, {Yal{\c{c}}{\i}nkaya},
  {Ivanov}, {Valyavin}, {Jehin}, {Gillon}, {Pak{\v{s}}tien{\.{e}}}, {Hentunen},
  {Shadick}, {Bretton}, {W{\"u}nsche}, {Garlitz}, {Jongen}, {Molina},
  {Girardin}, {Grau Horta}, {Naves}, {Benkhaldoun}, {Joner}, {Spencer},
  {Bieryla}, {Stevens}, {Jensen}, {Collins}, {Charbonneau}, {Quintana},
  {Mullally}, \& {Henze}}]{Benni2020}
{Benni}, P., {Burdanov}, A.~Y., {Krushinsky}, V.~V., {et~al.} 2021, \mnras,
  505, 4956

\bibitem[{{Blanco-Cuaresma} {et~al.}(2014){Blanco-Cuaresma}, {Soubiran},
  {Heiter}, \& {Jofr{\'e}}}]{Blanco-Cuaresma2014}
{Blanco-Cuaresma}, S., {Soubiran}, C., {Heiter}, U., \& {Jofr{\'e}}, P. 2014,
  \aap, 569, A111

\bibitem[{{Bonomo} {et~al.}(2015){Bonomo}, {Sozzetti}, {Santerne}, {Deleuil},
  {Almenara}, {Bruno}, {D{\'\i}az}, {H{\'e}brard}, \& {Moutou}}]{Bonomo2015}
{Bonomo}, A.~S., {Sozzetti}, A., {Santerne}, A., {et~al.} 2015, \aap, 575, A85

\bibitem[{{Borgniet} {et~al.}(2014){Borgniet}, {Boisse}, {Lagrange}, {Bouchy},
  {Arnold}, {D{\'\i}az}, {Galland}, {Delorme}, {H{\'e}brard}, {Santerne},
  {Ehrenreich}, {S{\'e}gransan}, {Bonfils}, {Delfosse}, {Santos}, {Forveille},
  {Moutou}, {Udry}, {Eggenberger}, {Pepe}, {Astudillo}, \&
  {Montagnier}}]{Borgniet2014}
{Borgniet}, S., {Boisse}, I., {Lagrange}, A.~M., {et~al.} 2014, \aap, 561, A65

\bibitem[{{Borgniet} {et~al.}(2017){Borgniet}, {Lagrange}, {Meunier}, \&
  {Galland}}]{Borgniet2017}
{Borgniet}, S., {Lagrange}, A.~M., {Meunier}, N., \& {Galland}, F. 2017, \aap,
  599, A57

\bibitem[{{Borgniet} {et~al.}(2019){Borgniet}, {Lagrange}, {Meunier},
  {Galland}, {Arnold}, {Astudillo-Defru}, {Beuzit}, {Boisse}, {Bonfils},
  {Bouchy}, {Debondt}, {Deleuil}, {Delfosse}, {Desort}, {D{\'\i}az},
  {Eggenberger}, {Ehrenreich}, {Forveille}, {H{\'e}brard}, {Loeillet}, {Lovis},
  {Montagnier}, {Moutou}, {Pepe}, {Perrier}, {Pont}, {Queloz}, {Santerne},
  {Santos}, {S{\'e}gransan}, {da Silva}, {Sivan}, {Udry}, \&
  {Vidal-Madjar}}]{Borgniet2019}
{Borgniet}, S., {Lagrange}, A.~M., {Meunier}, N., {et~al.} 2019, \aap, 621, A87

\bibitem[{{Boss}(1997)}]{Boss1997}
{Boss}, A.~P. 1997, Science, 276, 1836

\bibitem[{{Bouchy} {et~al.}(2011{\natexlab{a}}){Bouchy}, {Bonomo}, {Santerne},
  {Moutou}, {Deleuil}, {D{\'\i}az}, {Eggenberger}, {Ehrenreich}, {Gry},
  {Guillot}, {Havel}, {H{\'e}brard}, \& {Udry}}]{Bouchy18Mjup2011}
{Bouchy}, F., {Bonomo}, A.~S., {Santerne}, A., {et~al.} 2011{\natexlab{a}},
  \aap, 533, A83

\bibitem[{{Bouchy} {et~al.}(2011{\natexlab{b}}){Bouchy}, {Deleuil}, {Guillot},
  {Aigrain}, {Carone}, {Cochran}, {Almenara}, {Alonso}, {Auvergne}, {Baglin},
  {Barge}, {Bonomo}, {Bord{\'e}}, {Csizmadia}, {de Bondt}, {Deeg}, {D{\'\i}az},
  {Dvorak}, {Endl}, {Erikson}, {Ferraz-Mello}, {Fridlund}, {Gandolfi},
  {Gazzano}, {Gibson}, {Gillon}, {Guenther}, {Hatzes}, {Havel}, {H{\'e}brard},
  {Jorda}, {L{\'e}ger}, {Lovis}, {Llebaria}, {Lammer}, {MacQueen}, {Mazeh},
  {Moutou}, {Ofir}, {Ollivier}, {Parviainen}, {P{\"a}tzold}, {Queloz}, {Rauer},
  {Rouan}, {Santerne}, {Schneider}, {Tingley}, \& {Wuchterl}}]{Bouchy2011}
{Bouchy}, F., {Deleuil}, M., {Guillot}, T., {et~al.} 2011{\natexlab{b}}, \aap,
  525, A68

\bibitem[{{Bouchy} \& {the Sophie Team}(2006)}]{SophieTeam}
{Bouchy}, F. \& {the Sophie Team}. 2006, in Tenth Anniversary of 51 Peg-b:
  Status of and prospects for hot Jupiter studies, ed. L.~{Arnold},
  F.~{Bouchy}, \& C.~{Moutou}, 319

\bibitem[{{Brahm} {et~al.}(2017){Brahm}, {Jord{\'a}n}, \& {Espinoza}}]{ceres}
{Brahm}, R., {Jord{\'a}n}, A., \& {Espinoza}, N. 2017, \pasp, 129, 034002

\bibitem[{{Brown} {et~al.}(2013){Brown}, {Baliber}, {Bianco}, {Bowman},
  {Burleson}, {Conway}, {Crellin}, {Depagne}, {De Vera}, {Dilday}, {Dragomir},
  {Dubberley}, {Eastman}, {Elphick}, {Falarski}, {Foale}, {Ford}, {Fulton},
  {Garza}, {Gomez}, {Graham}, {Greene}, {Haldeman}, {Hawkins}, {Haworth},
  {Haynes}, {Hidas}, {Hjelstrom}, {Howell}, {Hygelund}, {Lister}, {Lobdill},
  {Martinez}, {Mullins}, {Norbury}, {Parrent}, {Paulson}, {Petry}, {Pickles},
  {Posner}, {Rosing}, {Ross}, {Sand}, {Saunders}, {Shobbrook}, {Shporer},
  {Street}, {Thomas}, {Tsapras}, {Tufts}, {Valenti}, {Vander Horst}, {Walker},
  {White}, \& {Willis}}]{Brown:2013}
{Brown}, T.~M., {Baliber}, N., {Bianco}, F.~B., {et~al.} 2013, \pasp, 125, 1031

\bibitem[{{Buchhave} {et~al.}(2012){Buchhave}, {Latham}, {Johansen},
  {Bizzarro}, {Torres}, {Rowe}, {Batalha}, {Borucki}, {Brugamyer}, {Caldwell},
  {Bryson}, {Ciardi}, {Cochran}, {Endl}, {Esquerdo}, {Ford}, {Geary},
  {Gilliland}, {Hansen}, {Isaacson}, {Laird}, {Lucas}, {Marcy}, {Morse},
  {Robertson}, {Shporer}, {Stefanik}, {Still}, \& {Quinn}}]{buchhave_2012}
{Buchhave}, L.~A., {Latham}, D.~W., {Johansen}, A., {et~al.} 2012, \nat, 486,
  375

\bibitem[{{Buchner} {et~al.}(2014){Buchner}, {Georgakakis}, {Nandra}, {Hsu},
  {Rangel}, {Brightman}, {Merloni}, {Salvato}, {Donley}, \&
  {Kocevski}}]{pymultinest}
{Buchner}, J., {Georgakakis}, A., {Nandra}, K., {et~al.} 2014, \aap, 564, A125

\bibitem[{{Ca{\~n}as} {et~al.}(2022){Ca{\~n}as}, {Mahadevan}, {Bender},
  {Salazar Rivera}, {Monson}, {Beard}, {Lubin}, {Robertson}, {Gupta},
  {Cochran}, {Fredrick}, {Hearty}, {Jones}, {Kanodia}, {Lin}, {Ninan},
  {Ramsey}, {Schwab}, \& {Stef{\'a}nsson}}]{toi2119}
{Ca{\~n}as}, C.~I., {Mahadevan}, S., {Bender}, C.~F., {et~al.} 2022, \aj, 163,
  89

\bibitem[{{Carmichael} {et~al.}(2019){Carmichael}, {Latham}, \& {Vand
  erburg}}]{Carmichael2019}
{Carmichael}, T.~W., {Latham}, D.~W., \& {Vand erburg}, A.~M. 2019, \aj, 158,
  38

\bibitem[{{Carmichael} {et~al.}(2020){Carmichael}, {Quinn}, {Mustill}, {Huang},
  {Zhou}, {Persson}, {Nielsen}, {Collins}, {Ziegler}, {Collins}, {Rodriguez},
  {Shporer}, {Brahm}, {Mann}, {Bouchy}, {Fridlund}, {Stassun}, {Hellier},
  {Seidel}, {Stalport}, {Udry}, {Pepe}, {Ireland}, {{\v{Z}}erjal},
  {Brice{\~n}o}, {Law}, {Jord{\'a}n}, {Espinoza}, {Henning}, {Sarkis}, \&
  {Latham}}]{Carmichael2020}
{Carmichael}, T.~W., {Quinn}, S.~N., {Mustill}, A.~J., {et~al.} 2020, \aj, 160,
  53

\bibitem[{{Carmichael} {et~al.}(2021){Carmichael}, {Quinn}, {Zhou}, {Grieves},
  {Irwin}, {Stassun}, {Vanderburg}, {Winn}, {Bouchy}, {Brasseur},
  {Brice{\~n}o}, {Caldwell}, {Charbonneau}, {Collins}, {Colon}, {Eastman},
  {Fausnaugh}, {Fong}, {F{\H{u}}r{\'e}sz}, {Huang}, {Jenkins}, {Kielkopf},
  {Latham}, {Law}, {Lund}, {Mann}, {Ricker}, {Rodriguez}, {Schwarz}, {Shporer},
  {Tenenbaum}, {Wood}, \& {Ziegler}}]{Carmichael2021}
{Carmichael}, T.~W., {Quinn}, S.~N., {Zhou}, G., {et~al.} 2021, \aj, 161, 97

\bibitem[{{Casewell} {et~al.}(2015){Casewell}, {Lawrie}, {Maxted}, {Marley},
  {Fortney}, {Rimmer}, {Littlefair}, {Wynn}, {Burleigh}, \&
  {Helling}}]{Casewell2015}
{Casewell}, S.~L., {Lawrie}, K.~A., {Maxted}, P.~F.~L., {et~al.} 2015, \mnras,
  447, 3218

\bibitem[{{Castelli} \& {Kurucz}(2004)}]{Castelli:2004}
{Castelli}, F. \& {Kurucz}, R.~L. 2004, \aap, 419, 725

\bibitem[{{Chelli}(2000)}]{Chelli2000}
{Chelli}, A. 2000, \aap, 358, L59

\bibitem[{{Collier Cameron} {et~al.}(2010){Collier Cameron}, {Guenther},
  {Smalley}, {McDonald}, {Hebb}, {Andersen}, {Augusteijn}, {Barros}, {Brown},
  {Cochran}, {Endl}, {Fossey}, {Hartmann}, {Maxted}, {Pollacco}, {Skillen},
  {Telting}, {Waldmann}, \& {West}}]{WASP-33}
{Collier Cameron}, A., {Guenther}, E., {Smalley}, B., {et~al.} 2010, \mnras,
  407, 507

\bibitem[{{Collins} {et~al.}(2017){Collins}, {Kielkopf}, {Stassun}, \&
  {Hessman}}]{Collins:2017}
{Collins}, K.~A., {Kielkopf}, J.~F., {Stassun}, K.~G., \& {Hessman}, F.~V.
  2017, \aj, 153, 77

\bibitem[{{Collins} {et~al.}(2018){Collins}, {Quinn}, {Latham}, {Christiansen},
  {Ciardi}, {Crossfield}, \& {Seager}}]{karen2018}
{Collins}, K.~A., {Quinn}, S.~N., {Latham}, D.~W., {et~al.} 2018, in American
  Astronomical Society Meeting Abstracts $\#$231. p. 439.08

\bibitem[{{Csizmadia} {et~al.}(2015){Csizmadia}, {Hatzes}, {Gandolfi},
  {Deleuil}, {Bouchy}, {Fridlund}, {Szabados}, {Parviainen}, {Cabrera},
  {Aigrain}, {Alonso}, {Almenara}, {Baglin}, {Bord{\'e}}, {Bonomo}, {Deeg},
  {D{\'\i}az}, {Erikson}, {Ferraz-Mello}, {Tadeu dos Santos}, {Guenther},
  {Guillot}, {Grziwa}, {H{\'e}brard}, {Klagyivik}, {Ollivier}, {P{\"a}tzold},
  {Rauer}, {Rouan}, {Santerne}, {Schneider}, {Mazeh}, {Wuchterl}, {Carpano}, \&
  {Ofir}}]{Csizmadia2015}
{Csizmadia}, S., {Hatzes}, A., {Gandolfi}, D., {et~al.} 2015, \aap, 584, A13

\bibitem[{{David} {et~al.}(2019){David}, {Hillenbrand}, {Gillen}, {Cody},
  {Howell}, {Isaacson}, \& {Livingston}}]{David2019}
{David}, T.~J., {Hillenbrand}, L.~A., {Gillen}, E., {et~al.} 2019, \apj, 872,
  161

\bibitem[{{Deleuil} {et~al.}(2008){Deleuil}, {Deeg}, {Alonso}, {Bouchy},
  {Rouan}, {Auvergne}, {Baglin}, {Aigrain}, {Almenara}, {Barbieri}, {Barge},
  {Bruntt}, {Bord{\'e}}, {Collier Cameron}, {Csizmadia}, {de La Reza},
  {Dvorak}, {Erikson}, {Fridlund}, {Gandolfi}, {Gillon}, {Guenther}, {Guillot},
  {Hatzes}, {H{\'e}brard}, {Jorda}, {Lammer}, {L{\'e}ger}, {Llebaria},
  {Loeillet}, {Mayor}, {Mazeh}, {Moutou}, {Ollivier}, {P{\"a}tzold}, {Pont},
  {Queloz}, {Rauer}, {Schneider}, {Shporer}, {Wuchterl}, \&
  {Zucker}}]{Deleuil2008}
{Deleuil}, M., {Deeg}, H.~J., {Alonso}, R., {et~al.} 2008, \aap, 491, 889

\bibitem[{{Demory} \& {Seager}(2011)}]{demoryandseager}
{Demory}, B.-O. \& {Seager}, S. 2011, \apjs, 197, 12

\bibitem[{{Desort} {et~al.}(2008){Desort}, {Lagrange}, {Galland}, {Beust},
  {Udry}, {Mayor}, \& {Lo Curto}}]{Desort2008}
{Desort}, M., {Lagrange}, A.~M., {Galland}, F., {et~al.} 2008, \aap, 491, 883

\bibitem[{{Desort} {et~al.}(2009){Desort}, {Lagrange}, {Galland}, {Udry},
  {Montagnier}, {Beust}, {Boisse}, {Bonfils}, {Bouchy}, {Delfosse},
  {Eggenberger}, {Ehrenreich}, {Forveille}, {H{\'e}brard}, {Loeillet}, {Lovis},
  {Mayor}, {Meunier}, {Moutou}, {Pepe}, {Perrier}, {Pont}, {Queloz}, {Santos},
  {S{\'e}gransan}, \& {Vidal-Madjar}}]{Desort2009}
{Desort}, M., {Lagrange}, A.~M., {Galland}, F., {et~al.} 2009, \aap, 506, 1469

\bibitem[{{D{\'\i}az} {et~al.}(2013){D{\'\i}az}, {Damiani}, {Deleuil},
  {Almenara}, {Moutou}, {Barros}, {Bonomo}, {Bouchy}, {Bruno}, {H{\'e}brard},
  {Montagnier}, \& {Santerne}}]{Diaz2013}
{D{\'\i}az}, R.~F., {Damiani}, C., {Deleuil}, M., {et~al.} 2013, \aap, 551, L9

\bibitem[{{D{\'\i}az} {et~al.}(2014){D{\'\i}az}, {Montagnier}, {Leconte},
  {Bonomo}, {Deleuil}, {Almenara}, {Barros}, {Bouchy}, {Bruno}, {Damiani},
  {H{\'e}brard}, {Moutou}, \& {Santerne}}]{Diaz2014}
{D{\'\i}az}, R.~F., {Montagnier}, G., {Leconte}, J., {et~al.} 2014, \aap, 572,
  A109

\bibitem[{{Dong} {et~al.}(2021){Dong}, {Huang}, {Zhou}, {Dawson}, {Rodriguez},
  {Eastman}, {Collins}, {Quinn}, {Shporer}, {Triaud}, {Wang}, {Beatty},
  {Jackson}, {Collins}, {Abe}, {Suarez}, {Crouzet}, {M{\'e}karnia},
  {Dransfield}, {Jensen}, {Stockdale}, {Barkaoui}, {Heitzmann}, {Wright},
  {Addison}, {Wittenmyer}, {Okumura}, {Bowler}, {Horner}, {Kane}, {Kielkopf},
  {Liu}, {Plavchan}, {Mengel}, {Ricker}, {Vanderspek}, {Latham}, {Seager},
  {Winn}, {Jenkins}, {Christiansen}, \& {Paegert}}]{TOI3362}
{Dong}, J., {Huang}, C.~X., {Zhou}, G., {et~al.} 2021, \apjl, 920, L16

\bibitem[{{Dorval} {et~al.}(2020){Dorval}, {Talens}, {Otten}, {Brahm},
  {Jord{\'a}n}, {Torres}, {Vanzi}, {Zapata}, {Henry}, {Paredes}, {Jao},
  {James}, {Hinojosa}, {Bakos}, {Csubry}, {Bhatti}, {Suc}, {Osip}, {Mamajek},
  {Mellon}, {Wyttenbach}, {Stuik}, {Kenworthy}, {Bailey}, {Ireland},
  {Crawford}, {Lomberg}, {Kuhn}, \& {Snellen}}]{mascara4}
{Dorval}, P., {Talens}, G.~J.~J., {Otten}, G.~P.~P.~L., {et~al.} 2020, \aap,
  635, A60

\bibitem[{{Doyle} {et~al.}(2014){Doyle}, {Davies}, {Smalley}, {Chaplin}, \&
  {Elsworth}}]{Doyle2014}
{Doyle}, A.~P., {Davies}, G.~R., {Smalley}, B., {Chaplin}, W.~J., \&
  {Elsworth}, Y. 2014, \mnras, 444, 3592

\bibitem[{{Espinoza} \& {Jord{\'a}n}(2015)}]{limb_espinoza}
{Espinoza}, N. \& {Jord{\'a}n}, A. 2015, \mnras, 450, 1879

\bibitem[{{Espinoza} {et~al.}(2019){Espinoza}, {Kossakowski}, \&
  {Brahm}}]{juliet}
{Espinoza}, N., {Kossakowski}, D., \& {Brahm}, R. 2019, \mnras, 490, 2262

\bibitem[{{Fabrycky} \& {Tremaine}(2007)}]{Fabrycky2007}
{Fabrycky}, D. \& {Tremaine}, S. 2007, \apj, 669, 1298

\bibitem[{{Feroz} {et~al.}(2009){Feroz}, {Hobson}, \& {Bridges}}]{multinest}
{Feroz}, F., {Hobson}, M.~P., \& {Bridges}, M. 2009, \mnras, 398, 1601

\bibitem[{{Findeisen} {et~al.}(2011){Findeisen}, {Hillenbrand}, \&
  {Soderblom}}]{Findeisen2011}
{Findeisen}, K., {Hillenbrand}, L., \& {Soderblom}, D. 2011, \aj, 142, 23

\bibitem[{{Foreman-Mackey} {et~al.}(2017){Foreman-Mackey}, {Agol},
  {Ambikasaran}, \& {Angus}}]{Foreman:2017}
{Foreman-Mackey}, D., {Agol}, E., {Ambikasaran}, S., \& {Angus}, R. 2017, \aj,
  154, 220

\bibitem[{{Fulton} {et~al.}(2018){Fulton}, {Petigura}, {Blunt}, \&
  {Sinukoff}}]{radvel}
{Fulton}, B.~J., {Petigura}, E.~A., {Blunt}, S., \& {Sinukoff}, E. 2018, \pasp,
  130, 044504

\bibitem[{{Gaia Collaboration} {et~al.}(2018){Gaia Collaboration}, {Brown},
  {Vallenari}, {Prusti}, {de Bruijne}, {Babusiaux}, {Bailer-Jones}, {Biermann},
  {Evans}, {Eyer}, {Jansen}, {Jordi}, {Klioner}, {Lammers}, {Lindegren},
  {Luri}, {Mignard}, {Panem}, {Pourbaix}, {Randich}, {Sartoretti}, {Siddiqui},
  {Soubiran}, {van Leeuwen}, {Walton}, {Arenou}, {Bastian}, {Cropper},
  {Drimmel}, {Katz}, {Lattanzi}, {Bakker}, {Cacciari}, {Casta{\~n}eda},
  {Chaoul}, {Cheek}, {De Angeli}, {Fabricius}, {Guerra}, {Holl}, {Masana},
  {Messineo}, {Mowlavi}, {Nienartowicz}, {Panuzzo}, {Portell}, {Riello},
  {Seabroke}, {Tanga}, {Th{\'e}venin}, {Gracia-Abril}, {Comoretto},
  {Garcia-Reinaldos}, {Teyssier}, {Altmann}, {Andrae}, {Audard},
  {Bellas-Velidis}, {Benson}, {Berthier}, {Blomme}, {Burgess}, {Busso},
  {Carry}, {Cellino}, {Clementini}, {Clotet}, {Creevey}, {Davidson}, {De
  Ridder}, {Delchambre}, {Dell'Oro}, {Ducourant},
  {Fern{\'a}ndez-Hern{\'a}ndez}, {Fouesneau}, {Fr{\'e}mat}, {Galluccio},
  {Garc{\'\i}a-Torres}, {Gonz{\'a}lez-N{\'u}{\~n}ez}, {Gonz{\'a}lez-Vidal},
  {Gosset}, {Guy}, {Halbwachs}, {Hambly}, {Harrison}, {Hern{\'a}ndez},
  {Hestroffer}, {Hodgkin}, {Hutton}, {Jasniewicz}, {Jean-Antoine-Piccolo},
  {Jordan}, {Korn}, {Krone-Martins}, {Lanzafame}, {Lebzelter}, {L{\"o}ffler},
  {Manteiga}, {Marrese}, {Mart{\'\i}n-Fleitas}, {Moitinho}, {Mora}, {Muinonen},
  {Osinde}, {Pancino}, {Pauwels}, {Petit}, {Recio-Blanco}, {Richards},
  {Rimoldini}, {Robin}, {Sarro}, {Siopis}, {Smith}, {Sozzetti}, {S{\"u}veges},
  {Torra}, {van Reeven}, {Abbas}, {Abreu Aramburu}, {Accart}, {Aerts},
  {Altavilla}, {{\'A}lvarez}, {Alvarez}, {Alves}, {Anderson}, {Andrei},
  {Anglada Varela}, {Antiche}, {Antoja}, {Arcay}, {Astraatmadja}, {Bach},
  {Baker}, {Balaguer-N{\'u}{\~n}ez}, {Balm}, {Barache}, {Barata}, {Barbato},
  {Barblan}, {Barklem}, {Barrado}, {Barros}, {Barstow}, {Bartholom{\'e}
  Mu{\~n}oz}, {Bassilana}, {Becciani}, {Bellazzini}, {Berihuete}, {Bertone},
  {Bianchi}, {Bienaym{\'e}}, {Blanco-Cuaresma}, {Boch}, {Boeche}, {Bombrun},
  {Borrachero}, {Bossini}, {Bouquillon}, {Bourda}, {Bragaglia}, {Bramante},
  {Breddels}, {Bressan}, {Brouillet}, {Br{\"u}semeister}, {Brugaletta},
  {Bucciarelli}, {Burlacu}, {Busonero}, {Butkevich}, {Buzzi}, {Caffau},
  {Cancelliere}, {Cannizzaro}, {Cantat-Gaudin}, {Carballo}, {Carlucci},
  {Carrasco}, {Casamiquela}, {Castellani}, {Castro-Ginard}, {Charlot},
  {Chemin}, {Chiavassa}, {Cocozza}, {Costigan}, {Cowell}, {Crifo}, {Crosta},
  {Crowley}, {Cuypers}, {Dafonte}, {Damerdji}, {Dapergolas}, {David}, {David},
  {de Laverny}, {De Luise}, {De March}, {de Martino}, {de Souza}, {de Torres},
  {Debosscher}, {del Pozo}, {Delbo}, {Delgado}, {Delgado}, {Di Matteo},
  {Diakite}, {Diener}, {Distefano}, {Dolding}, {Drazinos}, {Dur{\'a}n},
  {Edvardsson}, {Enke}, {Eriksson}, {Esquej}, {Eynard Bontemps}, {Fabre},
  {Fabrizio}, {Faigler}, {Falc{\~a}o}, {Farr{\`a}s Casas}, {Federici},
  {Fedorets}, {Fernique}, {Figueras}, {Filippi}, {Findeisen}, {Fonti},
  {Fraile}, {Fraser}, {Fr{\'e}zouls}, {Gai}, {Galleti}, {Garabato},
  {Garc{\'\i}a-Sedano}, {Garofalo}, {Garralda}, {Gavel}, {Gavras}, {Gerssen},
  {Geyer}, {Giacobbe}, {Gilmore}, {Girona}, {Giuffrida}, {Glass}, {Gomes},
  {Granvik}, {Gueguen}, {Guerrier}, {Guiraud}, {Guti{\'e}rrez-S{\'a}nchez},
  {Haigron}, {Hatzidimitriou}, {Hauser}, {Haywood}, {Heiter}, {Helmi}, {Heu},
  {Hilger}, {Hobbs}, {Hofmann}, {Holland}, {Huckle}, {Hypki}, {Icardi},
  {Jan{\ss}en}, {Jevardat de Fombelle}, {Jonker}, {Juh{\'a}sz}, {Julbe},
  {Karampelas}, {Kewley}, {Klar}, {Kochoska}, {Kohley}, {Kolenberg},
  {Kontizas}, {Kontizas}, {Koposov}, {Kordopatis}, {Kostrzewa-Rutkowska},
  {Koubsky}, {Lambert}, {Lanza}, {Lasne}, {Lavigne}, {Le Fustec}, {Le
  Poncin-Lafitte}, {Lebreton}, {Leccia}, {Leclerc}, {Lecoeur-Taibi},
  {Lenhardt}, {Leroux}, {Liao}, {Licata}, {Lindstr{\o}m}, {Lister}, {Livanou},
  {Lobel}, {L{\'o}pez}, {Managau}, {Mann}, {Mantelet}, {Marchal}, {Marchant},
  {Marconi}, {Marinoni}, {Marschalk{\'o}}, {Marshall}, {Martino}, {Marton},
  {Mary}, {Massari}, {Matijevi{\v{c}}}, {Mazeh}, {McMillan}, {Messina},
  {Michalik}, {Millar}, {Molina}, {Molinaro}, {Moln{\'a}r}, {Montegriffo},
  {Mor}, {Morbidelli}, {Morel}, {Morris}, {Mulone}, {Muraveva}, {Musella},
  {Nelemans}, {Nicastro}, {Noval}, {O'Mullane}, {Ord{\'e}novic},
  {Ord{\'o}{\~n}ez-Blanco}, {Osborne}, {Pagani}, {Pagano}, {Pailler},
  {Palacin}, {Palaversa}, {Panahi}, {Pawlak}, {Piersimoni}, {Pineau}, {Plachy},
  {Plum}, {Poggio}, {Poujoulet}, {Pr{\v{s}}a}, {Pulone}, {Racero}, {Ragaini},
  {Rambaux}, {Ramos-Lerate}, {Regibo}, {Reyl{\'e}}, {Riclet}, {Ripepi}, {Riva},
  {Rivard}, {Rixon}, {Roegiers}, {Roelens}, {Romero-G{\'o}mez}, {Rowell},
  {Royer}, {Ruiz-Dern}, {Sadowski}, {Sagrist{\`a} Sell{\'e}s}, {Sahlmann},
  {Salgado}, {Salguero}, {Sanna}, {Santana-Ros}, {Sarasso}, {Savietto},
  {Schultheis}, {Sciacca}, {Segol}, {Segovia}, {S{\'e}gransan}, {Shih},
  {Siltala}, {Silva}, {Smart}, {Smith}, {Solano}, {Solitro}, {Sordo}, {Soria
  Nieto}, {Souchay}, {Spagna}, {Spoto}, {Stampa}, {Steele},
  {Steidelm{\"u}ller}, {Stephenson}, {Stoev}, {Suess}, {Surdej}, {Szabados},
  {Szegedi-Elek}, {Tapiador}, {Taris}, {Tauran}, {Taylor}, {Teixeira},
  {Terrett}, {Teyssandier}, {Thuillot}, {Titarenko}, {Torra Clotet}, {Turon},
  {Ulla}, {Utrilla}, {Uzzi}, {Vaillant}, {Valentini}, {Valette}, {van Elteren},
  {Van Hemelryck}, {van Leeuwen}, {Vaschetto}, {Vecchiato}, {Veljanoski},
  {Viala}, {Vicente}, {Vogt}, {von Essen}, {Voss}, {Votruba}, {Voutsinas},
  {Walmsley}, {Weiler}, {Wertz}, {Wevers}, {Wyrzykowski}, {Yoldas},
  {{\v{Z}}erjal}, {Ziaeepour}, {Zorec}, {Zschocke}, {Zucker}, {Zurbach}, \&
  {Zwitter}}]{GaiaCollaboration2018}
{Gaia Collaboration}, {Brown}, A.~G.~A., {Vallenari}, A., {et~al.} 2018, \aap,
  616, A1

\bibitem[{{Galland} {et~al.}(2006){Galland}, {Lagrange}, {Udry}, {Beuzit},
  {Pepe}, \& {Mayor}}]{GallandBD2006}
{Galland}, F., {Lagrange}, A.~M., {Udry}, S., {et~al.} 2006, \aap, 452, 709

\bibitem[{{Galland} {et~al.}(2005{\natexlab{a}}){Galland}, {Lagrange}, {Udry},
  {Chelli}, {Pepe}, {Beuzit}, \& {Mayor}}]{Galland2005b}
{Galland}, F., {Lagrange}, A.~M., {Udry}, S., {et~al.} 2005{\natexlab{a}},
  \aap, 444, L21

\bibitem[{{Galland} {et~al.}(2005{\natexlab{b}}){Galland}, {Lagrange}, {Udry},
  {Chelli}, {Pepe}, {Queloz}, {Beuzit}, \& {Mayor}}]{Galland2005a}
{Galland}, F., {Lagrange}, A.~M., {Udry}, S., {et~al.} 2005{\natexlab{b}},
  \aap, 443, 337

\bibitem[{{Gaudi} {et~al.}(2017){Gaudi}, {Stassun}, {Collins}, {Beatty},
  {Zhou}, {Latham}, {Bieryla}, {Eastman}, {Siverd}, {Crepp}, {Gonzales},
  {Stevens}, {Buchhave}, {Pepper}, {Johnson}, {Colon}, {Jensen}, {Rodriguez},
  {Bozza}, {Novati}, {D'Ago}, {Dumont}, {Ellis}, {Gaillard}, {Jang-Condell},
  {Kasper}, {Fukui}, {Gregorio}, {Ito}, {Kielkopf}, {Manner}, {Matt}, {Narita},
  {Oberst}, {Reed}, {Scarpetta}, {Stephens}, {Yeigh}, {Zambelli}, {Fulton},
  {Howard}, {James}, {Penny}, {Bayliss}, {Curtis}, {Depoy}, {Esquerdo},
  {Gould}, {Joner}, {Kuhn}, {Labadie-Bartz}, {Lund}, {Marshall}, {McLeod},
  {Pogge}, {Relles}, {Stockdale}, {Tan}, {Trueblood}, \&
  {Trueblood}}]{Gaudi:2017}
{Gaudi}, B.~S., {Stassun}, K.~G., {Collins}, K.~A., {et~al.} 2017, \nat, 546,
  514

\bibitem[{{Gavel} {et~al.}(2014){Gavel}, {Kupke}, {Dillon}, {Norton},
  {Ratliff}, {Cabak}, {Phillips}, {Rockosi}, {McGurk}, {Srinath}, {Peck},
  {Deich}, {Lanclos}, {Gates}, {Saylor}, {Ward}, \&
  {Pfister}}]{2014SPIE.9148E..05G}
{Gavel}, D., {Kupke}, R., {Dillon}, D., {et~al.} 2014, in Society of
  Photo-Optical Instrumentation Engineers (SPIE) Conference Series, Vol. 9148,
  Adaptive Optics Systems IV, ed. E.~{Marchetti}, L.~M. {Close}, \& J.-P.
  {Vran}, 914805

\bibitem[{{Gillen} {et~al.}(2017){Gillen}, {Hillenbrand}, {David}, {Aigrain},
  {Rebull}, {Stauffer}, {Cody}, \& {Queloz}}]{Gillen2017}
{Gillen}, E., {Hillenbrand}, L.~A., {David}, T.~J., {et~al.} 2017, \apj, 849,
  11

\bibitem[{{Gray} \& {Corbally}(1994)}]{Gray1994}
{Gray}, R.~O. \& {Corbally}, C.~J. 1994, \aj, 107, 742

\bibitem[{{Grieves} {et~al.}(2021){Grieves}, {Bouchy}, {Lendl}, {Carmichael},
  {Mireles}, {Shporer}, {McLeod}, {Collins}, {Brahm}, {Stassun}, {Gill},
  {Bouma}, {Guillot}, {Cointepas}, {Dos Santos}, {Casewell}, {Jenkins},
  {Henning}, {Nielsen}, {Psaridi}, {Udry}, {S{\'e}gransan}, {Eastman}, {Zhou},
  {Abe}, {Agabi}, {Bakos}, {Charbonneau}, {Collins}, {Colon}, {Crouzet},
  {Dransfield}, {Evans}, {Goeke}, {Hart}, {Irwin}, {Jensen}, {Jord{\'a}n},
  {Kielkopf}, {Latham}, {Marie-Sainte}, {M{\'e}karnia}, {Nelson}, {Quinn},
  {Radford}, {Rodriguez}, {Rowden}, {Schmider}, {Schwarz}, {Smith},
  {Stockdale}, {Suarez}, {Tan}, {Triaud}, {Waalkes}, \&
  {Wingham}}]{Grieves2021_FiveHb}
{Grieves}, N., {Bouchy}, F., {Lendl}, M., {et~al.} 2021, \aap, 652, A127

\bibitem[{{Guillot} {et~al.}(2015){Guillot}, {Abe}, {Agabi}, {Rivet}, {Daban},
  {M{\'e}karnia}, {Aristidi}, {Schmider}, {Crouzet}, {Gon{\c{c}}alves},
  {Gouvret}, {Ottogalli}, {Faradji}, {Blanc}, {Bondoux}, \&
  {Valbousquet}}]{guillot2015}
{Guillot}, T., {Abe}, L., {Agabi}, A., {et~al.} 2015, Astronomische
  Nachrichten, 336, 638

\bibitem[{{Hennebelle} \& {Chabrier}(2008)}]{HennebelleChabrier2008}
{Hennebelle}, P. \& {Chabrier}, G. 2008, \apj, 684, 395

\bibitem[{{Hod{\v{z}}i{\'c}} {et~al.}(2018){Hod{\v{z}}i{\'c}}, {Triaud},
  {Anderson}, {Bouchy}, {Collier Cameron}, {Delrez}, {Gillon}, {Hellier},
  {Jehin}, {Lendl}, {Maxted}, {Pepe}, {Pollacco}, {Queloz}, {S{\'e}gransan},
  {Smalley}, {Udry}, \& {West}}]{Hodzic2018}
{Hod{\v{z}}i{\'c}}, V., {Triaud}, A. H.~M.~J., {Anderson}, D.~R., {et~al.}
  2018, \mnras, 481, 5091

\bibitem[{{H{\o}g} {et~al.}(2000){H{\o}g}, {Fabricius}, {Makarov}, {Urban},
  {Corbin}, {Wycoff}, {Bastian}, {Schwekendiek}, \& {Wicenec}}]{Tycho}
{H{\o}g}, E., {Fabricius}, C., {Makarov}, V.~V., {et~al.} 2000, \aap, 355, L27

\bibitem[{{Hooton} {et~al.}(2022){Hooton}, {Hoyer}, {Kitzmann}, {Morris},
  {Smith}, {Collier Cameron}, {Futyan}, {Maxted}, {Queloz}, {Demory}, {Heng},
  {Lendl}, {Cabrera}, {Csizmadia}, {Deline}, {Parviainen}, {Salmon}, {Sulis},
  {Wilson}, {Bonfanti}, {Brandeker}, {Demangeon}, {Oshagh}, {Persson},
  {Scandariato}, {Alibert}, {Alonso}, {Anglada Escud{\'e}}, {B{\'a}rczy},
  {Barrado}, {Barros}, {Baumjohann}, {Beck}, {Beck}, {Benz}, {Billot},
  {Bonfils}, {Bourrier}, {Broeg}, {Busch}, {Charnoz}, {Davies}, {Deleuil},
  {Delrez}, {Ehrenreich}, {Erikson}, {Farinato}, {Fortier}, {Fossati},
  {Fridlund}, {Gandolfi}, {Gillon}, {G{\"u}del}, {Isaak}, {Jones}, {Kiss},
  {Laskar}, {Lecavelier des Etangs}, {Lovis}, {Luntzer}, {Magrin},
  {Nascimbeni}, {Olofsson}, {Ottensamer}, {Pagano}, {Pall{\'e}}, {Peter},
  {Piotto}, {Pollacco}, {Ragazzoni}, {Rando}, {Ratti}, {Rauer}, {Ribas},
  {Santos}, {S{\'e}gransan}, {Simon}, {Sousa}, {Steller}, {Szab{\'o}},
  {Thomas}, {Udry}, {Ulmer}, {Van Grootel}, \& {Walton}}]{Hooton2022}
{Hooton}, M.~J., {Hoyer}, S., {Kitzmann}, D., {et~al.} 2022, \aap, 658, A75

\bibitem[{{Huang} {et~al.}(2020){Huang}, {Vanderburg}, {P{\'a}l}, {Sha}, {Yu},
  {Fong}, {Fausnaugh}, {Shporer}, {Guerrero}, {Vanderspek}, \& {Ricker}}]{QLP}
{Huang}, C.~X., {Vanderburg}, A., {P{\'a}l}, A., {et~al.} 2020, Research Notes
  of the American Astronomical Society, 4, 204

\bibitem[{{Irwin} {et~al.}(2010){Irwin}, {Buchhave}, {Berta}, {Charbonneau},
  {Latham}, {Burke}, {Esquerdo}, {Everett}, {Holman}, {Nutzman}, {Berlind},
  {Calkins}, {Falco}, {Winn}, {Johnson}, \& {Gazak}}]{Irwin2010}
{Irwin}, J., {Buchhave}, L., {Berta}, Z.~K., {et~al.} 2010, \apj, 718, 1353

\bibitem[{{Irwin} {et~al.}(2018){Irwin}, {Charbonneau}, {Esquerdo}, {Latham},
  {Winters}, {Dittmann}, {Newton}, {Berta-Thompson}, {Berlind}, \&
  {Calkins}}]{Irwin2018}
{Irwin}, J.~M., {Charbonneau}, D., {Esquerdo}, G.~A., {et~al.} 2018, \aj, 156,
  140

\bibitem[{{Jackman} {et~al.}(2019){Jackman}, {Wheatley}, {Bayliss}, {Gill},
  {Hodgkin}, {Burleigh}, {Braker}, {G{\"u}nther}, {Louden}, {Turner},
  {Anderson}, {Belardi}, {Bouchy}, {Briegal}, {Bryant}, {Cabrera}, {Casewell},
  {Chaushev}, {Costes}, {Csizmadia}, {Eigm{\"u}ller}, {Erikson},
  {G{\"a}nsicke}, {Gillen}, {Goad}, {Jenkins}, {McCormac}, {Moyano}, {Nielsen},
  {Pollacco}, {Poppenhaeger}, {Queloz}, {Rauer}, {Raynard}, {Smith}, {Udry},
  {Vines}, {Watson}, \& {West}}]{Jackman2019}
{Jackman}, J. A.~G., {Wheatley}, P.~J., {Bayliss}, D., {et~al.} 2019, \mnras,
  489, 5146

\bibitem[{{Jenkins} {et~al.}(2016){Jenkins}, {Twicken}, {McCauliff},
  {Campbell}, {Sanderfer}, {Lung}, {Mansouri-Samani}, {Girouard}, {Tenenbaum},
  {Klaus}, {Smith}, {Caldwell}, {Chacon}, {Henze}, {Heiges}, {Latham},
  {Morgan}, {Swade}, {Rinehart}, \& {Vanderspek}}]{Jenkins2016}
{Jenkins}, J.~M., {Twicken}, J.~D., {McCauliff}, S., {et~al.} 2016, in Society
  of Photo-Optical Instrumentation Engineers (SPIE) Conference Series, Vol.
  9913, Software and Cyberinfrastructure for Astronomy IV, ed. G.~{Chiozzi} \&
  J.~C. {Guzman}, 99133E

\bibitem[{{Jensen}(2013)}]{Jensen:2013}
{Jensen}, E. 2013, {Tapir: A web interface for transit/eclipse observability},
  Astrophysics Source Code Library, record ascl:1306.007

\bibitem[{{Johnson} {et~al.}(2011){Johnson}, {Apps}, {Gazak}, {Crepp},
  {Crossfield}, {Howard}, {Marcy}, {Morton}, {Chubak}, \&
  {Isaacson}}]{Johnson2011}
{Johnson}, J.~A., {Apps}, K., {Gazak}, J.~Z., {et~al.} 2011, \apj, 730, 79

\bibitem[{{Jones} {et~al.}(2019){Jones}, {Brahm}, {Espinoza}, {Wang},
  {Shporer}, {Henning}, {Jord{\'a}n}, {Sarkis}, {Paredes}, {Hodari-Sadiki},
  {Henry}, {Cruz}, {Nielsen}, {Bouchy}, {Pepe}, {S{\'e}gransan}, {Turner},
  {Udry}, {Marmier}, {Lovis}, {Bakos}, {Osip}, {Suc}, {Ziegler}, {Tokovinin},
  {Law}, {Mann}, {Relles}, {Collins}, {Bayliss}, {Sedaghati}, {Latham},
  {Seager}, {Winn}, {Jenkins}, {Smith}, {Davies}, {Tenenbaum}, {Dittmann},
  {Vanderburg}, {Christiansen}, {Haworth}, {Doty}, {Fur{\'e}sz}, {Laughlin},
  {Matthews}, {Crossfield}, {Howell}, {Ciardi}, {Gonzales}, {Matson},
  {Beichman}, \& {Schlieder}}]{TOI2685}
{Jones}, M.~I., {Brahm}, R., {Espinoza}, N., {et~al.} 2019, \aap, 625, A16

\bibitem[{{Kennedy} \& {Kenyon}(2008)}]{KennedyKenyon}
{Kennedy}, G.~M. \& {Kenyon}, S.~J. 2008, \apj, 673, 502

\bibitem[{{Kounkel} {et~al.}(2020){Kounkel}, {Covey}, \&
  {Stassun}}]{Kounkel:2020}
{Kounkel}, M., {Covey}, K., \& {Stassun}, K.~G. 2020, \aj, 160, 279

\bibitem[{{Kraft}(1967)}]{kraft}
{Kraft}, R.~P. 1967, \apj, 150, 551

\bibitem[{{Kreidberg}(2015)}]{batman}
{Kreidberg}, L. 2015, \pasp, 127, 1161

\bibitem[{{Kupke} {et~al.}(2012){Kupke}, {Gavel}, {Roskosi}, {Cabak}, {Cowley},
  {Dillon}, {Gates}, {McGurk}, {Norton}, {Peck}, {Ratliff}, \&
  {Reinig}}]{2012SPIE.8447E..3GK}
{Kupke}, R., {Gavel}, D., {Roskosi}, C., {et~al.} 2012, in Society of
  Photo-Optical Instrumentation Engineers (SPIE) Conference Series, Vol. 8447,
  Adaptive Optics Systems III, ed. B.~L. {Ellerbroek}, E.~{Marchetti}, \& J.-P.
  {V{\'e}ran}, 84473G

\bibitem[{{Kurucz}(1992)}]{Kurucz1992}
{Kurucz}, R.~L. 1992, in The Stellar Populations of Galaxies, ed. B.~{Barbuy}
  \& A.~{Renzini}, Vol. 149, 225

\bibitem[{{Lagrange} {et~al.}(2020){Lagrange}, {Rubini}, {Nowak}, {Lacour},
  {Grandjean}, {Boccaletti}, {Langlois}, {Delorme}, {Gratton}, {Wang},
  {Flasseur}, {Galicher}, {Kral}, {Meunier}, {Beust}, {Babusiaux}, {Le
  Coroller}, {Thebault}, {Kervella}, {Zurlo}, {Maire}, {Wahhaj}, {Amorim},
  {Asensio-Torres}, {Benisty}, {Berger}, {Bonnefoy}, {Brandner}, {Cantalloube},
  {Charnay}, {Chauvin}, {Choquet}, {Cl{\'e}net}, {Christiaens}, {Coud{\'e} Du
  Foresto}, {de Zeeuw}, {Desidera}, {Duvert}, {Eckart}, {Eisenhauer},
  {Galland}, {Gao}, {Garcia}, {Garcia Lopez}, {Gendron}, {Genzel}, {Gillessen},
  {Girard}, {Hagelberg}, {Haubois}, {Henning}, {Heissel}, {Hippler},
  {Horrobin}, {Janson}, {Kammerer}, {Kenworthy}, {Keppler}, {Kreidberg},
  {Lapeyr{\`e}re}, {Le Bouquin}, {L{\'e}na}, {M{\'e}rand}, {Messina},
  {Molli{\`e}re}, {Monnier}, {Ott}, {Otten}, {Paumard}, {Paladini}, {Perraut},
  {Perrin}, {Pueyo}, {Pfuhl}, {Rodet}, {Rodriguez-Coira}, {Rousset}, {Samland},
  {Shangguan}, {Schmidt}, {Straub}, {Straubmeier}, {Stolker}, {Vigan},
  {Vincent}, {Widmann}, {Woillez}, \& {Gravity Collaboration}}]{Lagrange2020}
{Lagrange}, A.~M., {Rubini}, P., {Nowak}, M., {et~al.} 2020, \aap, 642, A18

\bibitem[{{Li} {et~al.}(2019){Li}, {Tenenbaum}, {Twicken}, {Burke}, {Jenkins},
  {Quintana}, {Rowe}, \& {Seader}}]{Li2019}
{Li}, J., {Tenenbaum}, P., {Twicken}, J.~D., {et~al.} 2019, \pasp, 131, 024506

\bibitem[{{Lucy} \& {Sweeney}(1971)}]{Lucy1971}
{Lucy}, L.~B. \& {Sweeney}, M.~A. 1971, \aj, 76, 544

\bibitem[{{Ma} \& {Ge}(2014)}]{maandge}
{Ma}, B. \& {Ge}, J. 2014, \mnras, 439, 2781

\bibitem[{{Mamajek} \& {Hillenbrand}(2008)}]{Mamajek2008}
{Mamajek}, E.~E. \& {Hillenbrand}, L.~A. 2008, \apj, 687, 1264

\bibitem[{{Mayor} \& {Queloz}(1995)}]{mayordidier}
{Mayor}, M. \& {Queloz}, D. 1995, \nat, 378, 355

\bibitem[{{McCully} {et~al.}(2018){McCully}, {Volgenau}, {Harbeck}, {Lister},
  {Saunders}, {Turner}, {Siiverd}, \& {Bowman}}]{McCully:2018}
{McCully}, C., {Volgenau}, N.~H., {Harbeck}, D.-R., {et~al.} 2018, in Society
  of Photo-Optical Instrumentation Engineers (SPIE) Conference Series, Vol.
  10707, \procspie, 107070K

\bibitem[{{McGurk} {et~al.}(2014){McGurk}, {Rockosi}, {Gavel}, {Kupke}, {Peck},
  {Pfister}, {Ward}, {Deich}, {Gates}, {Gates}, {Alcott}, {Cowley}, {Dillon},
  {Lanclos}, {Sandford}, {Saylor}, {Srinath}, {Weiss}, \& {Norton}}]{ShARCS}
{McGurk}, R., {Rockosi}, C., {Gavel}, D., {et~al.} 2014, in Society of
  Photo-Optical Instrumentation Engineers (SPIE) Conference Series, Vol. 9148,
  Adaptive Optics Systems IV, ed. E.~{Marchetti}, L.~M. {Close}, \& J.-P.
  {Vran}, 91483A

\bibitem[{{M{\'e}karnia} {et~al.}(2016){M{\'e}karnia}, {Guillot}, {Rivet},
  {Schmider}, {Abe}, {Gon{\c{c}}alves}, {Agabi}, {Crouzet}, {Fruth},
  {Barbieri}, {Bayliss}, {Zhou}, {Aristidi}, {Szulagyi}, {Daban},
  {Fante{\"\i}-Caujolle}, {Gouvret}, {Erikson}, {Rauer}, {Bouchy}, {Gerakis},
  \& {Bouchez}}]{mekarnia2016}
{M{\'e}karnia}, D., {Guillot}, T., {Rivet}, J.~P., {et~al.} 2016, \mnras, 463,
  45

\bibitem[{Miller \& Fortney(2011)}]{Miller_2011}
Miller, N. \& Fortney, J.~J. 2011, The Astrophysical Journal, 736, L29

\bibitem[{{Moutou} {et~al.}(2013){Moutou}, {Bonomo}, {Bruno}, {Montagnier},
  {Bouchy}, {Almenara}, {Barros}, {Deleuil}, {D{\'\i}az}, {H{\'e}brard}, \&
  {Santerne}}]{Moutou2013}
{Moutou}, C., {Bonomo}, A.~S., {Bruno}, G., {et~al.} 2013, \aap, 558, L6

\bibitem[{{Nowak} {et~al.}(2017){Nowak}, {Palle}, {Gandolfi}, {Dai}, {Lanza},
  {Hirano}, {Barrag{\'a}n}, {Fukui}, {Bruntt}, {Endl}, {Cochran}, {Prada
  Moroni}, {Prieto-Arranz}, {Kiilerich}, {Nespral}, {Hatzes}, {Albrecht},
  {Deeg}, {Winn}, {Yu}, {Kuzuhara}, {Grziwa}, {Smith}, {Guenther}, {Van Eylen},
  {Csizmadia}, {Fridlund}, {Cabrera}, {Eigm{\"u}ller}, {Erikson}, {Korth},
  {Narita}, {P{\"a}tzold}, {Rauer}, \& {Ribas}}]{Nowak2017}
{Nowak}, G., {Palle}, E., {Gandolfi}, D., {et~al.} 2017, \aj, 153, 131

\bibitem[{{Otegi} {et~al.}(2020){Otegi}, {Bouchy}, \& {Helled}}]{otegi2020}
{Otegi}, J.~F., {Bouchy}, F., \& {Helled}, R. 2020, \aap, 634, A43

\bibitem[{{Palle} {et~al.}(2021){Palle}, {Luque}, {Zapatero Osorio},
  {Parviainen}, {Ikoma}, {Tabernero}, {Zechmeister}, {Mustill}, {Bejar},
  {Narita}, \& {Murgas}}]{Palle2021}
{Palle}, E., {Luque}, R., {Zapatero Osorio}, M.~R., {et~al.} 2021, \aap, 650,
  A55

\bibitem[{{Paredes} {et~al.}(2021){Paredes}, {Henry}, {Quinn}, {Gies},
  {Hinojosa-Go{\~n}i}, {James}, {Jao}, \& {White}}]{2021AJ....162..176P}
{Paredes}, L.~A., {Henry}, T.~J., {Quinn}, S.~N., {et~al.} 2021, \aj, 162, 176

\bibitem[{{Pepe} {et~al.}(2002){Pepe}, {Mayor}, {Rupprecht}, {Avila},
  {Ballester}, {Beckers}, {Benz}, {Bertaux}, {Bouchy}, {Buzzoni}, {Cavadore},
  {Deiries}, {Dekker}, {Delabre}, {D'Odorico}, {Eckert}, {Fischer}, {Fleury},
  {George}, {Gilliotte}, {Gojak}, {Guzman}, {Koch}, {Kohler}, {Kotzlowski},
  {Lacroix}, {Le Merrer}, {Lizon}, {Lo Curto}, {Longinotti}, {Megevand},
  {Pasquini}, {Petitpas}, {Pichard}, {Queloz}, {Reyes}, {Richaud}, {Sivan},
  {Sosnowska}, {Soto}, {Udry}, {Ureta}, {van Kesteren}, {Weber}, {Weilenmann},
  {Wicenec}, {Wieland}, {Christensen-Dalsgaard}, {Dravins}, {Hatzes},
  {K{\"u}rster}, {Paresce}, \& {Penny}}]{Pepe2002}
{Pepe}, F., {Mayor}, M., {Rupprecht}, G., {et~al.} 2002, The Messenger, 110, 9

\bibitem[{{Persson} {et~al.}(2019){Persson}, {Csizmadia}, {Mustill},
  {Fridlund}, {Hatzes}, {Nowak}, {Georgieva}, {Gandolfi}, {Davies},
  {Livingston}, {Palle}, {Monta{\~n}es Rodr{\'\i}guez}, {Endl}, {Hirano},
  {Prieto-Arranz}, {Korth}, {Grziwa}, {Esposito}, {Albrecht}, {Johnson},
  {Barrag{\'a}n}, {Parviainen}, {Van Eylen}, {Alonso Sobrino}, {Beck},
  {Cabrera}, {Carleo}, {Cochran}, {Dai}, {Deeg}, {de Leon}, {Eigm{\"u}ller},
  {Erikson}, {Fukui}, {Gonz{\'a}lez-Cuesta}, {Guenther}, {Hidalgo}, {Hjorth},
  {Kabath}, {Knudstrup}, {Kusakabe}, {Lam}, {Lund}, {Luque}, {Mathur},
  {Murgas}, {Narita}, {Nespral}, {Niraula}, {Olofsson}, {P{\"a}tzold}, {Rauer},
  {Redfield}, {Ribas}, {Skarka}, {Smith}, {Subjak}, \& {Tamura}}]{Persson2019}
{Persson}, C.~M., {Csizmadia}, S., {Mustill}, A. e.~J., {et~al.} 2019, \aap,
  628, A64

\bibitem[{{Queloz} {et~al.}(2001){Queloz}, {Mayor}, {Udry}, {Burnet},
  {Carrier}, {Eggenberger}, {Naef}, {Santos}, {Pepe}, {Rupprecht}, {Avila},
  {Baeza}, {Benz}, {Bertaux}, {Bouchy}, {Cavadore}, {Delabre}, {Eckert},
  {Fischer}, {Fleury}, {Gilliotte}, {Goyak}, {Guzman}, {Kohler}, {Lacroix},
  {Lizon}, {Megevand}, {Sivan}, {Sosnowska}, \& {Weilenmann}}]{Queloz2001}
{Queloz}, D., {Mayor}, M., {Udry}, S., {et~al.} 2001, The Messenger, 105, 1

\bibitem[{{Ricker} {et~al.}(2015){Ricker}, {Winn}, {Vanderspek}, {Latham},
  {Bakos}, {Bean}, {Berta-Thompson}, {Brown}, {Buchhave}, {Butler}, {Butler},
  {Chaplin}, {Charbonneau}, {Christensen-Dalsgaard}, {Clampin}, {Deming},
  {Doty}, {De Lee}, {Dressing}, {Dunham}, {Endl}, {Fressin}, {Ge}, {Henning},
  {Holman}, {Howard}, {Ida}, {Jenkins}, {Jernigan}, {Johnson}, {Kaltenegger},
  {Kawai}, {Kjeldsen}, {Laughlin}, {Levine}, {Lin}, {Lissauer}, {MacQueen},
  {Marcy}, {McCullough}, {Morton}, {Narita}, {Paegert}, {Palle}, {Pepe},
  {Pepper}, {Quirrenbach}, {Rinehart}, {Sasselov}, {Sato}, {Seager},
  {Sozzetti}, {Stassun}, {Sullivan}, {Szentgyorgyi}, {Torres}, {Udry}, \&
  {Villasenor}}]{Ricker2015}
{Ricker}, G.~R., {Winn}, J.~N., {Vanderspek}, R., {et~al.} 2015, Journal of
  Astronomical Telescopes, Instruments, and Systems, 1, 014003

\bibitem[{{Santerne} {et~al.}(2013){Santerne}, {Fressin}, {D{\'\i}az},
  {Figueira}, {Almenara}, \& {Santos}}]{SanterneSecondary}
{Santerne}, A., {Fressin}, F., {D{\'\i}az}, R.~F., {et~al.} 2013, \aap, 557,
  A139

\bibitem[{{Santerne} {et~al.}(2016){Santerne}, {Moutou}, {Tsantaki}, {Bouchy},
  {H{\'e}brard}, {Adibekyan}, {Almenara}, {Amard}, {Barros}, {Boisse},
  {Bonomo}, {Bruno}, {Courcol}, {Deleuil}, {Demangeon}, {D{\'\i}az}, {Guillot},
  {Havel}, {Montagnier}, {Rajpurohit}, {Rey}, \& {Santos}}]{Santerne2016EBLM}
{Santerne}, A., {Moutou}, C., {Tsantaki}, M., {et~al.} 2016, \aap, 587, A64

\bibitem[{{Santos} {et~al.}(2017){Santos}, {Adibekyan}, {Figueira},
  {Andreasen}, {Barros}, {Delgado-Mena}, {Demangeon}, {Faria}, {Oshagh},
  {Sousa}, {Viana}, \& {Ferreira}}]{Santos2017}
{Santos}, N.~C., {Adibekyan}, V., {Figueira}, P., {et~al.} 2017, \aap, 603, A30

\bibitem[{{Savel} {et~al.}(2020){Savel}, {Dressing}, {Hirsch}, {Ciardi},
  {Fleming}, {Giacalone}, {Mayo}, \& {Christiansen}}]{2020AJ....160..287S}
{Savel}, A.~B., {Dressing}, C.~D., {Hirsch}, L.~A., {et~al.} 2020, \aj, 160,
  287

\bibitem[{Schlaufman(2018)}]{Schlaufman2018}
Schlaufman, K.~C. 2018, The Astrophysical Journal, 853, 37

\bibitem[{{Schlegel} {et~al.}(1998){Schlegel}, {Finkbeiner}, \&
  {Davis}}]{Schlegel:1998}
{Schlegel}, D.~J., {Finkbeiner}, D.~P., \& {Davis}, M. 1998, \apj, 500, 525

\bibitem[{{Sestovic} {et~al.}(2018){Sestovic}, {Demory}, \&
  {Queloz}}]{Sest2018}
{Sestovic}, M., {Demory}, B.-O., \& {Queloz}, D. 2018, \aap, 616, A76

\bibitem[{{Siverd} {et~al.}(2012){Siverd}, {Beatty}, {Pepper}, {Eastman},
  {Collins}, {Bieryla}, {Latham}, {Buchhave}, {Jensen}, {Crepp}, {Street},
  {Stassun}, {Gaudi}, {Berlind}, {Calkins}, {DePoy}, {Esquerdo}, {Fulton},
  {F{\H{u}}r{\'e}sz}, {Geary}, {Gould}, {Hebb}, {Kielkopf}, {Marshall},
  {Pogge}, {Stanek}, {Stefanik}, {Szentgyorgyi}, {Trueblood}, {Trueblood},
  {Stutz}, \& {van Saders}}]{Siverd2012}
{Siverd}, R.~J., {Beatty}, T.~G., {Pepper}, J., {et~al.} 2012, \apj, 761, 123

\bibitem[{{Skrutskie} {et~al.}(2006){Skrutskie}, {Cutri}, {Stiening},
  {Weinberg}, {Schneider}, {Carpenter}, {Beichman}, {Capps}, {Chester},
  {Elias}, {Huchra}, {Liebert}, {Lonsdale}, {Monet}, {Price}, {Seitzer},
  {Jarrett}, {Kirkpatrick}, {Gizis}, {Howard}, {Evans}, {Fowler}, {Fullmer},
  {Hurt}, {Light}, {Kopan}, {Marsh}, {McCallon}, {Tam}, {Van Dyk}, \&
  {Wheelock}}]{Skrutskie2006}
{Skrutskie}, M.~F., {Cutri}, R.~M., {Stiening}, R., {et~al.} 2006, \aj, 131,
  1163

\bibitem[{{Smith} {et~al.}(2012){Smith}, {Stumpe}, {Van Cleve}, {Jenkins},
  {Barclay}, {Fanelli}, {Girouard}, {Kolodziejczak}, {McCauliff}, {Morris}, \&
  {Twicken}}]{Smith2012}
{Smith}, J.~C., {Stumpe}, M.~C., {Van Cleve}, J.~E., {et~al.} 2012, \pasp, 124,
  1000

\bibitem[{{Speagle}(2020)}]{dynesty}
{Speagle}, J.~S. 2020, \mnras, 493, 3132

\bibitem[{{Stassun} {et~al.}(2017){Stassun}, {Collins}, \&
  {Gaudi}}]{Stassun:2017}
{Stassun}, K.~G., {Collins}, K.~A., \& {Gaudi}, B.~S. 2017, \aj, 153, 136

\bibitem[{{Stassun} {et~al.}(2018){Stassun}, {Corsaro}, {Pepper}, \&
  {Gaudi}}]{Stassun:2018}
{Stassun}, K.~G., {Corsaro}, E., {Pepper}, J.~A., \& {Gaudi}, B.~S. 2018, \aj,
  155, 22

\bibitem[{{Stassun} {et~al.}(2006){Stassun}, {Mathieu}, \&
  {Valenti}}]{Stassun2006}
{Stassun}, K.~G., {Mathieu}, R.~D., \& {Valenti}, J.~A. 2006, \nat, 440, 311

\bibitem[{{Stassun} {et~al.}(2019){Stassun}, {Oelkers}, {Paegert}, {Torres},
  {Pepper}, {De Lee}, {Collins}, {Latham}, {Muirhead}, {Chittidi},
  {Rojas-Ayala}, {Fleming}, {Rose}, {Tenenbaum}, {Ting}, {Kane}, {Barclay},
  {Bean}, {Brassuer}, {Charbonneau}, {Ge}, {Lissauer}, {Mann}, {McLean},
  {Mullally}, {Narita}, {Plavchan}, {Ricker}, {Sasselov}, {Seager}, {Sharma},
  {Shiao}, {Sozzetti}, {Stello}, {Vanderspek}, {Wallace}, \&
  {Winn}}]{Stassun2019}
{Stassun}, K.~G., {Oelkers}, R.~J., {Paegert}, M., {et~al.} 2019, \aj, 158, 138

\bibitem[{{Stassun} \& {Torres}(2016)}]{Stassun:2016}
{Stassun}, K.~G. \& {Torres}, G. 2016, \aj, 152, 180

\bibitem[{{Stassun} \& {Torres}(2021)}]{StassunTorres:2021}
{Stassun}, K.~G. \& {Torres}, G. 2021, \apjl, 907, L33

\bibitem[{{Stumpe} {et~al.}(2014){Stumpe}, {Smith}, {Catanzarite}, {Van Cleve},
  {Jenkins}, {Twicken}, \& {Girouard}}]{Stumpe2014}
{Stumpe}, M.~C., {Smith}, J.~C., {Catanzarite}, J.~H., {et~al.} 2014, \pasp,
  126, 100

\bibitem[{{Stumpe} {et~al.}(2012){Stumpe}, {Smith}, {Van Cleve}, {Twicken},
  {Barclay}, {Fanelli}, {Girouard}, {Jenkins}, {Kolodziejczak}, {McCauliff}, \&
  {Morris}}]{Stumpe2012}
{Stumpe}, M.~C., {Smith}, J.~C., {Van Cleve}, J.~E., {et~al.} 2012, \pasp, 124,
  985

\bibitem[{{Talens} {et~al.}(2017){Talens}, {Albrecht}, {Spronck}, {Lesage},
  {Otten}, {Stuik}, {Van Eylen}, {Van Winckel}, {Pollacco}, {McCormac},
  {Grundahl}, {Fredslund Andersen}, {Antoci}, \& {Snellen}}]{mascara1}
{Talens}, G.~J.~J., {Albrecht}, S., {Spronck}, J.~F.~P., {et~al.} 2017, \aap,
  606, A73

\bibitem[{{Temple} {et~al.}(2017){Temple}, {Hellier}, {Albrow}, {Anderson},
  {Bayliss}, {Beatty}, {Bieryla}, {Brown}, {Cargile}, {Collier Cameron},
  {Collins}, {Col{\'o}n}, {Curtis}, {D'Ago}, {Delrez}, {Eastman}, {Gaudi},
  {Gillon}, {Gregorio}, {James}, {Jehin}, {Joner}, {Kielkopf}, {Kuhn},
  {Labadie-Bartz}, {Latham}, {Lendl}, {Lund}, {Malpas}, {Maxted}, {Myers},
  {Oberst}, {Pepe}, {Pepper}, {Pollacco}, {Queloz}, {Rodriguez},
  {S{\'e}gransan}, {Siverd}, {Smalley}, {Stassun}, {Stevens}, {Stockdale},
  {Tan}, {Triaud}, {Udry}, {Villanueva}, {West}, \& {Zhou}}]{WASP-167}
{Temple}, L.~Y., {Hellier}, C., {Albrow}, M.~D., {et~al.} 2017, \mnras, 471,
  2743

\bibitem[{{Tokovinin}(2018)}]{SOAR}
{Tokovinin}, A. 2018, \pasp, 130, 035002

\bibitem[{{Tokovinin} {et~al.}(2013){Tokovinin}, {Fischer}, {Bonati},
  {Giguere}, {Moore}, {Schwab}, {Spronck}, \&
  {Szymkowiak}}]{2013PASP..125.1336T}
{Tokovinin}, A., {Fischer}, D.~A., {Bonati}, M., {et~al.} 2013, \pasp, 125,
  1336

\bibitem[{{Torres} {et~al.}(2010){Torres}, {Andersen}, \&
  {Gim{\'e}nez}}]{Torres:2010}
{Torres}, G., {Andersen}, J., \& {Gim{\'e}nez}, A. 2010, \aapr, 18, 67

\bibitem[{{Tremblin} {et~al.}(2017){Tremblin}, {Chabrier}, {Mayne}, {Amundsen},
  {Baraffe}, {Debras}, {Drummond}, {Manners}, \& {Fromang}}]{Tremblin2017}
{Tremblin}, P., {Chabrier}, G., {Mayne}, N.~J., {et~al.} 2017, \apj, 841, 30

\bibitem[{{Triaud} {et~al.}(2020){Triaud}, {Burgasser}, {Burdanov}, {Kunovac
  Hod{\v{z}}i{\'c}}, {Alonso}, {Bardalez Gagliuffi}, {Delrez}, {Demory}, {de
  Wit}, {Ducrot}, {Hessman}, {Husser}, {Jehin}, {Pedersen}, {Queloz},
  {McCormac}, {Murray}, {Sebastian}, {Thompson}, {Van Grootel}, \&
  {Gillon}}]{Triaud2020}
{Triaud}, A. H.~M.~J., {Burgasser}, A.~J., {Burdanov}, A., {et~al.} 2020,
  Nature Astronomy, 4, 650

\bibitem[{{Triaud} {et~al.}(2013){Triaud}, {Hebb}, {Anderson}, {Cargile},
  {Collier Cameron}, {Doyle}, {Faedi}, {Gillon}, {Gomez Maqueo Chew},
  {Hellier}, {Jehin}, {Maxted}, {Naef}, {Pepe}, {Pollacco}, {Queloz},
  {S{\'e}gransan}, {Smalley}, {Stassun}, {Udry}, \& {West}}]{Triaud2013}
{Triaud}, A.~H.~M.~J., {Hebb}, L., {Anderson}, D.~R., {et~al.} 2013, \aap, 549,
  A18

\bibitem[{{Twicken} {et~al.}(2018){Twicken}, {Catanzarite}, {Clarke},
  {Girouard}, {Jenkins}, {Klaus}, {Li}, {McCauliff}, {Seader}, {Tenenbaum},
  {Wohler}, {Bryson}, {Burke}, {Caldwell}, {Haas}, {Henze}, \&
  {Sanderfer}}]{Twicken2018}
{Twicken}, J.~D., {Catanzarite}, J.~H., {Clarke}, B.~D., {et~al.} 2018, \pasp,
  130, 064502

\bibitem[{{Udry} {et~al.}(2003){Udry}, {Mayor}, \& {Santos}}]{Udry2003}
{Udry}, S., {Mayor}, M., \& {Santos}, N.~C. 2003, \aap, 407, 369

\bibitem[{{Vanderburg} {et~al.}(2016){Vanderburg}, {Plavchan}, {Johnson},
  {Ciardi}, {Swift}, \& {Kane}}]{vanderburg2016}
{Vanderburg}, A., {Plavchan}, P., {Johnson}, J.~A., {et~al.} 2016, \mnras, 459,
  3565

\bibitem[{{{\v{S}}ubjak} {et~al.}(2020){{\v{S}}ubjak}, {Sharma}, {Carmichael},
  {Johnson}, {Gonzales}, {Matthews}, {Boffin}, {Brahm}, {Chaturvedi},
  {Chakraborty}, {Ciardi}, {Collins}, {Esposito}, {Fridlund}, {Gan},
  {Gandolfi}, {Garc{\'\i}a}, {Guenther}, {Hatzes}, {Latham}, {Mathis},
  {Mathur}, {Persson}, {Relles}, {Schlieder}, {Barclay}, {Dressing},
  {Crossfield}, {Howard}, {Rodler}, {Zhou}, {Quinn}, {Esquerdo}, {Calkins},
  {Berlind}, {Stassun}, {Bla{\v{z}}ek}, {Skarka}, {{\v{S}}pokov{\'a}},
  {{\v{Z}}{\'a}k}, {Albrecht}, {Sobrino}, {Beck}, {Cabrera}, {Carleo},
  {Cochran}, {Csizmadia}, {Dai}, {Deeg}, {de Leon}, {Eigm{\"u}ller}, {Endl},
  {Erikson}, {Fukui}, {Georgieva}, {Gonz{\'a}lez-Cuesta}, {Grziwa}, {Hidalgo},
  {Hirano}, {Hjorth}, {Knudstrup}, {Korth}, {Lam}, {Livingston}, {Lund},
  {Luque}, {Rodr{\'\i}guez}, {Murgas}, {Narita}, {Nespral}, {Niraula}, {Nowak},
  {Pall{\'e}}, {P{\"a}tzold}, {Prieto-Arranz}, {Rauer}, {Redfield}, {Ribas},
  {Smith}, {Eylen}, \& {Kab{\'a}th}}]{Subjak2020}
{{\v{S}}ubjak}, J., {Sharma}, R., {Carmichael}, T.~W., {et~al.} 2020, \aj, 159,
  151

\bibitem[{{Winn} {et~al.}(2010){Winn}, {Fabrycky}, {Albrecht}, \&
  {Johnson}}]{winn2010}
{Winn}, J.~N., {Fabrycky}, D., {Albrecht}, S., \& {Johnson}, J.~A. 2010, \apjl,
  718, L145

\bibitem[{{Wong} {et~al.}(2021){Wong}, {Shporer}, {Zhou}, {Kitzmann},
  {Komacek}, {Tan}, {Tronsgaard}, {Buchhave}, {Vissapragada}, {Greklek-McKeon},
  {Rodriguez}, {Ahlers}, {Quinn}, {Furlan}, {Howell}, {Bieryla}, {Heng},
  {Knutson}, {Collins}, {McLeod}, {Berlind}, {Brown}, {Calkins}, {de Leon},
  {Esparza-Borges}, {Esquerdo}, {Fukui}, {Gan}, {Girardin}, {Gnilka}, {Ikoma},
  {Jensen}, {Kielkopf}, {Kodama}, {Kurita}, {Lester}, {Lewin}, {Marino},
  {Murgas}, {Narita}, {Pall{\'e}}, {Schwarz}, {Stassun}, {Tamura}, {Watanabe},
  {Benneke}, {Ricker}, {Latham}, {Vanderspek}, {Seager}, {Winn}, {Jenkins},
  {Caldwell}, {Fong}, {Huang}, {Mireles}, {Schlieder}, {Shiao}, \& {Noel
  Villase{\~n}or}}]{TOI2109}
{Wong}, I., {Shporer}, A., {Zhou}, G., {et~al.} 2021, \aj, 162, 256

\bibitem[{Yee {et~al.}(2017)Yee, Petigura, \& von Braun}]{specmatch}
Yee, S.~W., Petigura, E.~A., \& von Braun, K. 2017, The Astrophysical Journal,
  836, 77

\bibitem[{{Zhou} {et~al.}(2019){Zhou}, {Bakos}, {Bayliss}, {Bento}, {Bhatti},
  {Brahm}, {Csubry}, {Espinoza}, {Hartman}, {Henning}, {Jord{\'a}n}, {Mancini},
  {Penev}, {Rabus}, {Sarkis}, {Suc}, {de Val-Borro}, {Rodriguez}, {Osip},
  {Kedziora-Chudczer}, {Bailey}, {Tinney}, {Durkan}, {L{\'a}z{\'a}r}, {Papp},
  \& {S{\'a}ri}}]{Zhou2019}
{Zhou}, G., {Bakos}, G.~{\'A}., {Bayliss}, D., {et~al.} 2019, \aj, 157, 31

\end{thebibliography}
\begin{appendix} 

\appendix
\onecolumn

\newpage
\clearpage
\section{Light curves}\label{sec:lightcurves}

\begin{figure*}[ht]
  \centering
  \includegraphics[width=0.49\textwidth]{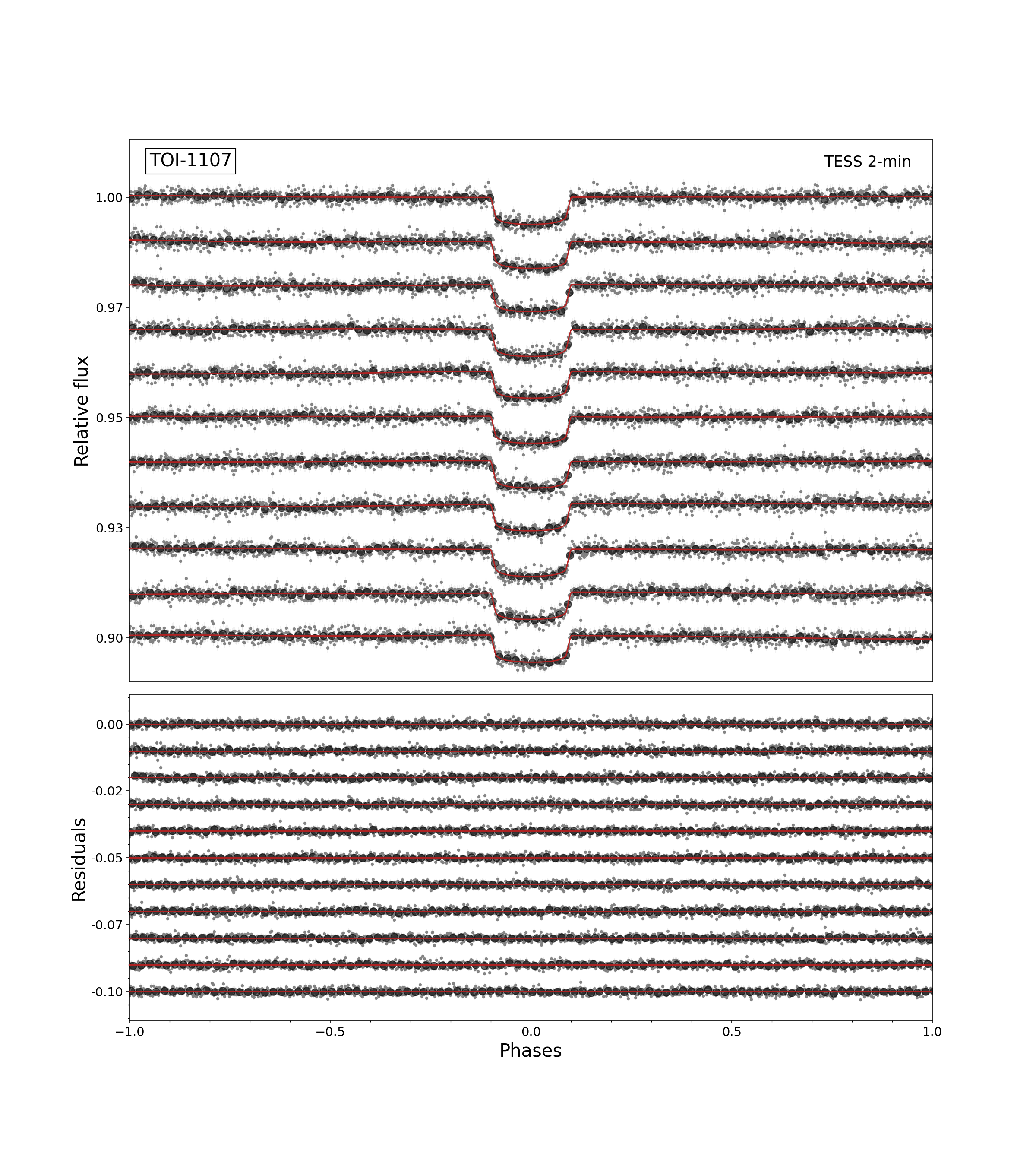}
  \includegraphics[width=0.49\textwidth]{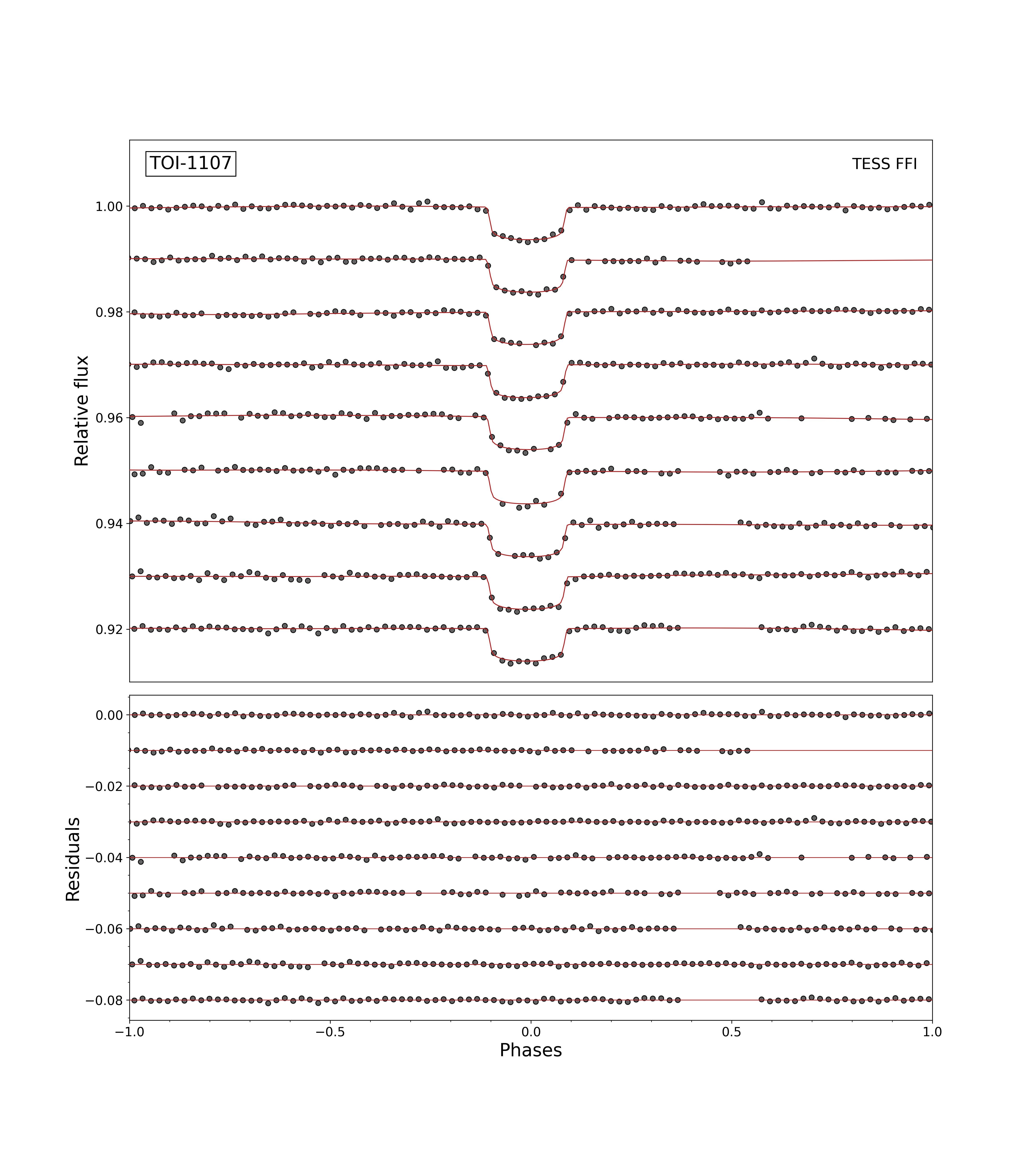}
  
   \includegraphics[width=0.37\textwidth]{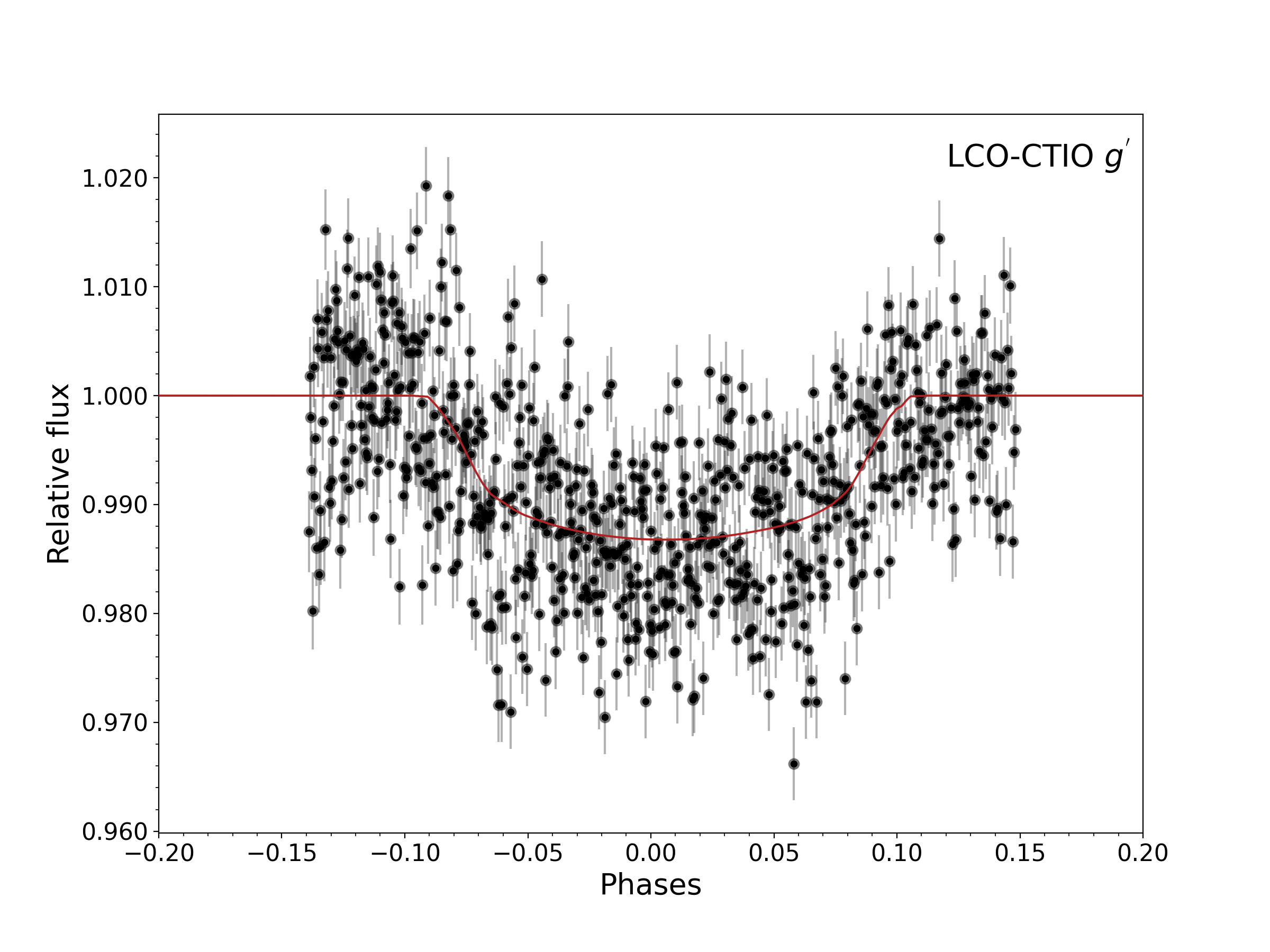}
   \includegraphics[width=0.37\textwidth]{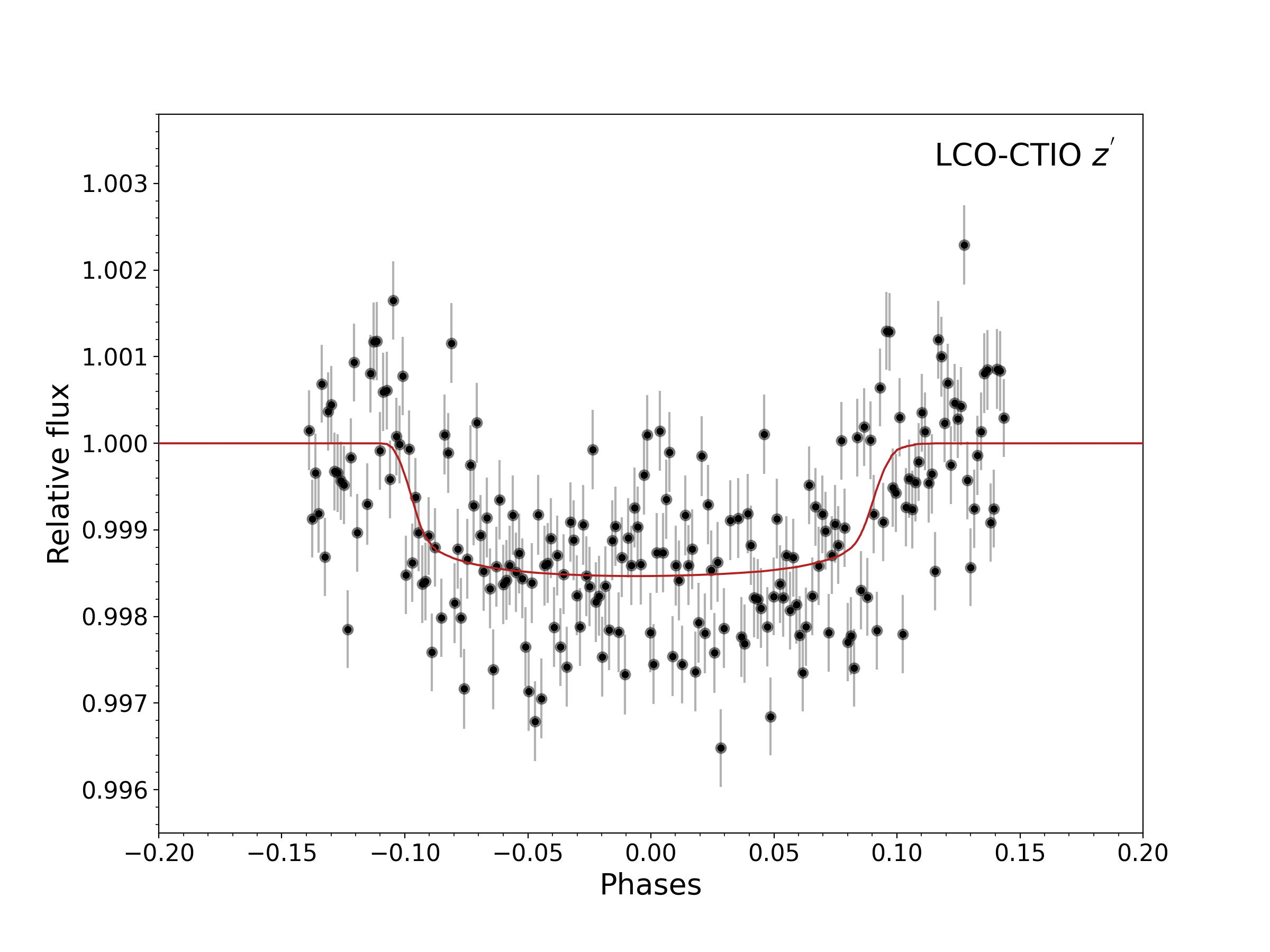}
   
    \includegraphics[width=0.37\textwidth]{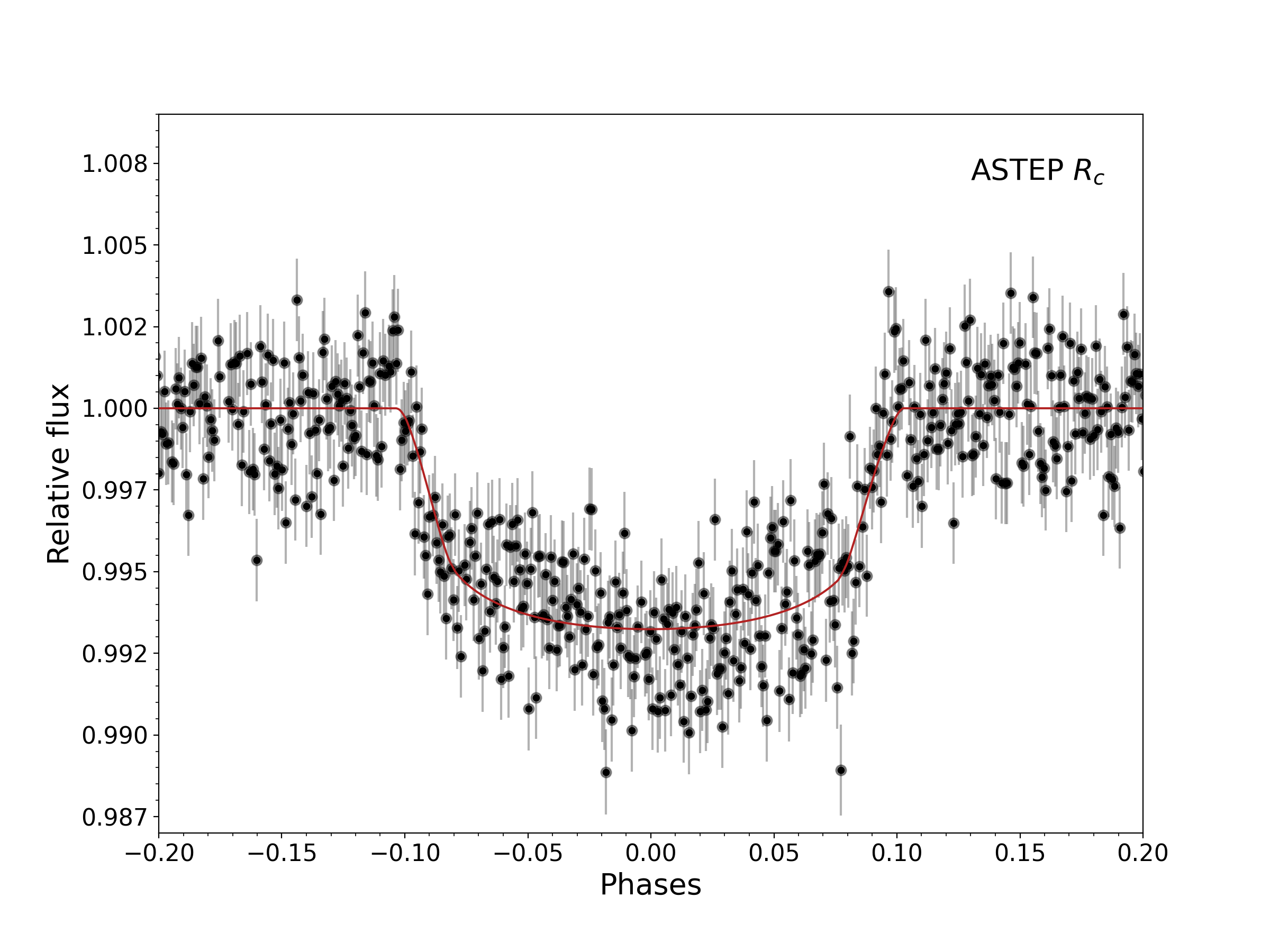}
   \includegraphics[width=0.37\textwidth]{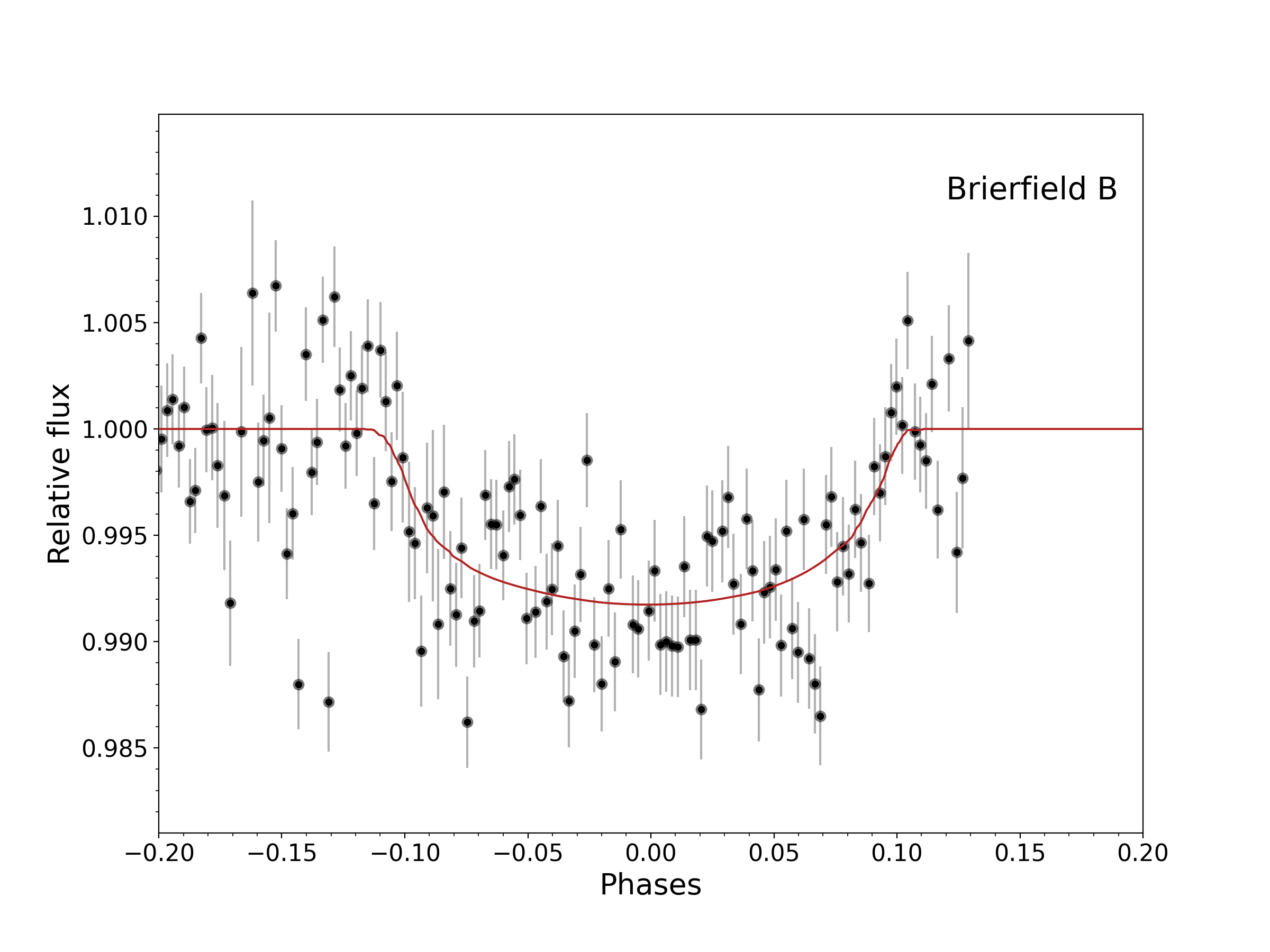}
  \caption{TESS and ground-based light curves of TOI-1107. \textit{Top}: TOI-1107 TESS 2-minute and FFI light curves. In the 2-minute cadence light curves the observed flux is shown in light gray circles and the 30-minute binned flux in black circles. The red line shows the GP and planet models that have been fit.
  \textit{Bottom}: TOI-1107 detrended ground-based light curves with the best-fit linear model shown in red.}
  \label{fig:LCsTOI1107}
\end{figure*}
\newpage

\begin{figure*}[ht]
  \centering
  \includegraphics[width=0.43\linewidth]{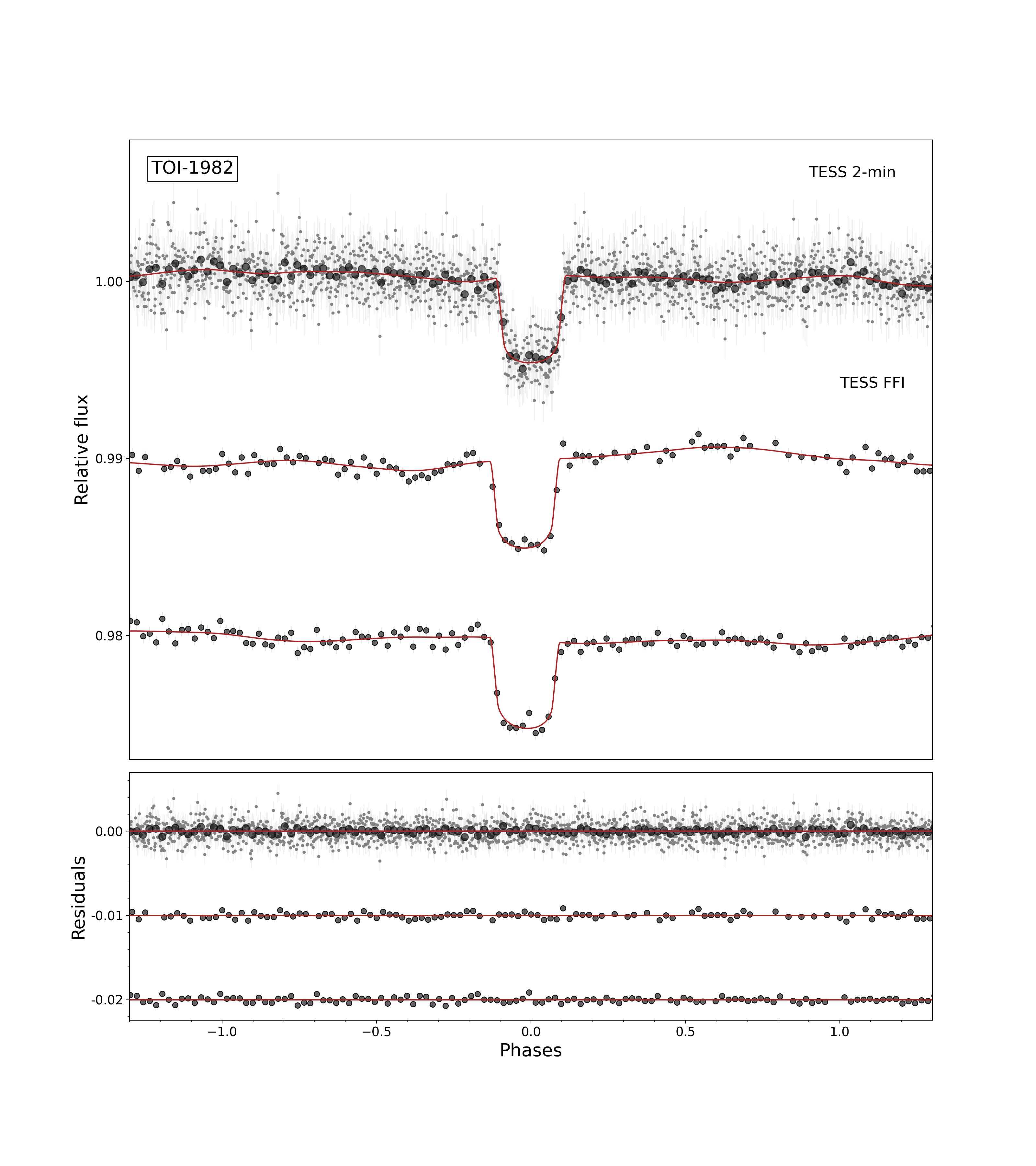}
 \hspace{4.00mm}
  \includegraphics[width=0.43\linewidth]{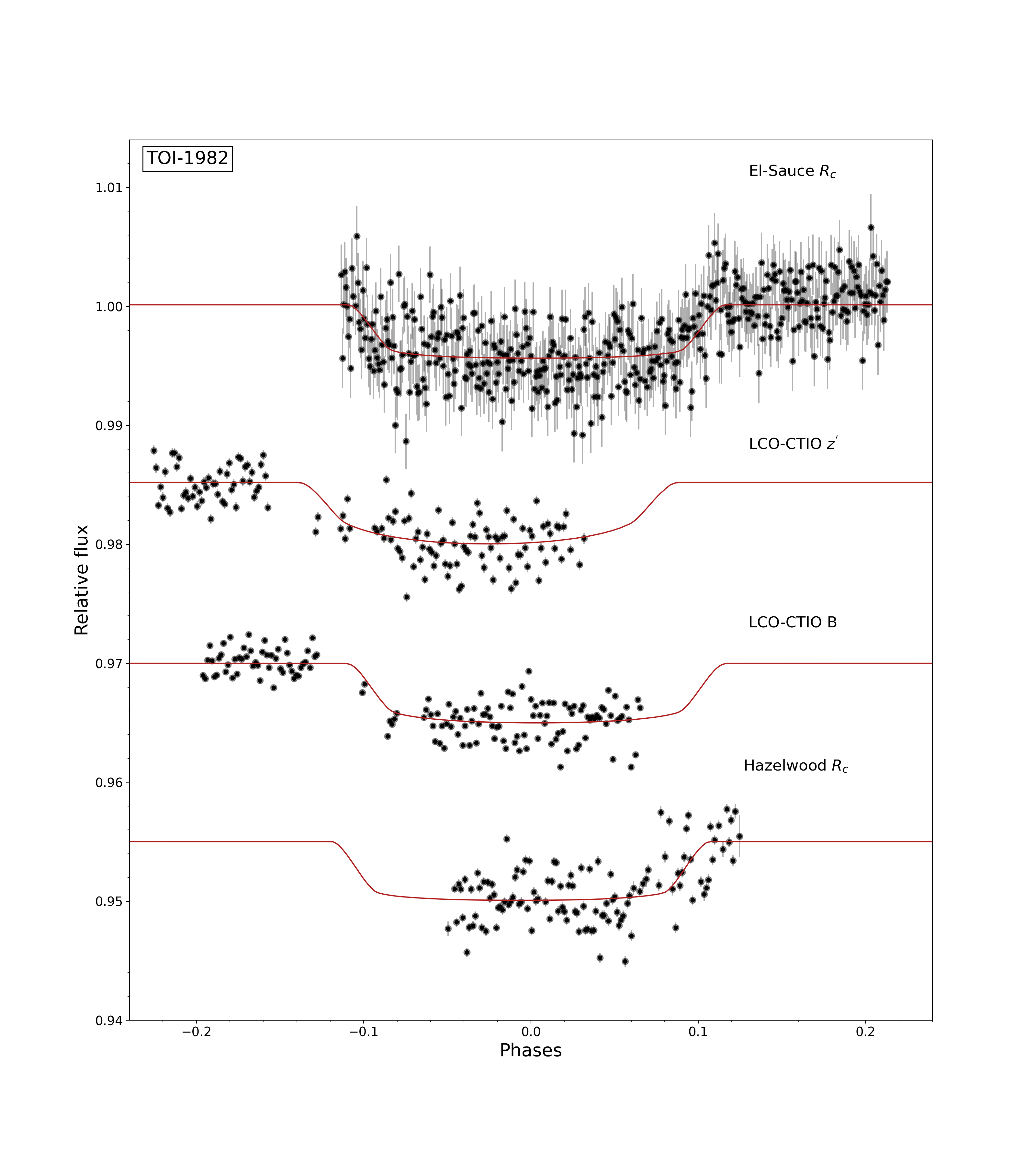}
  \caption{TESS and ground-based light curves of TOI-1982. \textit{Left}: TOI-1982 TESS 2-minute and FFI light curves. In the 2-minute cadence light curves the observed flux is shown in light gray circles and the 30-minute binned flux in black circles. The red line shows the GP and planet models that have been fit.
  \textit{Right}: TOI-1982 detrended ground-based light curves with the best-fit linear model shown in red.}
  \label{fig:LCTOI1982}
\end{figure*}

\begin{figure*}[h]
\centering
\begin{minipage}{.44\textwidth}
  \centering
  \includegraphics[width=1\linewidth]{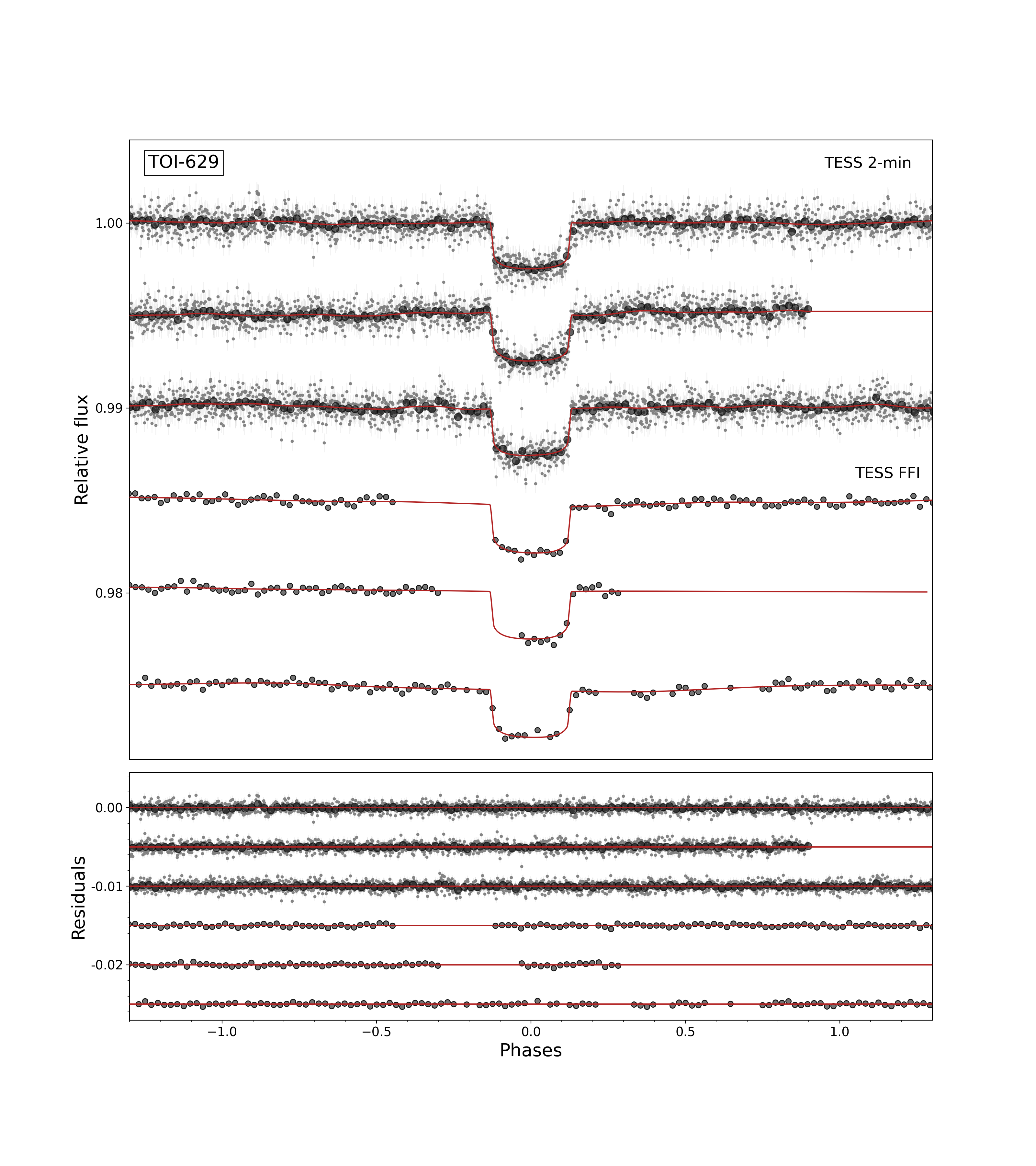}
  \captionof{figure}{TOI-629 TESS 2-minute and FFI 
  light curves. In the 2-minute cadence light curves the 
  observed flux is shown in light gray circles 
  and the 30-minute binned flux in black circles. The red line shows the GP and planet models
  that have been fit.}
  \label{fig:LCTOI629}
\end{minipage}
\hspace{9.00mm}
\begin{minipage}{.44\textwidth}
  \centering
  \includegraphics[width=1\linewidth]{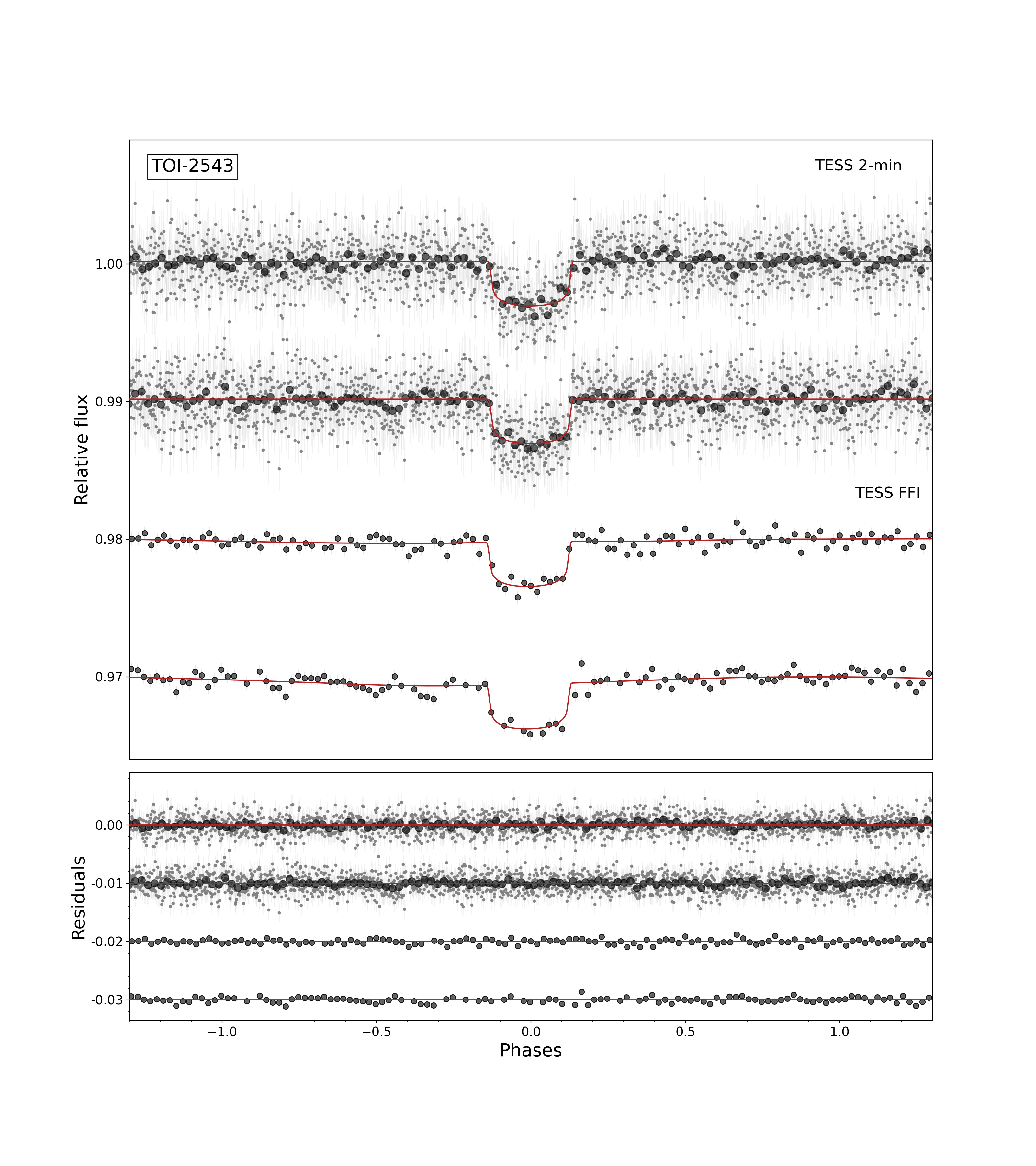}
  \captionof{figure}{TOI-2543 TESS 2-minute and FFI
  light curves. In the 2-minute cadence light curves the 
  observed flux is shown in light gray circles and the 30-minute binned flux in black circles. 
  The red line shows the GP and planet models 
  that have been fit.}
  \label{fig:LCTOI2543}
\end{minipage}
\end{figure*}

\newpage
\clearpage
\section{Global analysis with \juliet}\label{sec:julietglobal}
\begin{table}[ht]
     \caption{Prior and posterior values for the joint photometric and radial velocity fit with \juliet of TOI-629 (\textit{left}) and TOI-2543 (\textit{right}).}
\tiny
\renewcommand{\arraystretch}{1.4}
    \begin{subtable}[h]{0.52\textwidth}
        \centering
        \begin{tabular}{lll}
        \hline\hline
        Parameter$^{1}$ &Prior$^{2}$&Posterior estimate$^{3}$\\
        \hline 
        \textbf{TOI-629} & &  \\
        \textbf{Star:} & &  \\
        $\rho_*$ (cgs)  &$\mathcal{N}(0.240,0.031)$ & 0.240 $^{+0.017}_{-0.022}$ \\
        \textbf{Companion:} & &  \\
$P$ (days) & $\mathcal{U}(8.70,8.72)$ & 8.717993 $^{+0.000012}_{-0.000013}$ \\
$T_C$ (\bjdtdb) & $\mathcal{U}(2459204.2,2459204.3)$ & 2459204.2552 $^{+0.0005}_{-0.0005}$ \\
$r_{1}$ & $\mathcal{U}(0, 1)$ & 0.4686$^{+0.0856}_{-0.0005}$ \\
$r_{2}$& $\mathcal{U}(0, 1)$ & 0.04790 $^{+0.00043}_{-0.00042}$ \\
$\textit{e}$ &$\mathcal{U}(0, 1)$ & 0.2986$^{+0.0076}_{-0.0075}$\\
$\omega_*$ (deg) &$\mathcal{U}(0, 360)$ & 8.53 $^{+3.05}_{-3.21}$\\
K ($\kms$) &$\mathcal{U}(3, 5)$ & 3.954 $^{+0.073}_{-0.072}$\\
        \textbf{TESS 2-min:} \\
        $q_{1,TPF}$ &$\mathcal{N}(0.180,0.048)$ & 0.207 $^{+0.037}_{-0.036}$\\
        $q_{2,TPF}$ &$\mathcal{N}(0.182,0.034)$ & 0.181 $^{+0.036}_{-0.035}$\\   
        $D_{TPF}$ &$\mathcal{F}(1)$ & -  \\
        $M_{TPF}$ &$\mathcal{N}(0,0.1)$ & (-7.16 $^{+1.72}_{-1.75}$) $\times$ 10$^{-5}$ \\
        $\sigma_{w,TPF}$ (ppm) &log$\mathcal{U}(1, 1000)$ & 123 $^{+20}_{-25}$\\
        $\sigma_{GP,TPF}$ (relative flux) &log$\mathcal{U}(10^{-6},1)$ & (9.52 $^{+1.32}_{-1.17}$) $\times$ 10$^{-5}$\\
        $\rho_{GP,TPF}$ (days) &log$\mathcal{U}(0.005,0.26)$ & 0.102 $^{+0.040}_{-0.027}$\\
        \textbf{TESS QLP:} \\
        $q_{1,QLP}$ &$\mathcal{N}(0.180,0.048)$ & 0.207 $^{+0.037}_{-0.036}$\\
        $q_{2,QLP}$ &$\mathcal{N}(0.182,0.034)$ & 0.181 $^{+0.036}_{-0.035}$\\  
        $D_{QLP}$ &$\mathcal{F}(1)$ & - \\
        $M_{QLP}$ &$\mathcal{N}(0,0.1)$ & (-4.38 $^{+7.24}_{-7.49}$) $\times$ 10$^{-5}$ \\
        $\sigma_{w,QLP}$ (ppm) & $\mathcal{F}(0)$ & -\\  
        $\sigma_{GP,QLP}$ (relative flux) &log$\mathcal{U}(0.00005,1)$ & (19.74 $^{+5.93}_{-4.09}$) $\times$ 10$^{-5}$\\
        $\rho_{GP,QLP}$ (days) &log$\mathcal{U}(0.009,1.6)$ & 0.685 $^{+0.275}_{-0.192}$\\
        \textbf{RV:}  & & \\
        $\sigma_{w,CORALIE}$ ($\kms$) &log$\mathcal{U}(0.01, 0.6)$ & 0.316 $^{+0.062}_{-0.051}$ \\
        $\mu_{CORALIE}$ ($\kms$) &$\mathcal{U}(-13,-9)$ & -10.34 $^{+0.08}_{-0.07}$\\        
        $\sigma_{w,TRES}$ ($\kms$) &$\mathcal{F}(0)$ & - \\
        $\mu_{TRES}$ ($\kms$) &$\mathcal{U}(-3,0)$ & -2.12 $^{+0.06}_{-0.06}$\\ 
        &&\\
        &&\\
        \hline
       \end{tabular}
       \label{tab:Juliet629}
    \end{subtable}
    \hfill
\renewcommand{\arraystretch}{1.4}
    \begin{subtable}[h]{0.52\textwidth}
        \centering
        \begin{tabular}{lll}
        \hline\hline
        Parameter$^{1}$ &Prior$^{2}$ &Posterior estimate$^{3}$\\
        \hline 
        \textbf{TOI-2543} & &  \\
        \textbf{Star:} & &  \\
        $\rho_*$ (cgs)  &$\mathcal{N}(0.260,0.030)$ & 0.259 $^{+0.015}_{-0.022}$ \\
        \textbf{Companion:} & &  \\
$\it{P}$ (days) & $\mathcal{U}(7.4,7.7)$ & 7.542776 $^{+0.000031}_{-0.000031}$ \\
$T_C$ (\bjdtdb) & $\mathcal{U}(2459233.2,2459233.5)$ & 2459233.3450 $^{+0.0015}_{-0.0015}$ \\
$r_{1}$ & $\mathcal{U}(0, 1)$ & 0.4568 $^{+0.0883}_{-0.0813}$ \\
$r_{2}$& $\mathcal{U}(0, 1)$ & 0.05299 $^{+0.00077}_{-0.00076}$ \\
$\textit{e}$ &$\mathcal{U}(0, 1)$ & 0.0090 $^{+0.0028}_{-0.0022}$\\
$\omega_*$ (deg) &$\mathcal{U}(0, 360)$ & 328.1 $^{+19.7}_{-17.3}$\\
K (km$s^{-1}$) &$\mathcal{U}(4, 10)$ & 6.775 $^{+0.029}_{-0.029}$\\
        \textbf{TESS 2-min:} \\
        $q_{1,TPF}$ &$\mathcal{N}(0.303,0.065)$ & 0.316 $^{+0.061}_{-0.060}$\\
        $q_{2,TPF}$ &$\mathcal{N}(0.224,0.027)$ & 0.227 $^{+0.027}_{-0.027}$\\     
        $D_{TPF}$ &$\mathcal{F}(1)$ & -  \\
        $M_{TPF}$ &$\mathcal{N}(0,0.1)$ & (-1.71$^{+0.27}_{-0.26}$) $\times$ 10$^{-4}$ \\
        $\sigma_{w,TPF}$ (ppm) & $\mathcal{F}(0)$ & -\\ 
        $\sigma_{GP,TPF}$ (relative flux) &log$\mathcal{U}(10^{-7},1)$ & 0.305 $^{+0.323}_{-0.275}$\\
        $\rho_{GP,TPF}$ (days) &log$\mathcal{U}(0.0001,3.0)$ & 0.021 $^{+0.696}_{-0.020}$\\
        \textbf{TESS QLP:} \\
        $q_{1,QLP}$ &$\mathcal{N}(0.303,0.065)$ & 0.316 $^{+0.061}_{-0.060}$\\
        $q_{2,QLP}$ &$\mathcal{N}(0.224,0.027)$ & 0.227 $^{+0.027}_{-0.027}$\\   
        $D_{QLP}$ &$\mathcal{F}(1)$ & - \\
        $M_{QLP}$ &$\mathcal{N}(0,0.1)$ & (1.46 $^{+1.13}_{-1.26}$) $\times$ 10$^{-4}$ \\
        $\sigma_{w,QLP}$ (ppm) & $\mathcal{F}(0)$ & -\\  
        $\sigma_{GP,QLP}$ (relative flux) &log$\mathcal{U}(10^{-6},1)$ & (2.62 $^{+1.04}_{-0.70}$) $\times$ 10$^{-4}$\\
        $\rho_{GP,QLP}$ (days) &log$\mathcal{U}(0.001,2.0)$ & 0.712 $^{+0.367}_{-0.267}$\\
        \textbf{RV:}  & & \\
        $\sigma_{w,CORALIE}$ ($\kms$) &log$\mathcal{U}(0.001, 10.)$ & 0.184 $^{+0.064}_{-0.039}$ \\
        $\mu_{CORALIE}$ ($\kms$) &$\mathcal{U}(27.0, 45.0)$ & 34.17 $^{+0.06}_{-0.07}$\\        
        $\sigma_{w,CHIRON}$ ($\kms$) &$\mathcal{F}(0)$ & -\\
        $\mu_{CHIRON}$ ($\kms$) &$\mathcal{U}(28, 38)$ & 32.03 $^{+0.02}_{-0.02}$\\    
        $\sigma_{w,TRES}$ ($\kms$) &$\mathcal{F}(0)$ & - \\
        $\mu_{TRES}$ ($\kms$) &$\mathcal{U}(-10,3)$ & -6.91 $^{+0.07}_{-0.07}$\\ 
        \hline
        \end{tabular}
        \label{tab:Juliet2543}
     \end{subtable}
        \begin{tablenotes}
            \item\textbf{Notes:} $^{1}$ Parameter description: Density ($\rho_*$), Period (P), Time of conjunction ($T_C$), parametrization for p and b ($r_{1}$), parametrization for p and b ($r_{2}$), eccentricity of the orbit (e), argument of periastron ($\omega_*$), radial-velocity semi-amplitude of the companion (K), quadratic limb-darkening parametrization ($q_{1}$), quadratic limb-darkening parametrization ($q_{2}$), dilution factor (D), offset relative flux (M), jitter added in quadrature to the errorbars $\sigma_{w}$, amplitude of the GP ($\sigma_{GP}$), time/length-scale of the GP ($\rho_{GP}$), and systemic radial velocity ($\mu$).
            \\
         $^{2}$ For the priors, $\mathcal{N}$($\mu$, $\sigma^{2}$) indicates a normal distribution with mean $\mu$ and variance $\sigma^{2}$, $\mathcal{U}$(\textit{a}, \textit{b}) a uniform distribution between \textit{a} and \textit{b}, log$\mathcal{U}$(\textit{a}, \textit{b}) a log-uniform distribution between \textit{a} and \textit{b} and $\mathcal{F}$(\textit{a}) a parameter fixed to value \textit{a}.
         \\
         $^{3}$ The posterior estimate indicates the median value and then error bars the 68 $\%$ credibility intervals.
        \end{tablenotes}
     \label{tab:Juliet6292543}
\end{table}

\clearpage
\begin{table}[h]
\caption{Prior and posterior values for the joint photometric and radial velocity fit with \juliet of TOI-1107 (\textit{left}) and TOI-1982 (\textit{right}).}

\tiny
\renewcommand{\arraystretch}{1.3}
    \begin{subtable}[h]{0.52\textwidth}
        \centering
        \begin{tabular}{lll}
        \hline\hline
        Parameter$^{1}$ &Prior$^{2}$&Posterior estimate$^{3}$\\
        \hline 
        \textbf{TOI-1107} & &  \\
        \textbf{Star:} & &  \\
        $\rho_*$ (cgs)  &$\mathcal{N}(0.293,0.040)$ & 0.337 $^{+0.024}_{-0.026}$ \\
        \textbf{Companion:} & &  \\
$P$ (days) & $\mathcal{U}(4.0, 4.1)$ & 4.0782387 $^{+0.0000024}_{-0.0000025}$ \\
$T_C$ (\bjdtdb) & $\mathcal{U}(2459385.0,2459385.1)$ & 2459385.0123 $^{+0.0002}_{-0.0002}$ \\
$r_{1}$ & $\mathcal{U}(0, 1)$ & 0.4358 $^{+0.0709}_{-0.0662}$ \\
$r_{2}$& $\mathcal{U}(0, 1)$ & 0.07379 $^{+0.00033}_{-0.00030}$ \\
$\textit{e}$ &$\mathcal{U}(0, 1)$ & 0.0247 $^{+0.0234}_{-0.0164}$\\
$\omega_*$ (deg) &$\mathcal{U}(0, 360)$ & 71.97 $^{+20.48}_{-42.43}$\\
K (km$s^{-1}$) &$\mathcal{U}(0, 1)$ & 0.349 $^{+0.013}_{-0.013}$\\
        \textbf{TESS 2-min:} \\
        $q_{1,TPF}$ &$\mathcal{N}(0.306,0.064)$ & 0.219 $^{+0.032}_{-0.029}$\\
        $q_{2,TPF}$ &$\mathcal{N}(0.228,0.026)$ & 0.226 $^{+0.025}_{-0.025}$\\  
        $M_{TPF}$ &$\mathcal{N}(0,0.1)$ & (-1.30 $^{+0.36}_{-0.37}$) $\times$ 10$^{-4}$ \\
        $\sigma_{w,TPF}$ (ppm) &log$\mathcal{U}(10^{-4}, 1000)$ & 89.38 $^{+34.61}_{-26.09}$\\
        $\sigma_{GP,TPF}$ (relative flux) &log$\mathcal{U}(10^{-6},10^{-3})$ & (2.30 $^{+0.23}_{-0.19}$) $\times$ 10$^{-4}$\\
        $\rho_{GP,TPF}$ (days) &log$\mathcal{U}(0.1,1)$ & 0.435 $^{+0.071}_{-0.060}$\\
        \textbf{TESS QLP:} \\
        $q_{1,QLP}$ &$\mathcal{N}(0.306,0.064)$ & 0.219 $^{+0.032}_{-0.029}$\\
        $q_{2,QLP}$ &$\mathcal{N}(0.228,0.026)$ & 0.226 $^{+0.025}_{-0.025}$\\  
        $M_{QLP}$ &$\mathcal{N}(0,0.1)$ & (1.35 $^{+4.87}_{-4.65}$) $\times$ 10$^{-5}$ \\
        $\sigma_{w,QLP}$ (ppm) & $\mathcal{F}(0)$ & -\\  
        $\sigma_{GP,QLP}$ (relative flux) &log$\mathcal{U}(10^{-4},0.1)$ & (2.54 $^{+0.34}_{-0.27}$) $\times$ 10$^{-4}$\\
        $\rho_{GP,QLP}$ (days) &log$\mathcal{U}(0.1,1)$ & 0.511 $^{+0.110}_{-0.072}$\\
        \textbf{ASTEP $R_c$:} & &  \\       
        $q_{1,ASTEP}$ &$\mathcal{F}(0.3233)$ & -\\
        $q_{2,ASTEP}$ &$\mathcal{F}(0.2987)$ & -\\   
        $M_{ASTEP}$ &$\mathcal{N}(0,0.1)$ & 0.815 $^{+0.294}_{-0.215}$ \\
        $\sigma_{w,ASTEP}$ (ppm) &log$\mathcal{U}(0,3000)$ & 1021 $^{+76}_{-75}$\\ 
        \textbf{LCO-CTIO$_{Z}$:} & &  \\
        $q_{1,LCO_{Z}}$ &$\mathcal{F}(0.1849)$ & - \\
        $q_{2,LCO_{Z}}$ &$\mathcal{F}(0.2921)$ & - \\   
        $M_{LCO_{Z}}$ &$\mathcal{N}(0,0.1)$ & 1.978 $^{+0.004}_{-0.004}$\\
        $\sigma_{w,LCO_{Z}}$ (ppm) & log$\mathcal{U}(0,2000)$ & 713 $^{+51}_{-53}$ \\ 
        \textbf{Brierfield B:} & &  \\       
        $q_{1,Brierfield}$ &$\mathcal{F}(0.5976)$ & -\\
        $q_{2,Brierfield}$ &$\mathcal{F}(0.1935)$ & -\\   
        $M_{Brierfield}$ &$\mathcal{N}(0,0.1)$ & (-5.86 $^{+2.20}_{-1.94}$) $\times$ 10$^{2}$\\
        $\sigma_{w,Brierfield}$ (ppm) &log$\mathcal{U}(0,4000)$ & 2514 $^{+281}_{-275}$\\  
        \textbf{LCO-CTIO$_{g}$:} & &  \\
        $q_{1,LCO_{g}}$ &$\mathcal{F}(0.535962)$ & - \\
        $q_{2,LCO_{g}}$ &$\mathcal{F}(0.218820)$ & - \\   
        $M_{LCO_{g}}$ &$\mathcal{N}(0,0.1)$ & (2.33 $^{+3.21}_{-3.17}$) $\times$ 10$^{2}$\\
        $\sigma_{w,LCO_{g}}$ (ppm) & log$\mathcal{U}(0, 8000)$ & 6353 $^{+219}_{-208}$ \\ 
        $D_{All}$$^{4}$ &$\mathcal{F}(1)$ & -  \\
        \textbf{RV:}  & & \\
        $\sigma_{w,CORALIE}$ ($\kms$) &$\mathcal{F}(0)$ & - \\
        $\mu_{CORALIE}$ ($\kms$) &$\mathcal{U}(40,42)$ & 40.81 $^{+0.01}_{-0.01}$\\        
        $\sigma_{w,CHIRON}$ ($\kms$) & log$\mathcal{U}(0.001,1)$ & 0.046 $^{+0.016}_{-0.012}$\\
        $\mu_{CHIRON}$ ($\kms$) &$\mathcal{U}(-3,1)$ & 0.024 $^{+0.016}_{-0.015}$\\
        $\sigma_{w,MINERVA}$ ($\kms$) & $\mathcal{F}(0)$ & - \\
        $\mu_{MINERVA}$ ($\kms$) &$\mathcal{U}(39,41)$ & 40.19 $^{+0.02}_{-0.02}$\\
        \hline
       \end{tabular}
       \label{tab:Juliet1107}
    \end{subtable}
    \hfill
\renewcommand{\arraystretch}{1.3}
    \begin{subtable}[h]{0.52\textwidth}
        \centering
        \begin{tabular}{lll}
        \hline\hline
        Parameter$^{1}$ &Prior$^{2}$ &Posterior estimate$^{3}$\\
        \hline 
        \textbf{TOI-1982} & &  \\
        \textbf{Star:} & &  \\
        $\rho_*$ (cgs)  &$\mathcal{N}(0.549, 0.061)$ & 0.571 $^{+0.048}_{-0.047}$ \\
        \textbf{Companion:} & &  \\
$P$ (days) & $\mathcal{U}(17.0, 17.3)$ & 17.172446 $^{+0.000043}_{-0.000044}$ \\
$T_C$ (\bjdtdb) & $\mathcal{U}(2459323.7, 2459323.9)$ & 2459323.8230 $^{+0.0013}_{-0.0014}$ \\
$r_{1}$ & $\mathcal{U}(0, 1)$ & 0.8797 $^{+0.0083}_{-0.0085}$ \\
$r_{2}$& $\mathcal{U}(0, 1)$ & 0.06914 $^{+0.00095}_{-0.00096}$ \\
$\textit{e}$ &$\mathcal{U}(0, 1)$ & 0.2725 $^{+0.0138}_{-0.0143}$\\
$\omega_*$ (deg) &$\mathcal{U}(0, 360)$ & 268.15 $^{+1.24}_{-1.45}$\\
K (km$s^{-1}$) &$\mathcal{U}(0, 20)$ & 4.121 $^{+0.073}_{-0.069}$\\
        \textbf{TESS 2-min:} \\
        $q_{1,TPF}$ &$\mathcal{N}(0.299,0.066)$ & 0.291 $^{+0.061}_{-0.057}$\\
        $q_{2,TPF}$ &$\mathcal{N}(0.218,0.027)$ & 0.218 $^{+0.026}_{-0.027}$\\     
        $M_{TPF}$ &$\mathcal{N}(0,0.1)$ & (-2.33 $^{+1.13}_{-1.12}$) $\times$ 10$^{-4}$ \\
        $\sigma_{w,TPF}$ (ppm) & $\mathcal{F}(0)$ & - \\
        $\sigma_{GP,TPF}$ (relative flux) &log$\mathcal{U}(1, 6)$ $\times$ 10$^{-4}$ & (2.98 $^{+0.87}_{-0.62}$) $\times$ 10$^{-4}$ \\
        $\rho_{GP,TPF}$ (days) &log$\mathcal{U}(0.01,1)$ & 0.205 $^{+0.086}_{-0.060}$\\
        \textbf{TESS QLP:} \\
        $q_{1,QLP}$ &$\mathcal{N}(0.299,0.066)$ & 0.291 $^{+0.061}_{-0.057}$\\
        $q_{2,QLP}$ &$\mathcal{N}(0.218,0.027)$ & 0.218 $^{+0.026}_{-0.027}$\\    
        $M_{QLP}$ &$\mathcal{N}(0,0.1)$ & (1.0624 $^{+1.08}_{-1.10}$) $\times$ 10$^{-4}$ \\
        $\sigma_{w,QLP}$ (ppm) & $\mathcal{F}(0)$ & -\\  
        $\sigma_{GP,QLP}$ (relative flux) &log$\mathcal{U}(2, 5)$ $\times$ 10$^{-4}$ & (3.37 $^{+0.73}_{-0.57}$) $\times$ 10$^{-4}$\\
        $\rho_{GP,QLP}$ (days) &log$\mathcal{U}(0.1,0.5)$ & 0.299 $^{+0.093}_{-0.077}$\\
        \textbf{Hazelwood $R_c$:} & &  \\       
        $q_{1,Hazelwood}$ &$\mathcal{F}(0.12601)$ & -\\
        $q_{2,Hazelwood}$ &$\mathcal{F}(0.05544)$ & -\\   
        $M_{Hazelwood}$ &$\mathcal{N}(0,0.1)$ & (-2.73 $^{+1.08}_{-1.08}$) $\times$ 10$^{-3}$ \\
        $\sigma_{w,Hazelwood}$ (ppm) &log$\mathcal{U}(1900, 3000)$ & 2497 $^{+175}_{-155}$\\  
        \textbf{El Sauce $R_c$:} & &  \\       
        $q_{1,El-Sauce}$ &$\mathcal{F}(0.12601)$ &- \\
        $q_{2,El-Sauce}$ &$\mathcal{F}(0.05544)$ & - \\   
        $M_{El-Sauce}$ &$\mathcal{N}(0,0.1)$ & 0.093 $^{+0.002}_{-0.002}$ \\
        $\sigma_{w,El-Sauce}$ (ppm) & $\mathcal{F}(0)$& -\\  
        \textbf{LCO-CTIO$_{B}$:} & &  \\
        $q_{1,LCO_{B}}$ &$\mathcal{F}(0.61536)$ & - \\
        $q_{2,LCO_{B}}$ &$\mathcal{F}(0.35447)$ & - \\   
        $M_{LCO_{B}}$ &$\mathcal{N}(0,0.1)$ & (-3.74 $^{+2.68}_{-2.53}$) $\times$ 10$^{-3}$\\
        $\sigma_{w,LCO_{B}}$ (ppm) & log$\mathcal{U}(1000, 2400)$ & 1895 $^{+124}_{-110}$ \\  
        \textbf{LCO-CTIO$_{Z}$:} & &  \\
        $q_{1,LCO_{z}}$ &$\mathcal{F}(0.22016)$ & - \\
        $q_{2,LCO_{z}}$ &$\mathcal{F}(0.18173)$ & - \\   
        $M_{LCO_{z}}$ &$\mathcal{N}(0,0.1)$ &(-8.70 $^{+6.14}_{-6.39}$) $\times$ 10$^{-4}$\\
        $\sigma_{w,LCO_{z}}$ (ppm) & log$\mathcal{U}(1000, 3000)$ & 1480 $^{+94}_{-84}$\\    
        $D_{All}$$^{4}$ &$\mathcal{F}(1)$ & -  \\
        \textbf{RV:}  & & \\
        $\sigma_{w,CORALIE}$ ($\kms$) &log$\mathcal{U}(0.001, 2)$ & 0.170 $^{+0.057}_{-0.047}$ \\
        $\mu_{CORALIE}$ ($\kms$) &$\mathcal{U}(-46,-20)$ & -37.74 $^{+0.06}_{-0.06}$\\        
        $\sigma_{w,FEROS}$ ($\kms$) &$\mathcal{F}(0)$ & -\\
        $\mu_{FEROS}$ ($\kms$) &$\mathcal{U}(-45, -20)$ & -37.00 $^{+0.13}_{-0.14}$\\
        $\sigma_{w,TRES}$ ($\kms$) &$\mathcal{F}(0)$ & - \\
        $\mu_{TRES}$ ($\kms$) &$\mathcal{U}(-7, 8)$ & -0.113 $^{+0.211}_{-0.210}$\\
        \hline
        \end{tabular}
        \label{tab:Juliet1982}
     \end{subtable}
       \begin{tablenotes}
            \item\textbf{Notes:} The description of $^{1}$, $^{2}$ and $^{3}$ can be found in Table \ref{tab:Juliet6292543}.\\
            $^{4}$ The dilution factor was fixed to 1 for all the photometric data ($D_{All}$).
        \end{tablenotes}
     \label{tab:Juliet11071982}
\end{table}

\clearpage
\section{Spectroscopic binaries}\label{sec:falsepositives}
\begin{table*}[htp]
\small
\centering
\begin{tabular}{lcccc}
\hline\hline
Parameters & TOI-288 (TIC 47316976) & TOI-446 (TIC 1449640) & TOI-478 (TIC 172464366) & TOI-764 (TIC 181159386) \\
\hline
 M$_{\rm2}$ (\mjup) & 332.4 $\pm$ 41.8 & 178.8 $\pm$ 17.3 & 155.6 $\pm$ 12.0 &  264.3 $\pm$ 9.6\\
 R$_{\rm2}$ (\rjup) & - & 1.77 $\pm$ 0.15 & 1.78 $\pm$ 0.22 & 1.73 $\pm$ 0.20  \\
 $P$ (days) & 79.081 & 3.502 & 2.922 & 5.632\\
 K (\kms) & 13.047 $\pm$ 0.886 & 18.331 $\pm$ 0.459 & 16.861 $\pm$ 0.616 & 22.269 $\pm$ 0.175\\
 D (mmag) & - & 16.829 & 17.493 & 13.556\\
 ecc & 0.655 $\pm$ 0.010 & 0.039 $\pm$ 0.027 & 0.004 $\pm$ 0.009 & 0.175 $\pm$ 0.009\\
 M$_{\rm1}$ (\msol) & 1.38 $\pm$ 0.08 & 1.31 $\pm$ 0.08 & 1.30 $\pm$ 0.08 & 1.36 $\pm$ 0.08\\
 R$_{\rm1}$ (\rsol) & 1.86 $\pm$ 0.18 & 1.46 $\pm$ 0.18 & 1.44 $\pm$ 0.18 & 1.59 $\pm$ 0.18\\
 T$_{eff}$ (K) & 6380 $\pm$ 110 & 6477 $\pm$ 110 & 6452 $\pm$ 110 & 6557 $\pm$ 110\\
\hline
\end{tabular}
\caption{List of Single lined spectral Binary (SB1) systems detected with CORALIE spectrograph.}
\label{tab:sb1}
        \begin{tablenotes}
            \item\textbf{Notes:}  Parameters of the primary star are noted with $_{1}$ and of the secondary star with $_{2}$ (Section \ref{sec:EBLMs}).\\
            Parameter description: Stellar mass (M$_{\rm1,2}$), stellar radius (R$_{\rm1,2}$), orbital period (P), RV semi-amplitude (K), depth (D), eccentricity (e), and primary stellar effective temperature (\teff).
        \end{tablenotes}
\end{table*}

\begin{figure*}[h]
  \centering
  \includegraphics[width=0.45\textwidth]{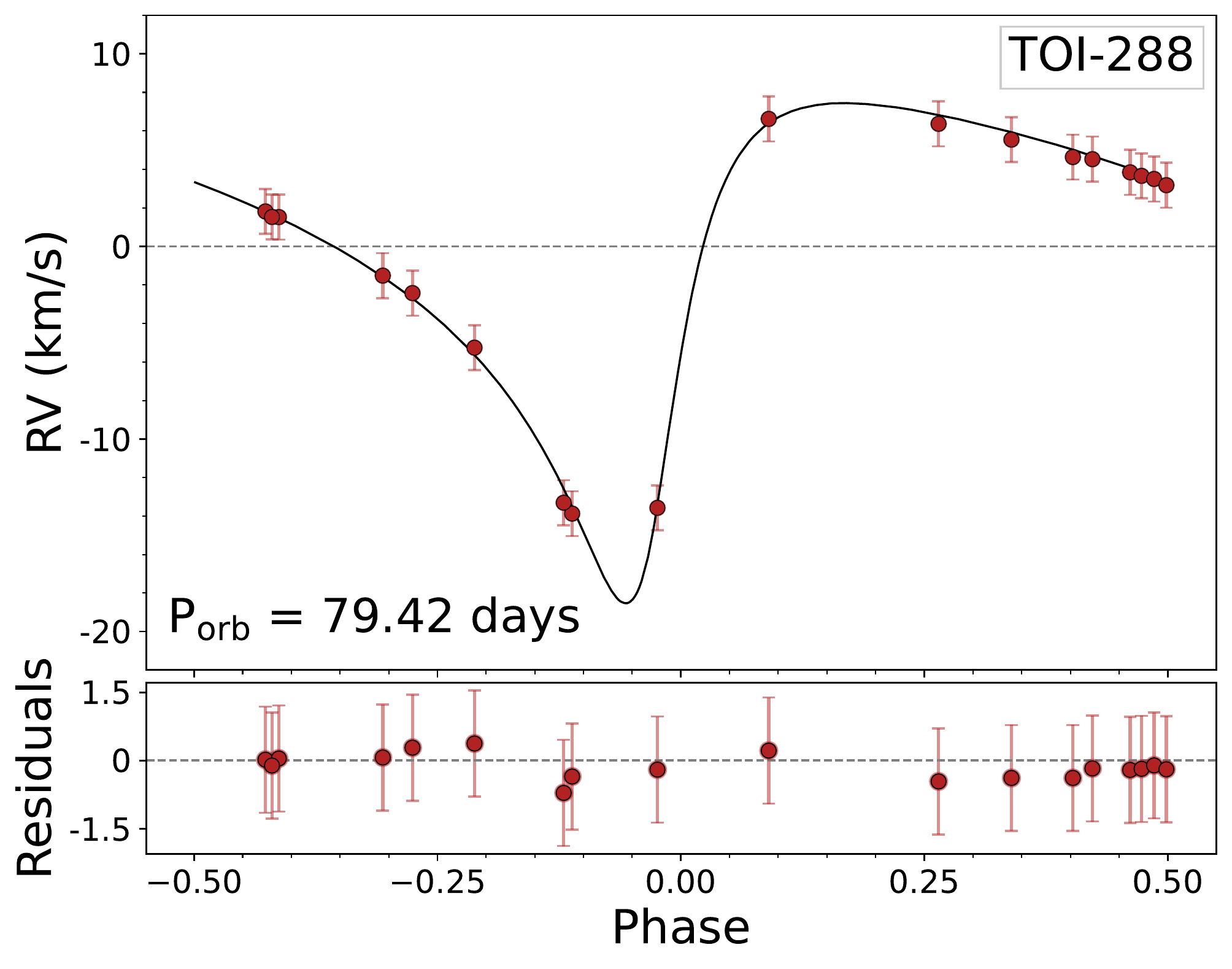}
  \includegraphics[width=0.45\textwidth]{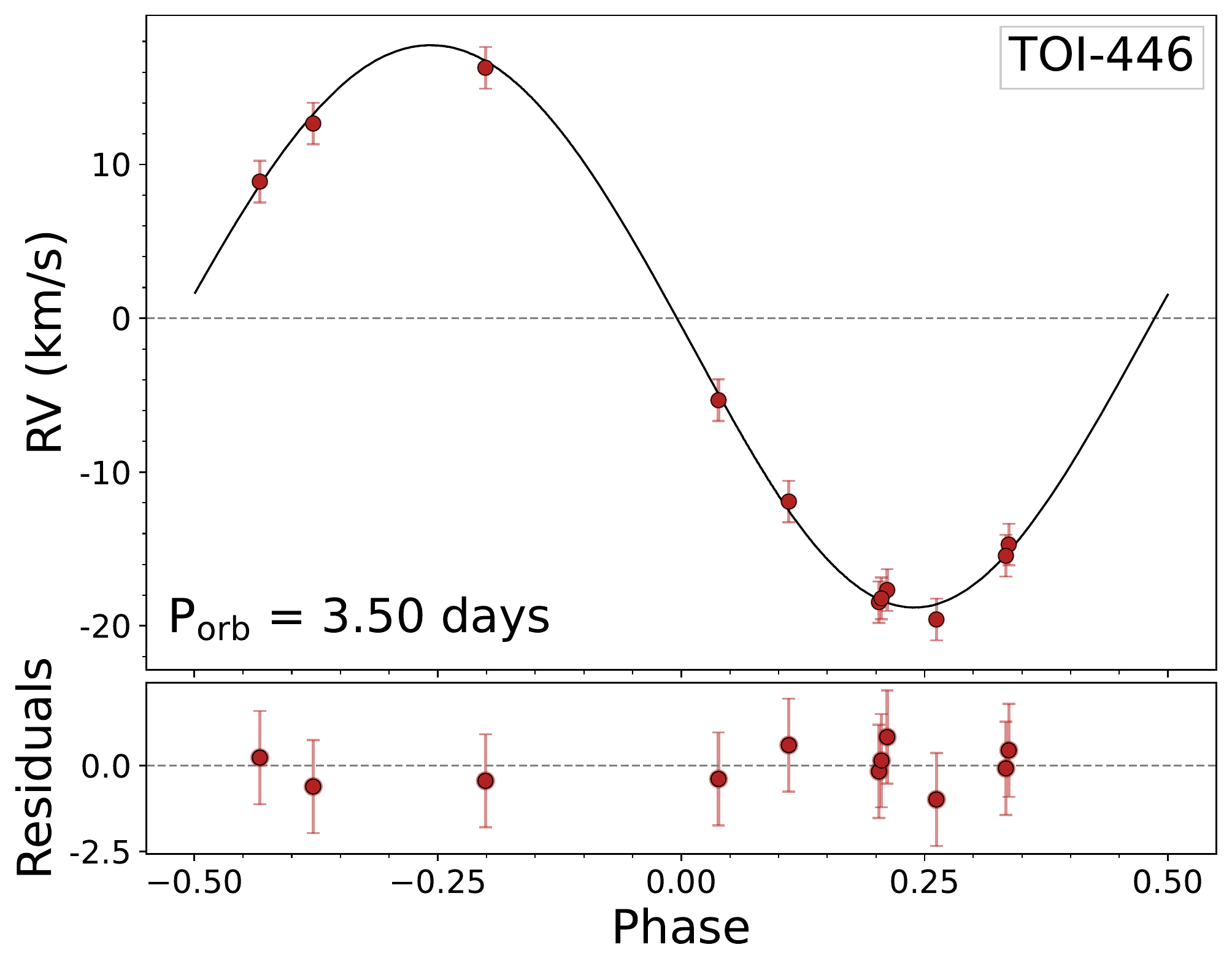}
  \\
  \includegraphics[width=0.45\textwidth]{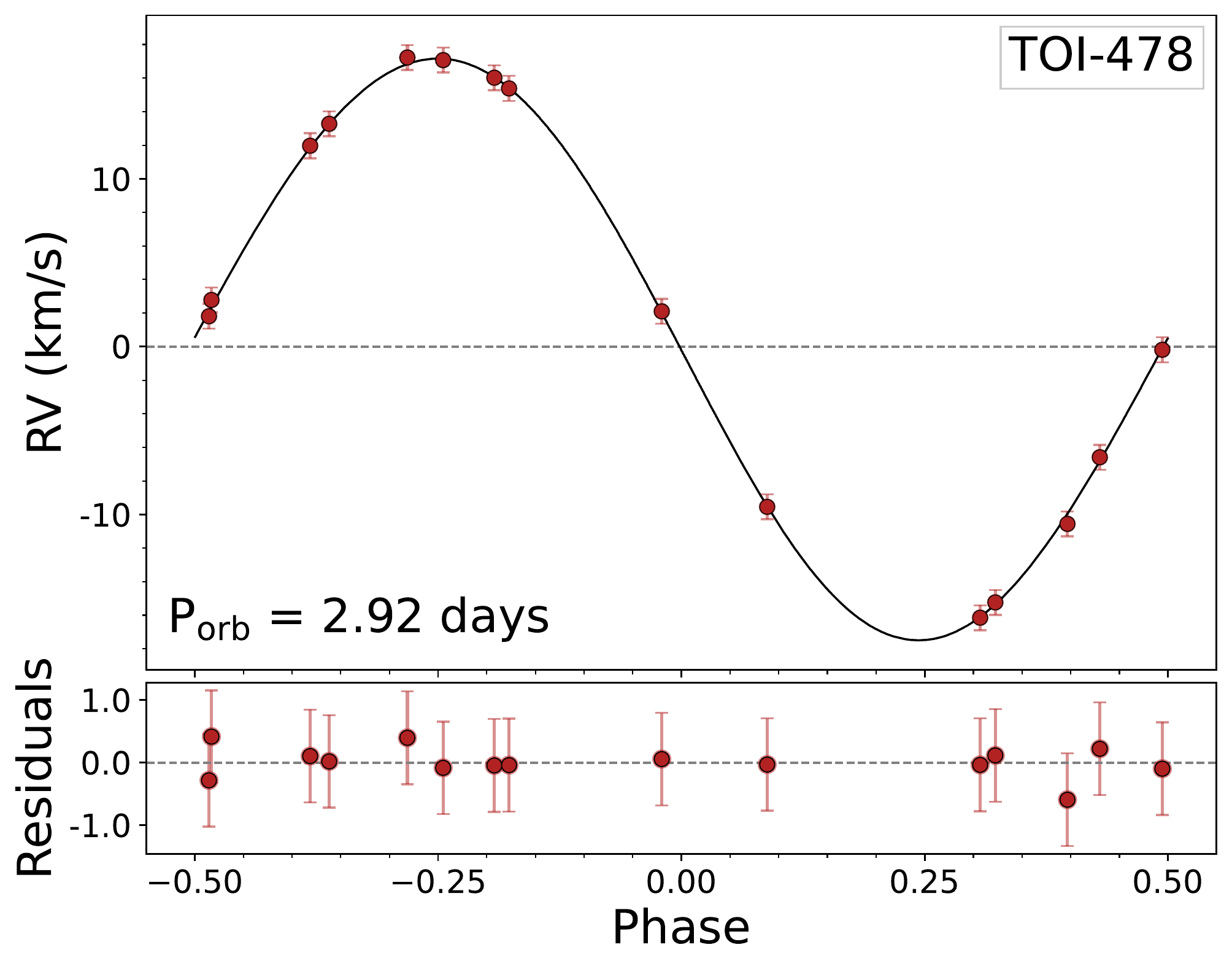}
  \includegraphics[width=0.45\textwidth]{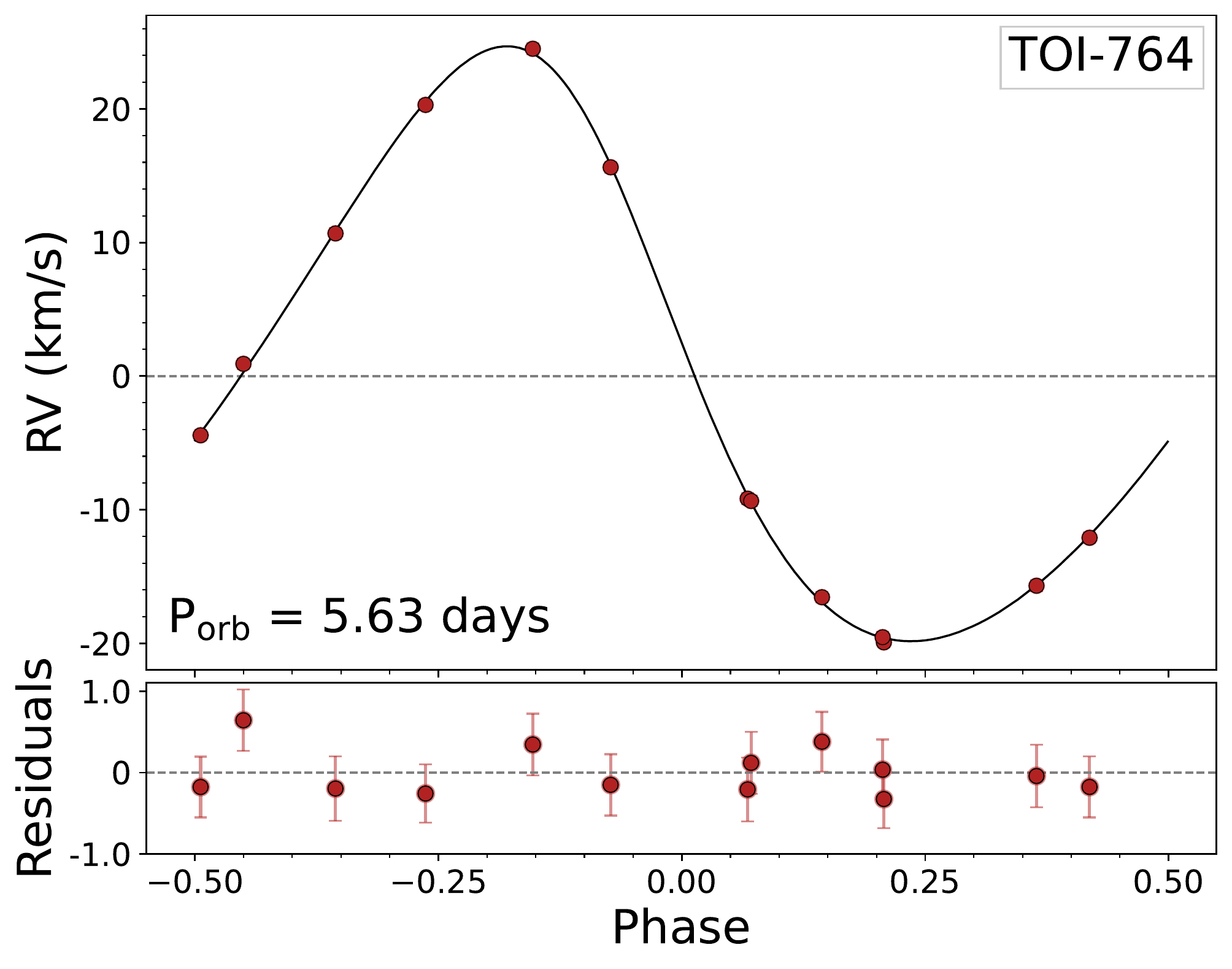}
  \caption{Phase-folded radial velocities of TOI-288, TOI-446, TOI-478, and TOI-764. The black line displays the model fit derived with \juliet. The residuals of the model fit are shown in the bottom panels.}
  \label{fig:SB1s}
\end{figure*}
\end{appendix}

\end{document}